\documentclass[reprint,superscriptaddress,aps,pra]{revtex4-2}

\usepackage{xcolor}
\usepackage{etoolbox} 
\usepackage{graphicx}
\usepackage{amsmath}
\usepackage{amssymb}
\usepackage{bm}

\begin{document}

\title{Semantic Segmentation of Anomalous Diffusion Using Deep Convolutional Networks}

\author{Xiang Qu}
\affiliation{School of Physics and Electronics, Hunan University, Changsha 410082, China}
\author{Yi Hu}
\affiliation{Hubei Medical Devices Quality Supervision and Test Institute, Wuhan 430075, China}
\author{Wenjie Cai}
\affiliation{School of Physics and Electronics, Hunan University, Changsha 410082, China}
\author{Yang Xu}
\affiliation{Hubei Medical Devices Quality Supervision and Test Institute, Wuhan 430075, China}
\author{Hu Ke}
\affiliation{Hubei Medical Devices Quality Supervision and Test Institute, Wuhan 430075, China}
\author{Guolong Zhu}
\affiliation{School of Physics and Electronics, Hunan University, Changsha 410082, China}
\author{Zihan Huang}
 \email{huangzih@hnu.edu.cn}
\affiliation{School of Physics and Electronics, Hunan University, Changsha 410082, China}


\begin{abstract}
Heterogeneous dynamics commonly emerges in anomalous diffusion with intermittent transitions of diffusion states but proves challenging to identify using conventional statistical methods. To effectively capture these transient changes of diffusion states, we propose a deep learning model (U-AnDi) for the semantic segmentation of anomalous diffusion trajectories. {This model is developed with the dilated causal convolution (DCC), gated activation unit (GAU), and U-Net architecture. The study addresses two key subtasks related to trajectory segmentation and changepoint detection, concentrating on variations in diffusion exponents and dynamic models. Additionally, extended analyses are conducted on the segmentation of single-model trajectories, multi-state biological trajectories, and anomalous diffusion with added long-time correlations. By rationally designing comparative models and evaluating the performance of U-AnDi against these models, we discover that U-AnDi consistently outperforms other models across all segmentation tasks, thereby affirming its superiority in the field. This performance edge also sheds light on the interpretability of U-AnDi's core components: DCC, GAU, and U-Net. The clarity with which these components contribute to U-AnDi's success underscores their congruence with the intrinsic physics underlying anomalous diffusion.} Furthermore, our model is examined using real-world anomalous diffusion data: the diffusion of transmembrane proteins on cell membrane surfaces, and the segmentation results are highly consistent with experimental observations. Our findings could offer a heuristic deep learning solution for the detection of heterogeneous dynamics in single-molecule/particle tracking experiments, and have the potential to be generalized as a universal scheme for time-series segmentation.
\end{abstract}

\maketitle

\section{Introduction}\label{introduction}

Anomalous diffusion \cite{Metzler, Klafter}, characterized by the non-linear transport phenomena that deviate from standard Brownian motion, has drawn significant attention due to its widespread occurrence in various scientific fields, such as physics \cite{Mason,Reis,Volpe,Sagi,Hartich}, chemistry \cite{Barkai,Ding,Song,Xu}, biology \cite{Hofling,Wu,Gonzalez,Wang,Chen}, and finance \cite{Plerou,Masoliver,Jiang}. Since anomalous diffusion can provide valuable insights into the underlying mechanisms of complex systems, it is of great importance to perform accurate and reliable analysis of this phenomenon \cite{Sposini,Vilk,Wang1,Seckler,Borja,Munoz-Gil,Manzo,Bo,Kowalek,Firbas,Munoz-Gil1,Gentili,Argun,LearningVerdier,Li,verdier2022,Simulation-based,Kabbech,multistep,ParticleM,seckler2023machine,MEIJERING}. However, due to the heterogeneous dynamics with intermittent transitions of diffusion states \cite{Chen1,Chen2,Persson,Monnier}, conventional statistical methods usually fail to precisely describe the anomalous diffusion processes in real-world experimental environments. This emphasizes the necessity of exploring effective techniques for the semantic segmentation of anomalous diffusion trajectories, which could offer a more fine-grained approach to detect state transitions and distinguish heterogeneous behaviors in diffusion dynamics.

On the other hand, as a pivotal technique in computer vision, semantic segmentation makes pixel-wise predictions in an image to enable the identification and delineation of distinct regions \cite{Mo}. With the advent of deep learning techniques, semantic segmentation methods have experienced a rapid development in recent years, significantly outperforming traditional approaches that rely on handcrafted features \cite{yu2015multi, He_2017_ICCV1,Lin_2017_CVPR1,Badrinarayanan,Li-Jia,Olaf,Antonelli,Getao}. These deep-learning-based methods have demonstrated remarkable success in a wide array of applications, including object recognition \cite{yu2015multi, He_2017_ICCV1,Lin_2017_CVPR1,Badrinarayanan}, scene understanding \cite{Li-Jia}, and medical image analysis \cite{Olaf,Antonelli,Getao}. In particular, convolutional neural networks (CNNs) have shown unparalleled capabilities in capturing complex patterns and hierarchical representations from raw data, leading to state-of-the-art results in various semantic segmentation tasks \cite{Mo}. One representative example is the U-Net architecture \cite{Olaf}, which is built on the basis of an encoder-decoder structure and the rational use of skip connections. U-Net has emerged as one of the most popular and effective frameworks for semantic segmentation and proven to be highly effective in diverse segmentation tasks \cite{Getao}.

In this work, motivated by the successful applications of CNNs in semantic segmentation, we explore the potential of employing deep convolutional networks for the semantic segmentation of anomalous diffusion. {We introduce a novel deep learning model, U-AnDi, which synergizes the dilated causal convolution (DCC), gated activation unit (GAU), and U-Net architecture. The model's capabilities are validated across two pivotal subtasks of trajectory segmentation, based on variable diffusion exponents and different dynamic models, respectively. Moreover, the adaptability of U-AnDi is demonstrated by thoroughly analyzing both single-model trajectories and multi-state biological trajectories, as well as diffusion exhibiting extended long-time correlations. The results manifest that U-AnDi excels in all these segmentation tasks, outperforming other comparative models. The superior performance of U-AnDi underscores its effectiveness and the interpretability of its core components: DCC, GAU, and U-Net, in reflecting the model's alignment with the nature of anomalous diffusion.} Furthermore, the generalization ability of U-AnDi for real-world data is evaluated through experimental observations of the diffusion of transmembrane proteins on cell membrane surfaces \cite{Granik}. The segmentation results exhibit a high degree of consistency with experimental observations, highlighting the applicability of our approach to characterizing anomalous diffusion in real complex systems.

The rest of this paper is organized as follows. Sec. \ref{task} describes the segmentation tasks for anomalous diffusion, methods for generating simulated trajectories, and main evaluation metrics used in this work. Sec. \ref{model} presents the architecture and core components of U-AnDi and its design principles. Sec. \ref{analysis} provides an analysis of U-AnDi's efficacy on simulated trajectories and compares its performance with that of other models. The interpretability of U-AnDi's core components is consolidated in Sec. \ref{inter}. In Sec. \ref{realworld}, we showcase the implementation of our method on experimental trajectories of transmembrane protein on membrane surfaces. Finally, we engage in a discussion and draw our conclusions in Sec. \ref{conclusion}.

\begin{figure*}
\centering
\includegraphics[width=16.6cm]{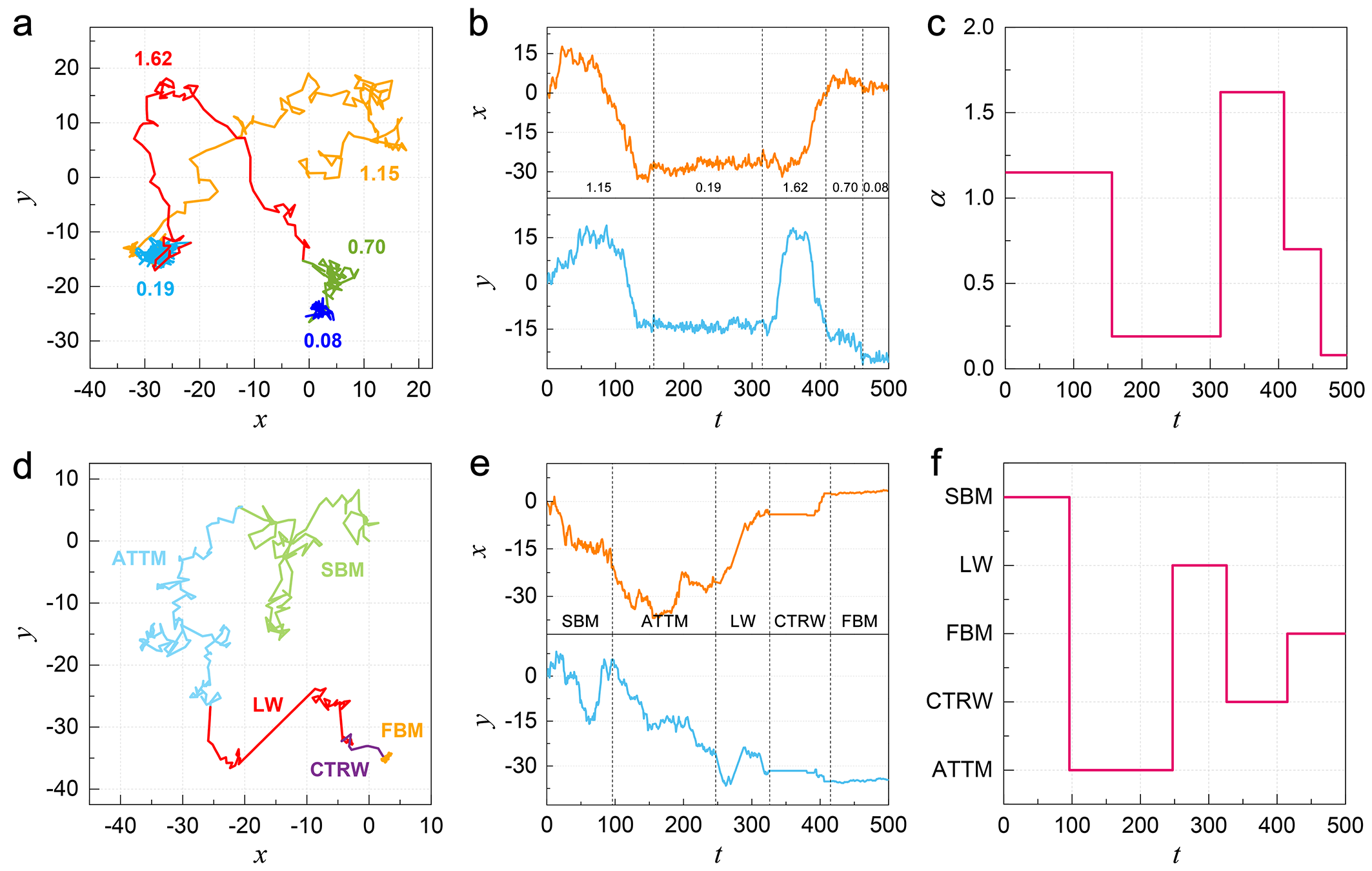}
\caption{[(a), (d)] Representative examples of trajectories with varying diffusion exponents (a) and different dynamics models (d). Different segments are marked in distinct colors. [(b), (e)] Time evolutions of trajectories in (a) and (d) for both $x$ and $y$ dimensions, respectively. Changepoints of diffusion states are indicated by dashed lines. [(c), (f)] Point-wise labels of trajectories in (a) and (d), respectively.}
\label{fig:fig1}
\end{figure*}

\section{Tasks and Datasets}\label{task}

In this section, we present the details of two subtasks for the semantic segmentation of anomalous diffusion and corresponding simulated datasets used for model training and evaluation. These tasks aim to effectively address the challenge of capturing the intermittent transitions among different diffusion states. Data generation for both tasks is facilitated by utilizing the open-source Python package, {\it andi-datasets}, which is specifically designed for generating, managing, and analyzing anomalous diffusion trajectories \cite{Munoz-Gil, Munoz-Gil2}. In particular, we primarily focus on 2D trajectory data in this work, as most experimental observations of anomalous diffusion are performed in 2D or quasi-2D environments.

\subsection{Subtask 1: Segmentation of trajectories with varying diffusion exponents}
The first subtask focuses on the segmentation of trajectories in which the sub-trajectories exhibit varying diffusion exponents. Here, the diffusion exponent $\alpha$ quantifies the relation between mean squared displacement (MSD) and time {$t$}, written as
{\begin{equation}
{\rm MSD} \sim t^\alpha.
\end{equation}
}This exponent equals 1 in the description of standard Brownian motion. For anomalous diffusion, it can be either less than or greater than 1, corresponding to sub-diffusion and super-diffusion, respectively. In real-world scenarios, the diffusion exponent may not remain constant due to inevitable changes of interactions between the random walker and its surroundings \cite{Chen2}. However, traditional statistical methods, which are based on the calculation of MSD, require sufficiently long trajectories to accurately determine the diffusion exponent. As a result, for anomalous diffusion trajectories with limited length scales, these conventional approaches face substantial difficulties when attempting to distinguish segments with different diffusion exponents.

To address this challenge, subtask 1 is designed to empower our model with the capability to identify these segments and predict their corresponding diffusion exponents. In detail, simulated segments are generated by the {\it andi-datasets} package within the theoretical framework of fractional Brownian motion (FBM) \cite{Mandelbrot}. This diffusion model describes an ergodic diffusion process that is driven by a fractional Gaussian noise, where the diffusion exponent $\alpha$ satisfies $0 < \alpha <2$. When generating segments, their diffusion coefficients are fixed at 1.0, while diffusion exponents $\alpha$ are drawn uniformly from the interval $[0.05, 2)$. {Here, to quantify the difference in diffusion exponents between two adjacent segments, we introduce $\delta \alpha=\alpha_2-\alpha_1$, where $\alpha_1$ and $\alpha_2$ represent the exponents of front and rear segments, respectively. In particular, a minimum value for $|\delta \alpha|$ is set as $\delta \alpha_{\min}$ to ensure a distinct transition between diffusion states. The default value of $\delta \alpha_{\min}$ is 0.5 in this work.}

After that, {$M$ segments} are combined to form a trajectory of length $L$. Here, $M$ is randomly selected from the values 2, 3, 4, and 5. {The length of each segment, $T$, is determined from a uniform distribution, with its bounds set between a minimum length $T_{\min}=10$ and a maximum length $T_{\max} = L-(M-1)T_{\min}$.} Representative example of a single simulated trajectory with $L=500$ and $M=5$ is illustrated in Fig. \ref{fig:fig1}(a), where different segments are marked in distinct colors. Time evolution of this trajectory is depicted in Fig. \ref{fig:fig1}(b) for both the $x$ and $y$ dimensions, indicating that the trajectory data exhibits typical time-series characteristics. We assign the diffusion exponents of segments to each point as point-wise labels, leading to a point-wise regression task for our model. As shown in Fig. \ref{fig:fig1}(c), these labels enable the clear identification of transition points (changepoints) among different diffusion states.

\subsection{Subtask 2: Segmentation of trajectories with different dynamics models}\label{task2}

The second subtask aims to segment trajectories composed of segments, each with dynamics originating from a distinct diffusion model. It has been proven that the dynamics of anomalous diffusion is multifaceted and can be described by a variety of theoretical stochastic process models. Similar to the varying diffusion exponents in subtask 1, the anomalous diffusion dynamics observed in the real world often undergoes transitions due to changes or fluctuations in the surrounding media \cite{Chen1,Granik}. For instance, the diffusion of transmembrane proteins on the cell membrane surface \cite{Granik,Chein} exhibits a duality of dynamics, encompassing both FBM and continuous-time random walk \cite{Scher}. However, traditional statistical methods are not particularly adept at identifying the dynamics model from raw trajectory data, and the detection of transitions among these dynamics proves even more challenging. Therefore, developing effective methods to address this issue bears considerable significance for the analysis of anomalous diffusion behaviors and the elucidation of underlying physical mechanisms.

For that purpose, we design the subtask 2 with the goal of enabling our model to detect state changepoints in the diffusion process and accurately identify the dynamics models of segments. Five theoretical diffusion models are considered:
\begin{itemize}
\item Annealed transient time motion (ATTM) \cite{Massignan}: Brownian motion with a diffusion coefficient that varies randomly in either time or space $(0.05 \leq \alpha \leq 1)$.
\item Continuous-time random walk (CTRW) \cite{Scher}: The waiting time between two consecutive steps is irregular and randomly chosen $(0.05 \leq \alpha \leq 1)$.
\item Fractional Brownian motion (FBM) \cite{Mandelbrot}: Diffusion process driven by a power-law correlated fractional Gaussian noise $(0.05 \leq \alpha < 2)$.
\item L\'{e}vy walk (LW) \cite{Klafter1}: The waiting time between subsequent steps is irregular, while the step length is not Gaussian distributed $(1 \leq \alpha \leq 2)$.
\item Scaled Brownian motion (SBM) \cite{Lim}: Brownian motion with a deterministically time-dependent diffusion coefficient $(0.05 \leq \alpha \leq 2)$.
\end{itemize}
Under the guidance of these 5 models, segments are generated using the {\it andi-datasets} package. The procedure for forming a trajectory in this subtask is the same as in subtask 1, {i.e., $M$ segments are joined to form a trajectory of length $L$}. The parameter values are also consistent with those in subtask 1. As an illustrative example, we present a typical sample of a trajectory with $L=500$ and $M=5$ in Fig. \ref{fig:fig1}(d) and its corresponding time evolution for both $x$ and $y$ dimensions in Fig. \ref{fig:fig1}(e). This trajectory with different dynamics models is also labeled in a point-wise manner, as depicted in Fig. \ref{fig:fig1}(f).

\begin{figure*}
\centering
\includegraphics[width=16.8cm]{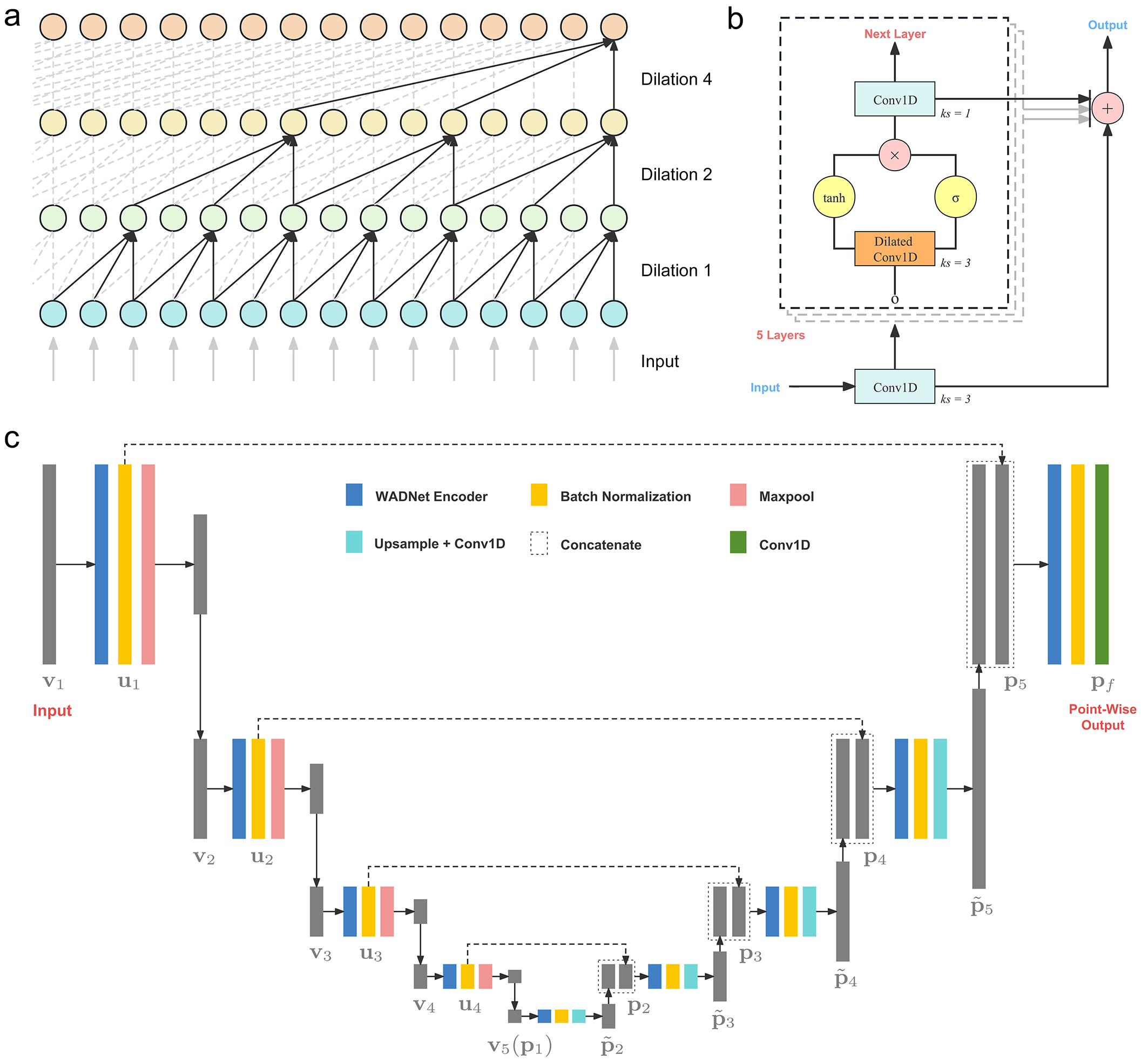}
\caption{(a) Schematic diagram of the dilated causal 1D convolution with a kernel size of 3. The dilation factor of convolution is doubled layer by layer. (b) Detailed structure of the WADNet encoder. (c) The architecture of our U-AnDi model. Skip connections are highlighted by the dashed lines.}
\label{fig:fig2}
\end{figure*}

\subsection{{Evaluation Metrics}}

The main evaluation metric for subtask1 is the MAE:
\begin{equation}
\mathrm{MAE}=\frac{1}{NL} \sum_{i=1}^N\sum_{t=1}^L\left|\alpha_{i, t}^{\rm Pred}-\alpha_{i, t}^{\rm GT}\right|.
\end{equation}
Here, $N$ is the number of trajectories to be evaluated. For the $i$th trajectory at time step $t$, $\alpha_{i, t}^{\rm Pred}$ and $\alpha_{i, t}^{\rm GT}$ denote the predicted and ground truth values of the diffusion exponent, respectively.

For subtask2, we select the mean Dice coefficient (mDice) as the main evaluation metric. The Dice coefficient, also known as the S{\o}rensen-Dice coefficient, can measure the similarity of two samples and is commonly utilized in segmentation tasks \cite{Dice}. This metric can be calculated in subtask 2 as:
\begin{equation}
{\rm mDice} = \frac{1}{N}\sum_{i=1}^N\frac{2{\sum}\left({\bf X}^{\rm Pred}_i\otimes{\bf X}^{\rm GT}_i\right)}{{\sum}{\bf X}^{\rm Pred}_i+{\sum}{\bf X}^{\rm GT}_i}.
\end{equation}
Here, ${\bf X}^{\rm Pred}_i$ is the predicted point-wise label and ${\bf X}^{\rm GT}_i$ refers to the ground truth of the $i$th trajectory. {Both are one-hot encoded as matrices of dimensions $[L, n_c]$ for the Dice coefficient calculation, where $n_c$ is the total number of diffusion-model types.} The symbol $\otimes$ denotes the element-wise multiplication operator, {and the small $\sum$ in the fraction represents the summation of all elements in the matrix.} The value of the mean Dice coefficient ranges from 0 to 1, with a higher value indicating a better segmentation performance.

{On the other hand, pinpointing the changepoints among diffusion states is of significant importance with respect to the segmentation of anomalous diffusion. However, when performing the changepoint detection, the model may generate some non-existent changepoints (false positives, FPs), while potentially missing genuine changepoints (false negatives, FNs). This inevitably leads to errors in the identification of state transitions. Hence, to systematically quantify the model's performance in detecting these transitions, we define $\delta c$ as:
\begin{equation}
\delta c = c^{\rm Pred} - c^{\rm GT},
\end{equation}
where $c^{\rm Pred}$ and $c^{\rm GT}$ are the predicted and ground truth changepoints, respectively. For the model-predicted changepoints, we categorize those with $|\delta c| < T_{\min}$ as true positives (TPs) and those with $|\delta c| \geq T_{\min}$ as false positives (FPs). The number of FNs can be obtained by subtracting the count of true positives (TPs) from the total number of actual changepoints. After that, we employ the metrics below to further evaluate model's performance in changepoint detection for both subtasks} {\cite{Li}}:
{\begin{gather}
{\text {precision}} = \frac{\rm TP}{\rm TP+FP}\\
{\text {recall}} = \frac{\rm TP}{\rm TP+FN}\\
{\text{F1-score}}  = 2\cdot\frac{ {\rm Precision}\cdot{\rm Recall}}{\rm Precision + Recall}
\end{gather}}

\section{Model Structure and Design Principles}\label{model}

The U-AnDi model proposed in this work is constructed by integrating the WADNet encoder \cite{Li} into the U-Net architecture. {This integration is primarily driven by the aim to effectively capture the intrinsic dynamics of anomalous diffusion, particularly the long-time correlations, with the help of dilated causal convolution (DCC) and gated activation unit (GAU). Another motivation is the utilization of U-Net's adeptness at capturing local features and nuances for semantic segmentation tasks, which complements the WADNet encoder's focus on long-range dependencies. In the following parts, we will introduce the design principles of U-AnDi in detail, and outline their connections with the underlying physics of anomalous diffusion.} The detailed implementation of U-AnDi can be found in Appendix \ref{modeldetail}, while the associated training scheme is outlined in Appendix \ref{training}.

\subsection{{Dilated causal convolution}}

{Considering that the motion of a random walker is solely determined by its current and past states, preserving causality to prevent using future information when processing the anomalous diffusion data is not just a technical requirement but a fundamental necessity. This crucial aspect is highlighted by Verdier {\it et al}. in their recent work \cite{verdier2022}, where the causality is maintained by selecting incoming edges of each node originate only from nodes in the past in a graph neural network (GNN). Unlike the specific wiring scheme used in the GNN model, U-AnDi employs causal convolution to preserve the causality inherent in anomalous diffusion.} Specifically, when applying the causal convolution to calculate the output at time step $t$, only the data from the preceding $t$ steps in the previous layer will be used. This can be achieved by applying padding on both sides of the sequence.

{On the other hand, as the long-time correlation is one of the essential properties of anomalous diffusion, our design should empower U-AnDi with the capability to capture these extended dependencies in trajectory data. However, the receptive field of a standard causal convolution is limited, posing challenges in capturing the long-range dependencies in longer trajectories. To address this issue, dilated convolution is concurrently employed with causal convolution within the WADNet encoder. As highlighted in Ref. \cite{yu2015multi}, dilated convolutions can systematically aggregate multi-scale contextual information without compromising resolution by harnessing the exponentially expanding receptive fields. This is particularly crucial when dealing with anomalous diffusion data, where discerning long-time tails and correlations is of utmost importance.} A schematic representation of a dilated causal 1D convolution is given in Fig. 2(a), where the kernel size ($ks$) is set as 3. Notably, by doubling the dilation factor layer by layer, after applying convolutions for $d$ (dilation depth) times, a single node can cover the information of $2^{d+1}-1$ nodes in the input layer. {This exponential expansion of receptive capability enables our model to encapsulate more long-range correlation information within the feature map. Consequently, it assists the model in appropriately balancing the sampling for measurements associated with long-time tails with respect to anomalous diffusion.}

\begin{table}
\caption{\label{tab:table1}
{Overview of comparative models, their respective abbreviations, and the presence (highlighted by $\checkmark$) or absence of key components: dilated causal convolution (DCC), gated activation unit (GAU), and U-Net architecture.}}
\begin{ruledtabular}
\begin{tabular}{lcccc}
\multicolumn{1}{c}{Model} & Abbr. & DCC & GAU & U-Net\\
\hline
Long short-term memory \cite{Hochreiter} & LSTM & ~ & \checkmark & ~\\
Gated recurrent unit \cite{Kyunghyun} & GRU & ~ & \checkmark & ~\\
Transformer \cite{Ashish} & TFM & ~ &~ &~ \\
{WADNet encoder} \cite{Li} & WE & \checkmark &\checkmark &~ \\
{U-CNN} & U-CNN & \checkmark &~ &\checkmark
\end{tabular}
\end{ruledtabular}
\end{table}

\subsection{{Gated activation unit}}

{While the exponentially expanding receptive field enhances U-AnDi's ability in collecting long-range correlation information, effectively processing this information and subsequently inferring the intrinsic properties of anomalous diffusion remains a challenge. Recognizing that anomalous diffusion often exhibits correlations where the influence of a past event decays slowly over time, the gated activation unit \cite{Hochreiter,Kyunghyun,Aaron}, with its capacity to maintain memory and manage information flow, is well-suited to capture these slowly decaying dependencies. Here, the gating mechanism within the GAU plays a pivotal role. Essentially, it acts as a regulatory system, determining which information should be allowed to pass through and which should be retained or discarded. Given the nature of anomalous diffusion, where certain events or states have prolonged effects, it's crucial for a model to discern which pieces of information are pertinent over extended periods. The gating mechanism allows the model to ``remember" significant events from the past and "forget" or downplay less relevant ones. This selective retention on certain data points aligns perfectly with the characteristics of anomalous diffusion, ensuring that the model remains sensitive to long-time correlations. Therefore, to adeptly process the long-range information gathered by the dilated causal convolution, we integrate the GAU into our network encoder.} The detailed structure of this encoder is depicted in Fig. 2(b), with a comprehensive description in a mathematical form provided in Appendix \ref{modeldetail}.

\subsection{{U-Net architecture}}

{For semantic segmentation tasks, the precise delineation of boundaries between distinct objects or states necessitates a model's capability to effectively capture local variations. This aspect is as crucial as apprehending long-time correlations when it comes to the semantic segmentation and transition detection in anomalous diffusion. The reason is that transitions in diffusion states typically don't manifest as slow and continuous processes but rather occur within short time intervals.}

{Therefore, to enhance the model's ability in detecting these local changes, we integrate the WADNet encoder into the U-Net architecture. Such an architecture, originally designed for biomedical image segmentation, has garnered significant attention due to its exceptional performance in various segmentation tasks. As illustrated in Fig. 2(c), U-Net utilizes a symmetric encoder-decoder architecture, ensuring that spatial information diminished during encoding is effectively reinstated during decoding. Moreover, skip connections in U-Net transfer fine-grained details from the encoder directly to the decoder, ensuring that local features and nuances are preserved and enhanced in the output. That is, U-Net captures fine details through skip connections and grasps the broader context with its encoder-decoder structure. This dual capability makes it well-suited for pinpointing state transition points in anomalous diffusion trajectories, where subtle local variations and long-range dependencies are closely interwoven.} Details of constructing the U-Net architecture in this work are also provided in Appendix A.

\begin{figure}
\centering
\includegraphics[width=8.6cm]{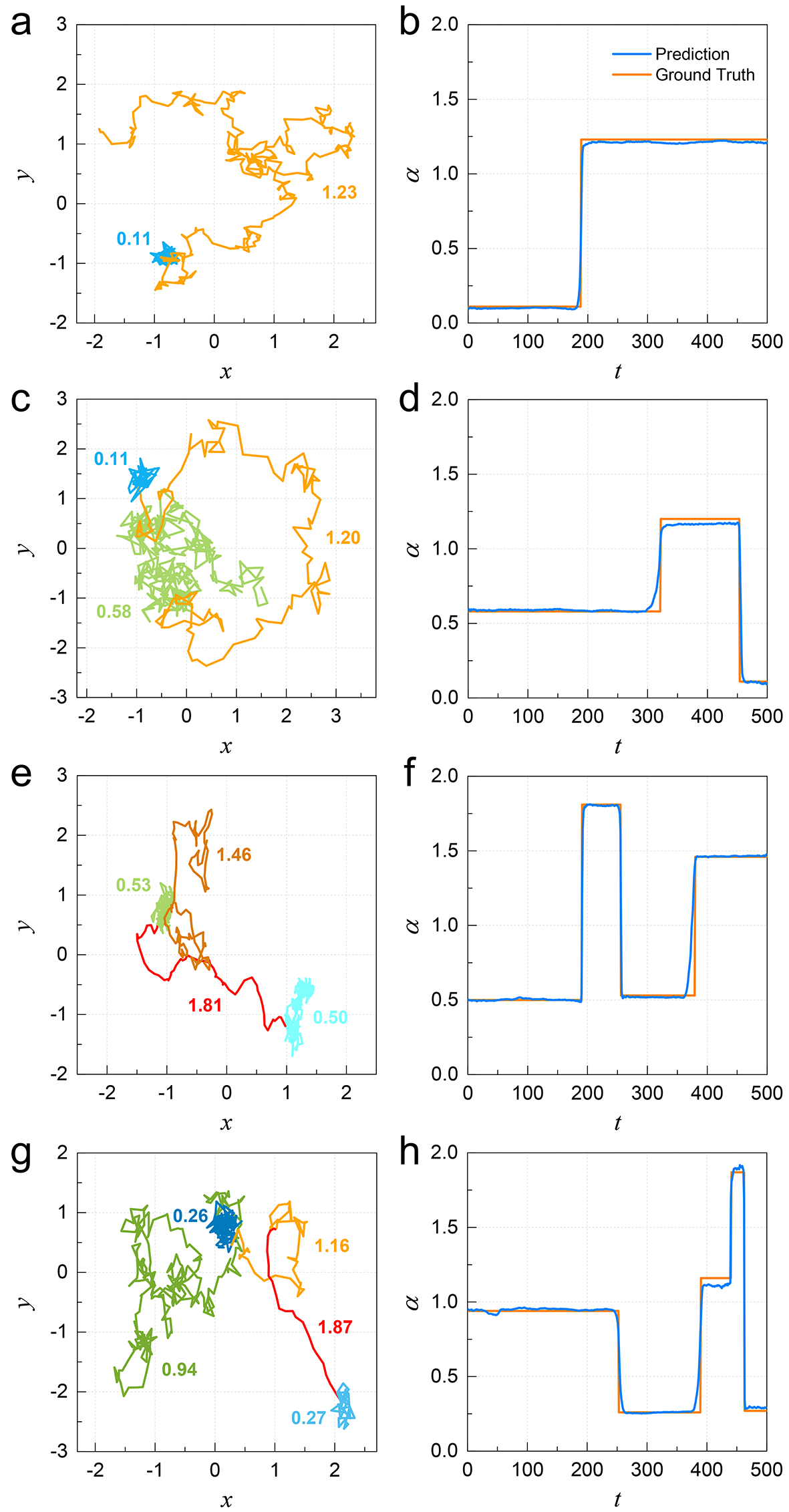}
\caption{Representative examples with $L=500$ demonstrating U-AnDi's performance on subtask 1 for various {number of segments} $M$: (a)-(b) $M=2$, (c)-(d) $M=3$, (e)-(f) $M=4$, (g)-(h) $M=5$. Left panel: Trajectory visualizations, with segments of different diffusion exponents marked in distinct colors. Right panel: Comparisons between the predicted (blue) and ground truth (orange) diffusion exponents.}
\label{fig:fig3}
\end{figure}

\section{Analysis of model performance on simulated trajectories}\label{analysis}

In this section, we present the performance of the U-AnDi model in segmenting simulated trajectories of varying lengths. The results are based on evaluations conducted on the validation sets for both subtasks, each comprising 200 000 trajectories. {Considering the ``black box" nature of machine learning models, it is imperative to provide controls and calibrations analogous to standard tests in traditional analytical methods.} {For this reason, and to gain a deeper understanding of the individual contributions of U-AnDi's core components: DCC, GAU, and U-Net architecture, we also evaluate the performance of several comparative models. Our choice of comparative models is both strategic and comprehensive, encompassing the models outlined in Table \ref{tab:table1}, with detailed specifications provided in Appendix \ref{compare}}. By contrasting U-AnDi with these baseline models, we can distinctly discern the individual and synergistic contributions of DCC, GAU, and U-Net to the semantic segmentation of anomalous diffusion.

\begin{figure*}
\centering
\includegraphics[width=17.0cm]{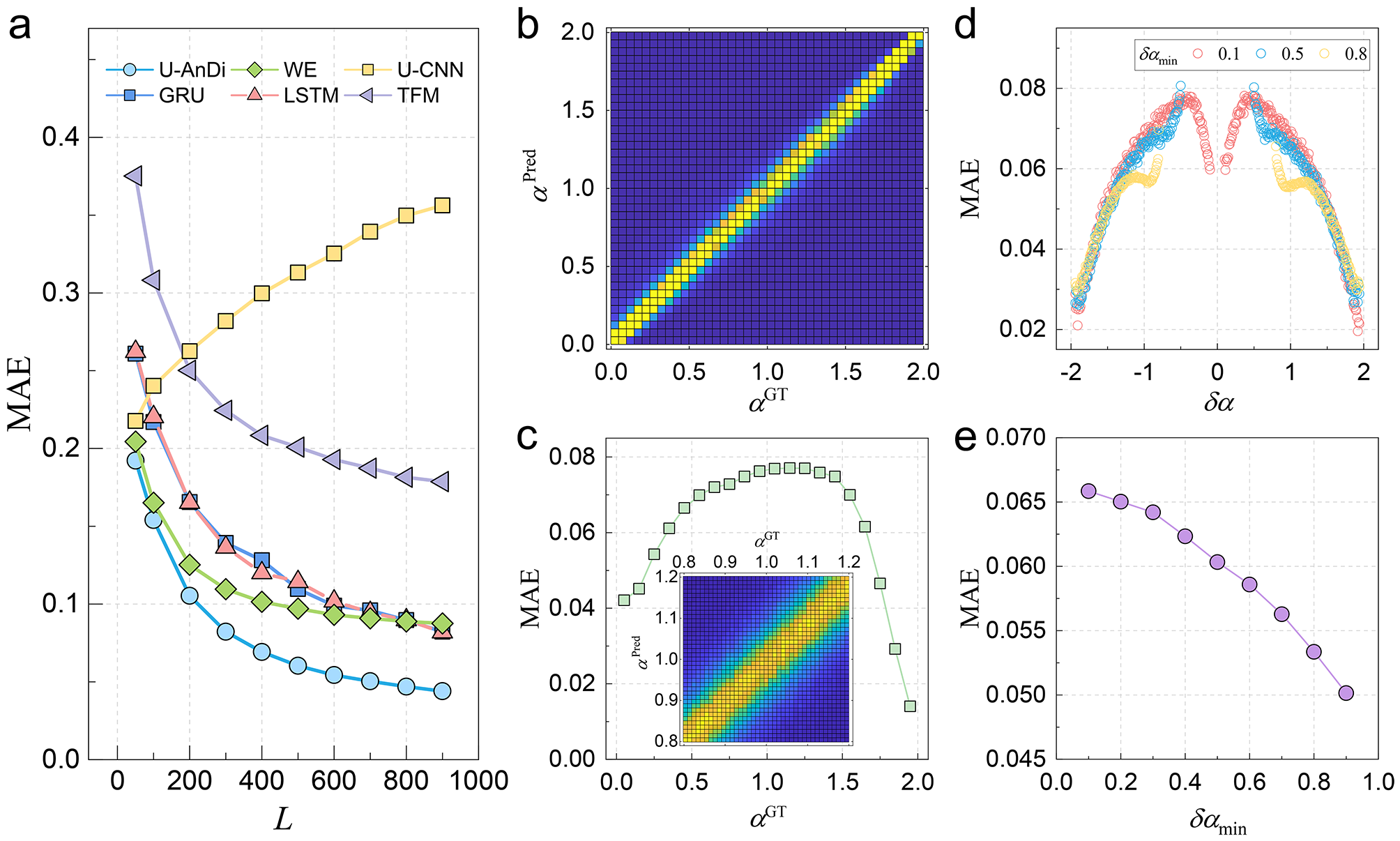}
\caption{{(a) MAE scores of U-AnDi on the validation set for various trajectory lengths from 50 to 900, with performances of other comparative models presented for reference.} (b) 2D histogram of prediction $\alpha^{\rm Pred}$ versus ground truth $\alpha^{\rm GT}$. (c) Distribution of MAE for $\alpha^{\rm GT}$ values ranging from 0.05 to 2. {Inset: higher-resolution 2D histogram of $\alpha^{\rm Pred}$ versus $\alpha^{\rm GT}$ spanning from 0.8 to 1.2.} {(d) MAE variations corresponding to different $\delta \alpha$ for three scenarios: $\delta \alpha_{\min} = 0.1, 0.5,$ and $0.8$. (e) Model performance, as characterized by MAE, across various values of $\delta \alpha_{\min}$.} Here, (b)-(e) display the results for trajectories of length 500.}
\label{fig:fig4}
\end{figure*}

\subsection{Performance of U-AnDi on subtask 1}

To provide a more intuitive demonstration of U-AnDi's performance on subtask 1, we present four representative trajectories of length $L=500$ and {number of segments} $M=2,3,4,5$ in the left panel of Fig. \ref{fig:fig3}. The right panel compares their predicted values with the ground truth values. It can be observed that the predicted diffusion exponents are quite close to the ground truth values, with only minor differences noticeable in a few short intervals. {Quantitatively, we show the MAE scores of U-AnDi on the validation set for various trajectory lengths spanning from 50 to 900 in Fig. \ref{fig:fig4}(a). The performances of other comparative models are also presented for reference. Evidently, across all trajectory lengths, U-AnDi consistently achieves the lowest MAE. This result underscores U-AnDi's robust capability, which is augmented by the synergistic effects of DCC, GAU, and U-Net, in discerning segments with different diffusion exponents.} {Interestingly, as the trajectory length increases, the MAE for all models, with the exception of U-CNN, consistently decreases.} This enhancement in segmentation performance with longer trajectories can be attributed to the extended average length of individual segments. Such elongation offers the model a richer feature information for analysis.

{Upon further analysis, models that integrate the GAU (U-AnDi, WE, LSTM, and GRU) outperform those without it (TFM and U-CNN). This pronounced difference highlights the pivotal role and efficacy of GAU in dealing with long-time correlations inherent in anomalous diffusion. Moreover, models equipped with DCC (U-AnDi and WE) exhibit an additional boost in performance, emphasizing the significance of DCC in capturing long-range dependence information for anomalous diffusion segmentation. However, the U-CNN, which integrates DCC but lacks GAU, experiences a decline in segmentation performance as trajectory length increases. This suggests that, in the absence of GAU, the DCC's ability to process long-time correlations is somewhat lacking. Additionally, it implies that the U-Net architecture, being more attuned to local variations, might misinterpret features without the ability to effectively handle long-time correlations, especially as trajectory lengths grow.}

\begin{figure*}
\centering
\includegraphics[width=17.0cm]{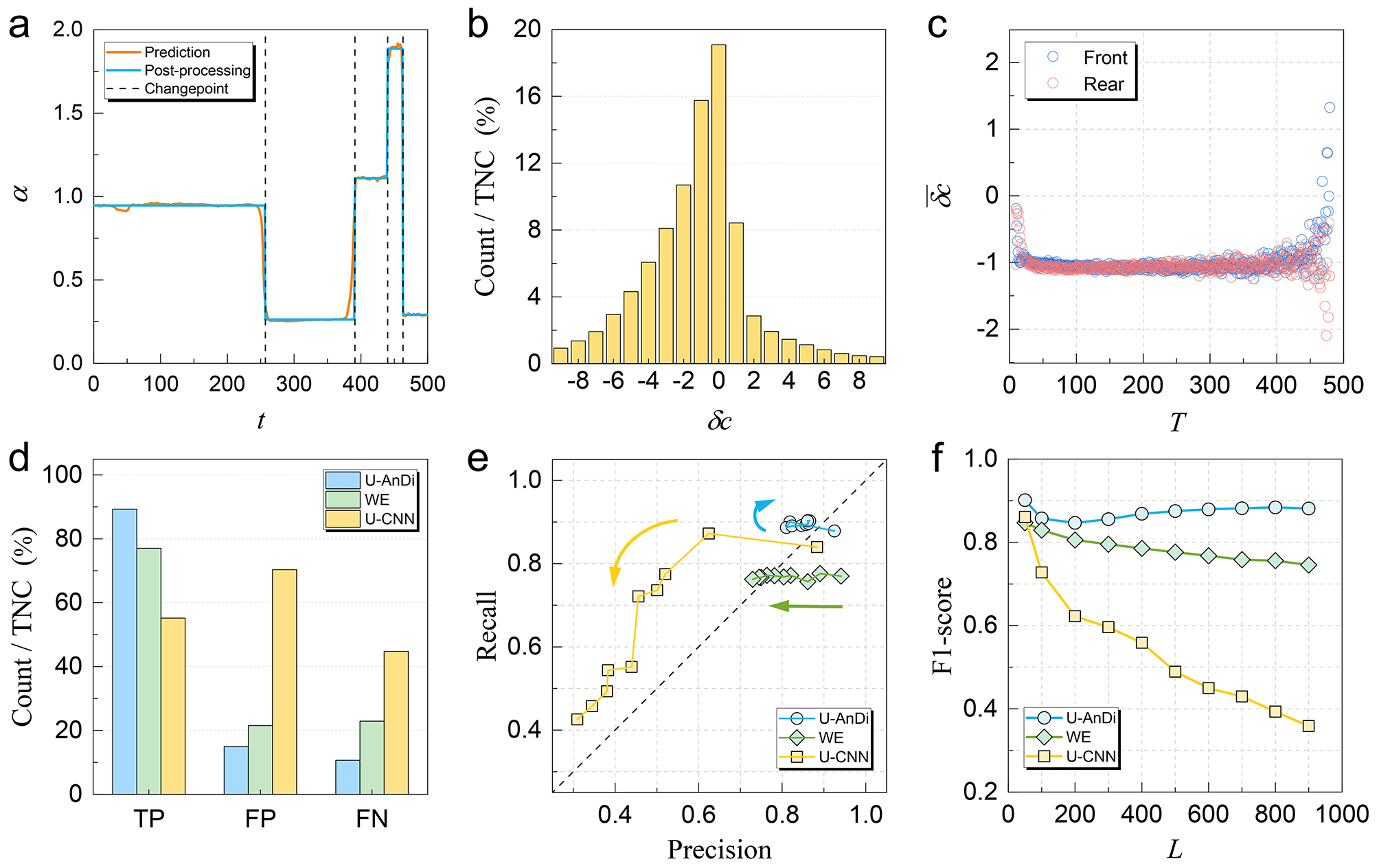}
\caption{(a) Representative example of the post-processing technique in subtask 1, which converts the continuous point-wise predictions into discrete segmented predictions. The dashlines denote the predicted locations of changepoints. (b) Distribution of $\delta c$ {as a percentage of TNC}, spanning from -9 to 9. {(c) Mean values of $\delta c$ ($\overline{\delta c}$) corresponding to various lengths of the front and rear segments adjacent to changepoints.} {(d) Distribution of TP, FP, and FN as a percentage of TNC, juxtaposed with results from WE and U-CNN.} Here, (b)-(d) show the results for trajectories of length 500. {(e) Plots of recall versus precision for U-AnDi, WE, and U-CNN across different trajectory lengths, with arrows indicating the direction of increasing trajectory length. (f) F1-score variations across trajectory lengths for the three models.}}
\label{fig:fig5}
\end{figure*}

Next, we use trajectories of length 500 to provide a detailed analysis of U-AnDi's segmentation performance on subtask 1. For a fine-grained comparison between predictions and ground truths of diffusion exponents, we present the 2D histogram of $\alpha^{\rm Pred}$ versus $\alpha^{\rm GT}$ in Fig. \ref{fig:fig4}(b). As expected, the high-frequency pairs are predominantly located near the line $\alpha^{\rm Pred} = \alpha^{\rm GT}$, suggesting that the prediction errors of are small for all $\alpha^{\rm GT}$ values ranging from 0.05 to 2. However, it should be noted that the prediction errors are not entirely uniform across different $\alpha^{\rm GT}$ values. As displayed in Fig. \ref{fig:fig4}(c), for $\alpha^{\rm GT}$ values close to 0 or 2, the errors are relatively smaller, while for values around 1, the errors tend to be slightly larger. This result can be attributed to the characteristics of FBM, which approaches standard Brownian motion without long-time correlations when $\alpha \approx 1$ \cite{Munoz-Gil}. Compared to scenarios with pronounced long-range dependencies (where $\alpha$ is close to 0 or 2), its identification becomes relatively more challenging. {Considering that $\alpha \approx 1$ might typify many experimental situations, we provide a higher-resolution histogram of $\alpha^{\rm Pred}$ versus $\alpha^{\rm GT}$ for $\alpha^{\rm GT}$ ranging from 0.8 to 1.2 in the inset of Fig. \ref{fig:fig4}(c). As observed, even though the MAE is generally larger around $\alpha^{\rm GT} \approx 1$, the error distribution of U-AnDi's predictions remains relatively uniform without any pronounced bias, highlighting the stability of model's prediction for $\alpha \approx 1$.}

{Given that the variation in diffusion states between two adjacent segments can be characterized by the difference in their diffusion exponents, denoted as $\delta \alpha$, we investigate the influences of $\delta \alpha$ on U-AnDi's segmentation performance to quantify the effects of state variations. MAEs corresponding to different values of $\delta \alpha$ are presented in Fig. \ref{fig:fig4}(d) for three scenarios: $\delta \alpha_{\min} = 0.1, 0.5,$ and $0.8$. A conspicuous feature of the MAE versus $\delta \alpha$ plots is their clear symmetry, indicating that whether the exponent $\alpha$ increases or decreases does not significantly affect U-AnDi's predictions. Moreover, in most cases, a larger magnitude of $|\delta \alpha|$ correlates with a lower MAE, suggesting that more pronounced differences in diffusion states facilitate the model's ability to discern transitions. However, an exception arises when $|\delta \alpha| \approx 0$, where this trend reverses. This is because a negligible $|\delta \alpha|$ might result in indistinct changes in diffusion states, prompting the model to treat two adjacent segments as a single diffusion state, thereby yielding errors commensurate with the magnitude of $|\delta \alpha|$. In addition, we showcase the model's overall performance across different $\delta \alpha_{\min}$ values in Fig. \ref{fig:fig4}(e). Notably, a larger $\delta \alpha_{\min}$, which guarantees more significant diffusion state variations, results in a smaller MAE, aligning well with our expectations.}

Besides the accuracy of predicting diffusion exponents, another crucial aspect is the identification of changepoints among different diffusion states. As shown in the right panel of Fig. \ref{fig:fig3}, the U-AnDi model, which generates continuous point-wise values as its output, enables an approximate visual determination of the changepoint ranges. However, this output cannot provide precise demarcations between successive distinct diffusion states. To address this issue, we introduce a post-processing technique that converts the continuous point-wise predictions into discrete segmented predictions (see Appendix \ref{postprocess} for details). An example of this process is illustrated in Fig. \ref{fig:fig5}(a). After applying the post-processing, the predictions for all points within each segment become consistent, allowing for the identification of changepoints through the abrupt changes in diffusion exponents, as denoted by the dashed lines in Fig. \ref{fig:fig5}(a). Based on this technique, we can further quantitatively investigate U-AnDi's performance on the detection of changepoints.

{We define the total number of changepoints in the validation set as TNC and display the distribution of $\delta c$ as a percentage of TNC in Fig. \ref{fig:fig5}(b), spanning from -9 to 9.} The majority of $\delta c$ values are clustered around 0, signifying a relatively small error in the prediction of changepoints. Interestingly, the distribution demonstrates a mild leftward skew, suggesting that the model has a slight inclination towards predicting changepoints earlier than their true positions. {In addition, the relationship between $\delta c$ and segment length $T$ is also examined. As demonstrated in Fig. \ref{fig:fig5}(c), we present the mean values of $\delta c$, denoted as $\overline{\delta c}$, corresponding to various lengths of the front and rear segments adjacent to the changepoints. It is evident that for the vast majority of segment lengths, $\overline{\delta c}$ consistently aligns close to -1. This trend reinforces our previous observation that U-AnDi has a propensity to slightly anticipate changepoints. In particular, when the segment length approaches the trajectory length, the prediction error increases significantly. Such an increase occurs when one segment dominates the majority of the trajectory, relegating other segments to near-minimal lengths. These short segments provide limited feature information, leading the model to yield less accurate predictions.}

{Expanding our analysis on the changepoint detection, we probe the distribution of TP, FP, and FN as a percentage of TNC and compare these metrics with the results from WE and U-CNN, as illustrated in Fig. \ref{fig:fig5}(d). Obviously, U-AnDi demonstrates a higher count of TP while exhibiting fewer instances of FP and FN, highlighting its superior precision and recall.} {To validate the generality of this observation, we explore the distribution of precision and recall for the three models across different trajectory lengths, with results summarized in Fig. \ref{fig:fig5}(e). Unsurprisingly, U-AnDi consistently outperforms both WE and U-CNN in terms of precision and recall across all lengths. Moreover, as indicated by the arrows, which denote the direction of increasing trajectory length, U-AnDi exhibits remarkably consistent changepoint detection performance, maintaining stable precision and recall across varying lengths. In contrast, as trajectory length increases, WE shows a slight decline in precision while its recall remains stable; U-CNN, devoid of GAU, witnesses a more pronounced drop in both precision and recall. The F1-score variations across trajectory lengths for the three models are illustrated in Fig. \ref{fig:fig5}(f). Mirroring the MAE trends seen in Fig. \ref{fig:fig4}(a), U-CNN, without GAU, significantly underperforms in changepoint detection for longer trajectories compared to GAU-equipped U-AnDi and WE. These findings reemphasize the pivotal role of GAU in effectively managing the long-time correlations inherent in anomalous diffusion.}

\subsection{Performance of U-AnDi on subtask 2}

\begin{figure*}
\centering
\includegraphics[width=17.2cm]{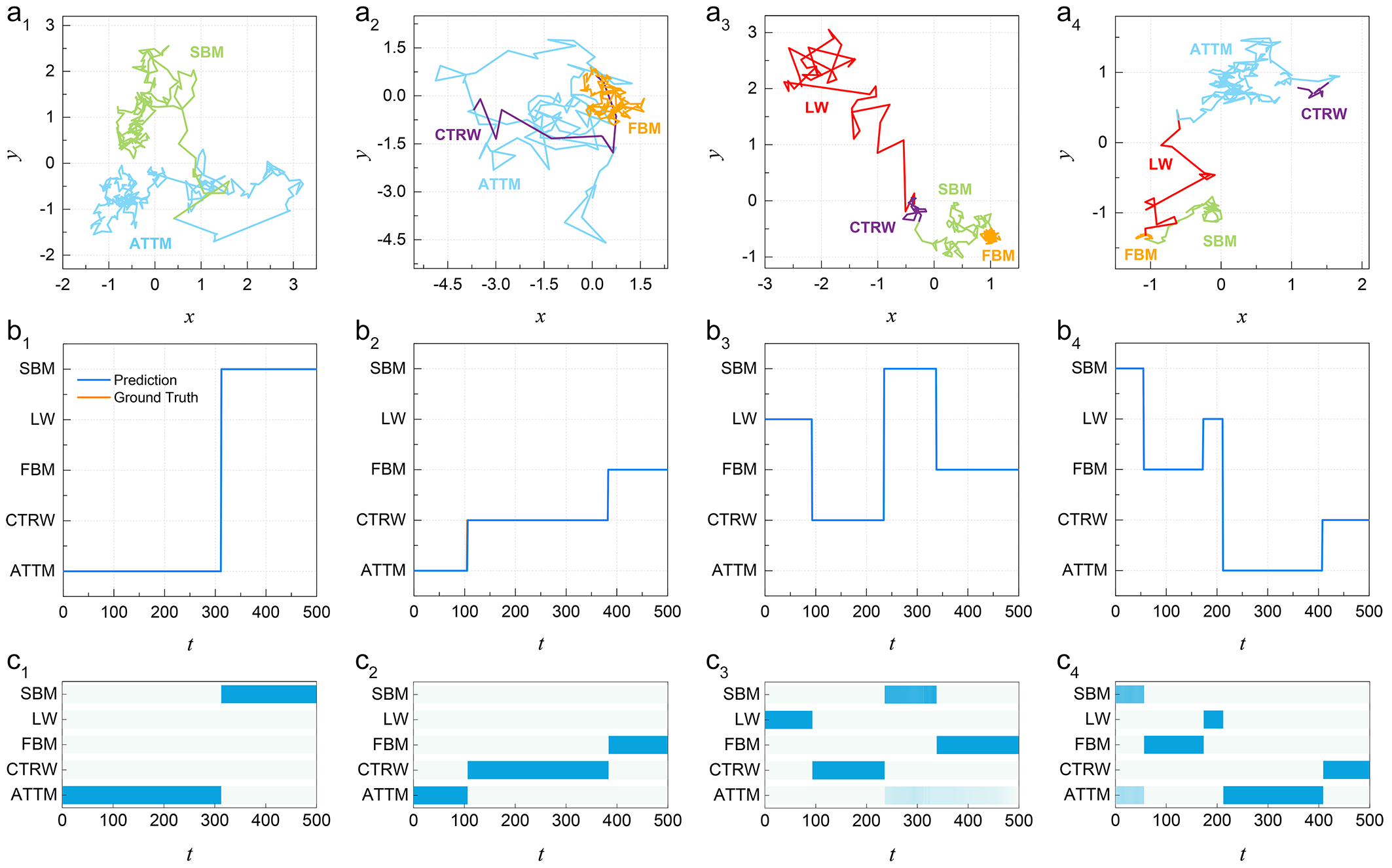}
\caption{Visualization of U-AnDi's performance on the segmentation task with different diffusion models. (${\rm a}_1$-${\rm a}_4$) Illustrations of four representative trajectories of length 500 with varying {number of segments} ($M = 2,3,4,5$), where segments with different diffusion models are marked in distinct colors. (${\rm b}_1$-${\rm b}_4$) Comparisons of predictions (blue) and ground truths (orange) for each trajectory. The strong alignment between them leads to the near invisibility of lines representing ground truth values. (${\rm c}_1$-${\rm c}_4$) Heatmaps of point-wise probability distribution for each trajectory, with darker colors indicating higher probabilities.}
\label{fig:fig6}
\end{figure*}

In a manner consistent with subtask 1, we initially seek to visually demonstrate U-AnDi's performance on the segmentation of trajectories with different dynamics models. For this purpose, the segmentation results of four representative trajectories of length 500 with {number of segments} $M = 2,3,4,5$ are summarized in Fig. \ref{fig:fig6}. In more detail, Fig. \ref{fig:fig6} is organized into three rows for a coherent presentation. The first row [Figs. \ref{fig:fig6}(${\rm a}_1$)-6(${\rm a}_4$)] presents the visualizations of these four trajectories, using unique colors to differentiate segments associated with distinct diffusion models. Subsequently, comparisons of predictions with ground truths for each of the four trajectories are displayed in the second row [Figs. \ref{fig:fig6}(${\rm b}_1$)-6(${\rm b}_4$)]. It is evident that the lines depicting the predicted values (blue) closely align with those representing the ground truth values (orange). This highlights the robust performance of U-AnDi in segmenting trajectories with different diffusion models. Moreover, to provide a more comprehensive understanding of the model predictions, we show the heatmaps of predicted point-wise probability distribution in the third row [Figs. \ref{fig:fig6}(${\rm c}_1$)-6(${\rm c}_4$)], with darker colors representing values closer to 1. As illustrated, in the majority of instances, the model is capable of predicting the specific model at a given step with a high degree of confidence (probability).

\begin{figure*}
\centering
\includegraphics[width=17.2cm]{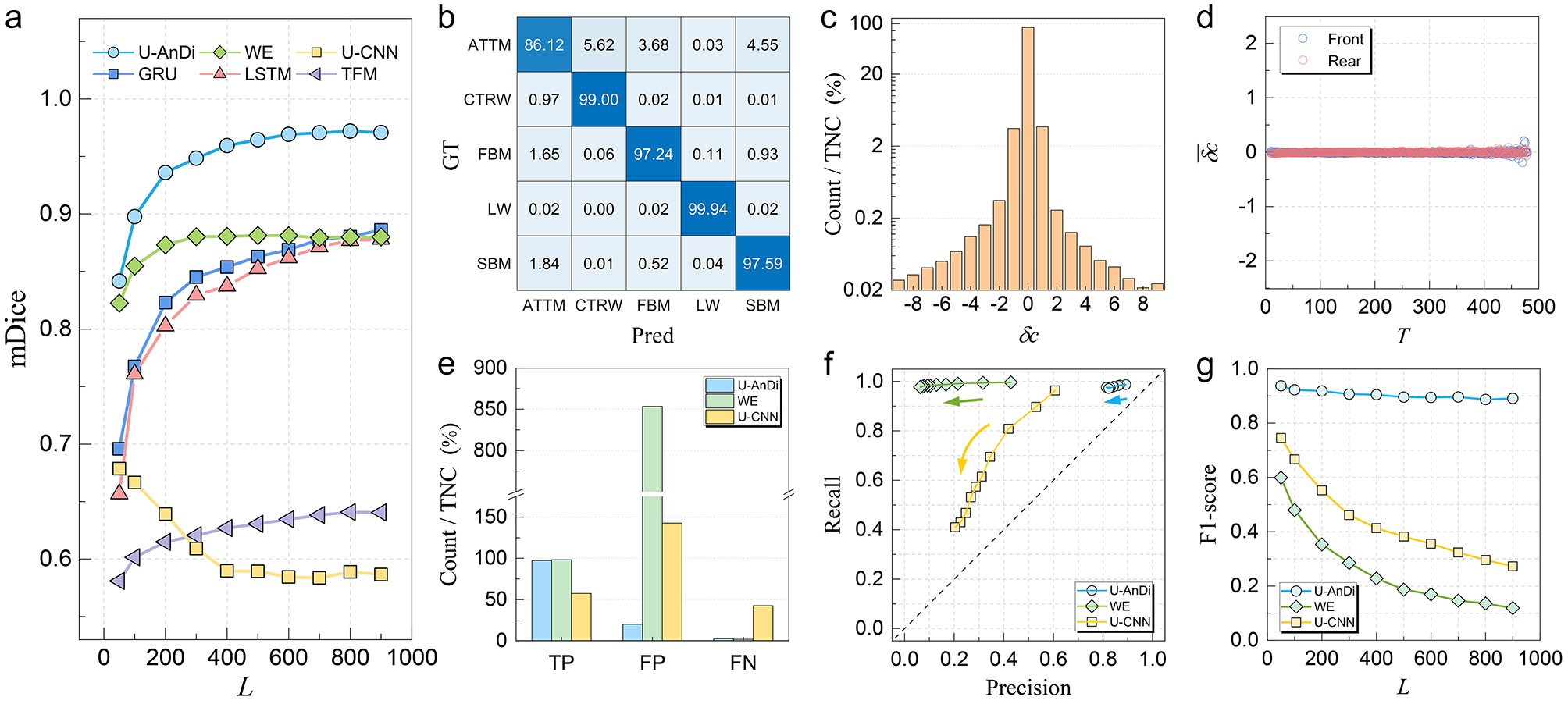}
\caption{{(a) mDice scores on the validation set of U-AnDi and comparative models across trajectory lengths from 50 to 900.} (b) Confusion matrix of the point-wise classification results of U-AnDi, where horizontal and vertical coordinates denote predicted and ground truth labels, respectively. (c) Distribution of $\delta c$ spanning from -9 to 9 {as a percentage of TNC}, which is plotted on a logarithmic scale. {(d) $\overline {\delta c}$ as a function of the lengths of the front and rear segments adjacent to changepoints.} {(e) Comparative distribution of TP, FP, and FN values for U-AnDi, WE, and U-CNN, expressed as a percentage of TNC.} Here, (b)-(e) illustrate the findings for trajectories of length 500. {(f) Recall against precision for U-AnDi, WE, and U-CNN across varying trajectory lengths, where arrows denote the direction of increasing trajectory length. (g) Dependence of F1-score on trajectory length for the three models.}}
\label{fig:fig7}
\end{figure*}

Building on the insights from subtask 1, we further quantitatively evaluate U-AnDi's performance on subtask 2. {As depicted in Fig. \ref{fig:fig7}(a), the mDice scores of U-AnDi and other comparative models across different trajectory lengths ranging from 50 to 900 align with the findings from subtask 1. This consistency not only underscores U-AnDi's robustness across different segmentation tasks of anomalous diffusion, but also reinforces our prior analysis about the functionalities of U-AnDi's core components: DCC, GAU, and U-Net.} Further, we continue to use trajectories of length 500 as our primary example. To gain a deep insight into the error origins regarding this point-wise classification task, the confusion matrix of classification results is displayed in Fig. 7(b). As observed, while the classification accuracy for the CTRW, FBM, LW, and SBM models exceeds 97\%, the accuracy for the ATTM model is merely around 86\%. This discrepancy can be attributed to the machine-learning features of ATTM extracted by the WADNet encoder, which bear resemblance to CTRW, FBM, or SBM, as elucidated in Ref. \cite{Li}. Consequently, this leads U-AnDi to occasionally misclassify ATTM steps as CTRW (5.62\%), FBM (3.68\%), or SBM (4.55\%).

On the other hand, detecting changepoints among diffusion states holds significant importance in subtask 2 as well. In contrast to subtask 1, no additional post-processing techniques are required for the detection in this task. We can efficiently determine the positions of changepoints by examining abrupt changes in the predicted labels. In Fig. \ref{fig:fig7}(c), the distribution of $\delta c$ values ranging from -9 to 9 is presented {as a percentage of TNC}. A distinction from subtask 1 is that 91.1\% of the cases have $\delta c$ values of 0, indicating a significant portion of the predicted changepoints aligning perfectly with the ground truths. Given the highly imbalanced distribution of $\delta c$ values, a logarithmic scale is employed in Fig. \ref{fig:fig7}(c) to enhance visualization. {Moreover, $\overline {\delta c}$ as a function of the lengths of segments adjacent to changepoints is explored, with the results depicted in Fig. \ref{fig:fig7}(d). Differing from subtask 1, the majority of data points for both front and rear segments are concentrated around 0. Even as the segment length nears the trajectory length, there is no significant escalation in error. This suggests that U-AnDi exhibits commendable stability in segmenting trajectories with varying diffusion models, even when dealing with very short segments.}

{Similar to subtask 1, regarding U-AnDi, WE, and U-CNN, the distribution of TP, FP, FN for $L=500$, along with the plots of recall versus precision across varying lengths and the F1-scores for different lengths are illustrated in Figs. \ref{fig:fig7}(e), \ref{fig:fig7}(f), and \ref{fig:fig7}(g) respectively. While the results largely mirror those from subtask 1, a stark deviation is observed in the FP count of WE, which have increased dramatically [Fig. \ref{fig:fig7}(f)]. This surge adversely impacts the WE's precision, rendering its F1-score lower than both U-AnDi and U-CNN [Fig. \ref{fig:fig7}(g)]. Considering that WE lacks the U-Net architecture, it can be deduced that such an architecture plays a crucial role in adeptly processing intricate local fluctuations intertwined with long-time correlations for the segmentation of anomalous diffusion.}

{Building on insights from both subtasks, we further investigate the case where the diffusion exponent and the dynamic model concurrently vary within a single trajectory. Detailed findings of this analysis are presented in Appendix \ref{bothchange}, which confirms U-AnDi's adaptability and robustness in the face of such complex conditions.}

\subsection{Extended analysis on the segmentation of anomalous diffusion}

{While U-AnDi has demonstrated commendable segmentation performance in discerning changes in diffusion exponents and transitions among diffusion models, it remains imperative to assess its capabilities in more intricate and broadly relevant anomalous diffusion or single-particle tracking scenarios. The robustness and widespread applicability of U-AnDi in diverse contexts are pivotal for its integration into practical applications, offering significant utility to potential researchers in the anomalous diffusion field who may not be well-versed in machine learning.}

{For that purpose, we expand our analysis to encompass three other distinct segmentation tasks in this section. Each task is designed to mirror the complexities often encountered in the domain of anomalous diffusion, serving as a basis for a rigorous examination of U-AnDi's versatility in handling complex anomalous diffusion scenarios.}

\subsubsection{{Semantic segmentation of single-model trajectories with constant parameters}}

\begin{figure}
\centering
\includegraphics[width=8.6cm]{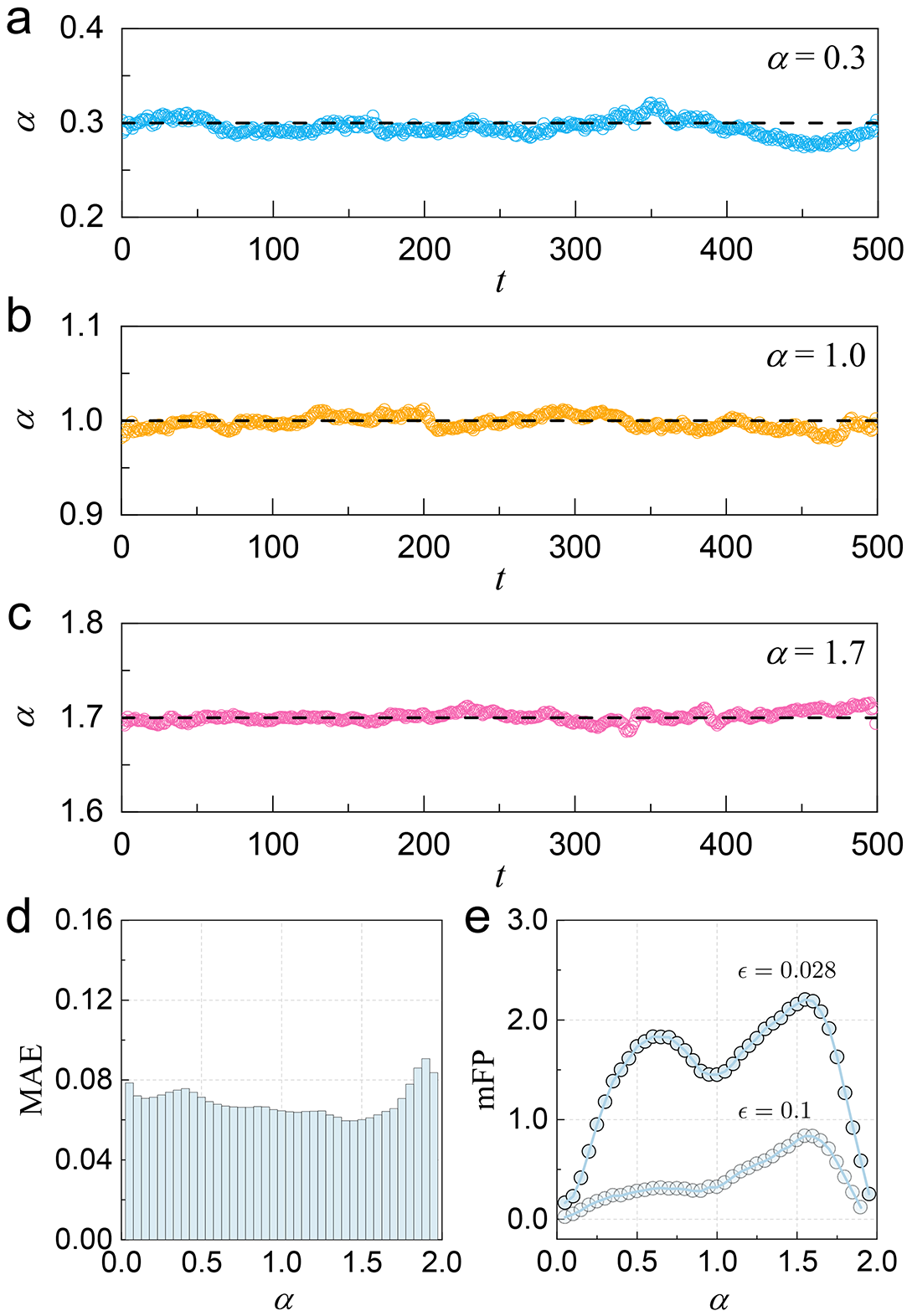}
\caption{{(a)-(c) Predictions of U-AnDi-Exponent for FBM trajectories with diffusion exponents $\alpha = 0.3$, $1.0$, and $1.7$, representing subdiffusion, Brownian diffusion, and superdiffusion, respectively. (d) MAE scores for various $\alpha$ values. (e) Distributions of the mFP for different $\alpha$ with threshold $\epsilon=0.028$ and $0.1$, respectively.}}
\label{fig:fig8}
\end{figure}

\begin{figure*}
\centering
\includegraphics[width=17.2cm]{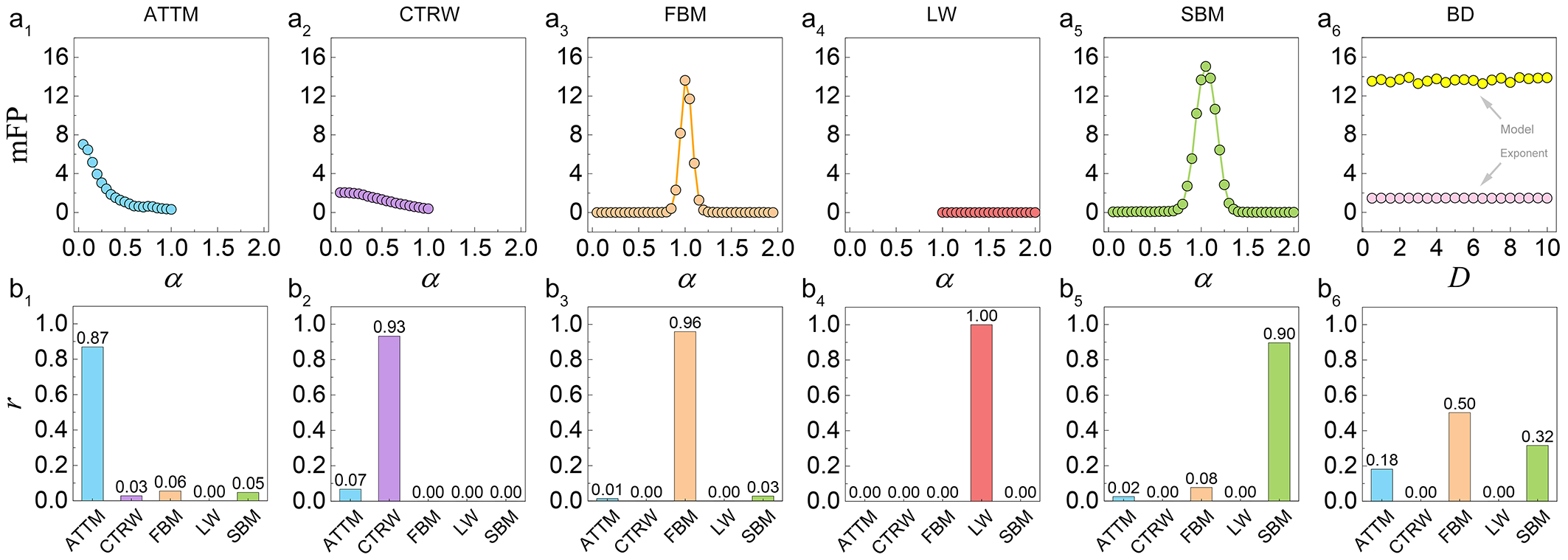}
\caption{{(${\rm a}_1$)-$({\rm a}_6$) Distribution of mFP predicted by U-AnDi-Model for trajectories generated by six different diffusion models: ATTM, CTRW, FBM, LW, SBM, and BD, respectively. The performance of U-AnDi-Exponent is also given in (${\rm a}_6$) for BD trajectories for comparison. (${\rm b}_1$)-$({\rm b}_6$) The mean proportion of predicted diffusion models within each individual trajectory for ATTM, CTRW, FBM, LW, SBM, and BD, respectively.}}
\label{fig:fig9}
\end{figure*}

{In the landscape of anomalous diffusion research, while transitions between diffusion states are frequently observed, it is equally common to encounter scenarios where a random walker moves exclusively within a singular diffusion model framework \cite{Metzler}. These single-model trajectories, devoid of state transitions, represent a fundamental and ubiquitous aspect of experimental studies. In our previous discussions, we center on trajectories that exhibit state transitions (with $M \geq 2$). This emphasis naturally leads to considerations about U-AnDi's performance on single-model trajectories. Such an evaluation is crucial as it offers potential users insights into U-AnDi's stability and reliability across an expansive range of diffusion scenarios.}

{For this investigation, we generate trajectories of length 500 using six diffusion models: ATTM, CTRW, FBM, LW, SBM, and the Brownian diffusion (BD). BD is included because it represents the most fundamental diffusion phenomenon in experimental observations. Here, it should be noted that BD is equivalent to the scenario of FBM and SBM when the diffusion exponent $\alpha=1$. To ensure the generality of our results, we extensively evaluate U-AnDi across a broad range of core parameters for the diffusion models. For the first five models, the selected parameter is $\alpha$, with its range consistent with the description in Sec. \ref{task2} for each model. For BD, the parameter is the diffusion coefficient $D$, which spans a range from 0.5 to 10. Further, to guarantee statistical robustness, we generate 10 000 trajectories for each parameter value for testing. For convenience, we introduce two nomenclatures in this section: ``U-AnDi-Exponent" refers to the model from subtask 1 that segments based on diffusion exponent, while ``U-AnDi-Model" pertains to the model from subtask 2 that segments based on the diffusion model.}

{We first assess the performance of U-AnDi-Exponent on FBM trajectories with fixed diffusion exponents $\alpha$. To visualize the variation between the predicted values and the fixed exponent, we display the predictions of U-AnDi-Exponent for three representative trajectories with $\alpha = 0.3$ (subdiffusion), $1.0$ (Brownian diffusion), and $1.7$ (superdiffusion) in Figs. \ref{fig:fig8}(a)-(c), respectively. Across these scenarios, the model's predictions closely cluster around the true values, indicating only slight variation. Quantitatively, we present the MAE scores for various $\alpha$ values in Fig. \ref{fig:fig8}(d). It is evident that MAE values are all approximately around 0.07, suggesting consistently minor fluctuations.}

{Furthermore, we apply the post-processing technique described in Appendix \ref{postprocess} to identify changepoints. Given that these trajectories inherently lack changepoints, our analysis concentrates on the FPs in the predictions. We define the mean number of FPs per trajectory within the dataset as mFP and illustrate the distribution of mFP for different $\alpha$ in Fig. \ref{fig:fig8}(e). Despite the fluctuation of mFP in response to changes in $\alpha$, the values consistently remain low. Notably, by increasing the threshold $\epsilon$ in the post-processing technique, the model's precision can be further improved, as evidenced by a reduction in mFP [Fig. \ref{fig:fig8}(e)]. Overall, U-AnDi-Exponent demonstrates a remarkable stability in dealing with FBM trajectories with constant parameters. In particular, the flexibility of our approach allows users to refine the precision of changepoint detection through threshold customization, thereby tailoring the model to their specific needs.}

{On the other hand, we employ U-AnDi-Model to test single-model trajectories generated by the six diffusion models. In addition to utilizing mFP as the evaluation metric [Figs. \ref{fig:fig9}(${\rm a}_1$)-\ref{fig:fig9}(${\rm a}_6$)], the mean proportion of predicted diffusion models within each individual trajectory, denoted as $r$, is displayed in Figs. \ref{fig:fig9}(${\rm b}_1$)-\ref{fig:fig9}(${\rm b}_6$) to manifest the variation between predicted and true values. It is noteworthy that for ATTM, CTRW, FBM, LW, and SBM, the true model accounts for at least 87\% of the predicted composition, underscoring the robustness of our method in this task. Since U-AnDi-Model primarily relies on the distinctive features of diffusion models to determine changepoints, the magnitude of mFP on single-model trajectories is largely influenced by the uniqueness of the diffusion model relative to others. For instance, the characteristics of LW are notably distinct from the other five models (Ref. \cite{Li}), allowing for a stable identification of LW without predicting state transitions, i.e., ${\rm mFP}=0$ for all $\alpha$ [Fig. \ref{fig:fig9}(${\rm a}_4$)]. Conversely, for FBM and SBM, as $\alpha$ approaches 1, the trajectories of these two models converge towards the characteristics of Brownian diffusion. This convergence results in an overlap that challenges U-AnDi's discriminative power, leading to an increased incidence of FPs [Figs. \ref{fig:fig9}(${\rm a}_3$) and \ref{fig:fig9}(${\rm a}_5$)]. This phenomenon is also evident when U-AnDi-Model is tasked with analyzing BD trajectories, where the mFPs closely mirror those observed for FBM and SBM with $\alpha=1$ and are considerably higher than the forecasts by U-AnDi-Exponent [Fig. \ref{fig:fig9}(${\rm a}_6$)]. As shown in Fig. \ref{fig:fig9}(${\rm b}_6$), predictions for BD trajectories show a significant proportion attributed to both FBM and SBM, emphasizing the indistinguishability between these two at $\alpha=1$.}

\subsubsection{{Segmentation of simulated multi-state trajectories in biological experiment scenarios}}

{In the realm of single-particle tracking experiments within biological systems \cite{Manzo1,Shen,Qian,Saxton,Torreno-Pina,TRamWAy}, the microdynamics of the observed entities are prone to transitions due to environmental influences, necessitating effective methods for segmenting the trajectory data. However, in such experimental systems, the diffusion dynamics often observed are not limited to specific anomalous diffusion models but include transitions among Brownian, confined, directed, and immobile states. These states correspond to:
\begin{itemize}
\item Brownian diffusion (BD): Free movement characterized by random walks, reflecting the absence of constraints in an open environment.
\item Confined diffusion (CD): Restricted movement within certain boundaries, such as within cellular compartments or membranes.
\item Directed Motion (DM): Motion driven by the external force or energy, such as bacteria propelled by flagellar motion or cells migrating in response to chemical signals.
\item Immobile state (IS): A state where particles are trapped or bound to structures within the cell, exhibiting little to no movement over time.
\end{itemize}
For instance, as described in Ref. \cite{multistep}, molecules like CD44 in macrophages can exhibit these states. CD44 molecules display confined diffusion when interacting with the actin cortex, free diffusion when such interactions are inhibited, and immobility possibly when tethered to the cortex.}

{Hence, to validate the effectiveness of U-AnDi in such biological experimental scenarios, we primarily adopt the simulation methods from Ref. \cite{multistep} to generate segments representing these four dynamic states. We then combine these segments using the same approach as in subtask 2 to create composite trajectories for the validation, with a representative example of length $L=500$ illustrated in Fig. \ref{fig:fig10}(a). The simulation methods for these segments are detailed in Appendix \ref{4statesgen}. The training of U-AnDi follows the same protocol as in subtask 2.}

{Furthermore, to observe the potential advancements of our model over conventional approaches, we benchmark the performance of U-AnDi against the traditional divide-and-conquer moment scaling spectrum (DC-MSS) analysis method \cite{multistep}. A brief introduction to DC-MSS is provided in Appendix \ref{compare}. As displayed in Fig. \ref{fig:fig10}(b), we present the segmentation results of both methods on the trajectory shown in (a). It is evident that U-AnDi's predictions align almost perfectly with the ground truth, whereas DC-MSS exhibits numerous misjudgments, including incorrect state identification and changepoint detection errors. Quantitatively, we compare the mDice scores of both models for trajectories of varying lengths, ranging from 50 to 900, and summarize the results in Fig. \ref{fig:fig10}(c). Notably, U-AnDi decisively outperforms DC-MSS across all lengths with mDice close to 1, indicating a significantly higher accuracy in state identification.}

\begin{figure}
\centering
\includegraphics[width=8.6cm]{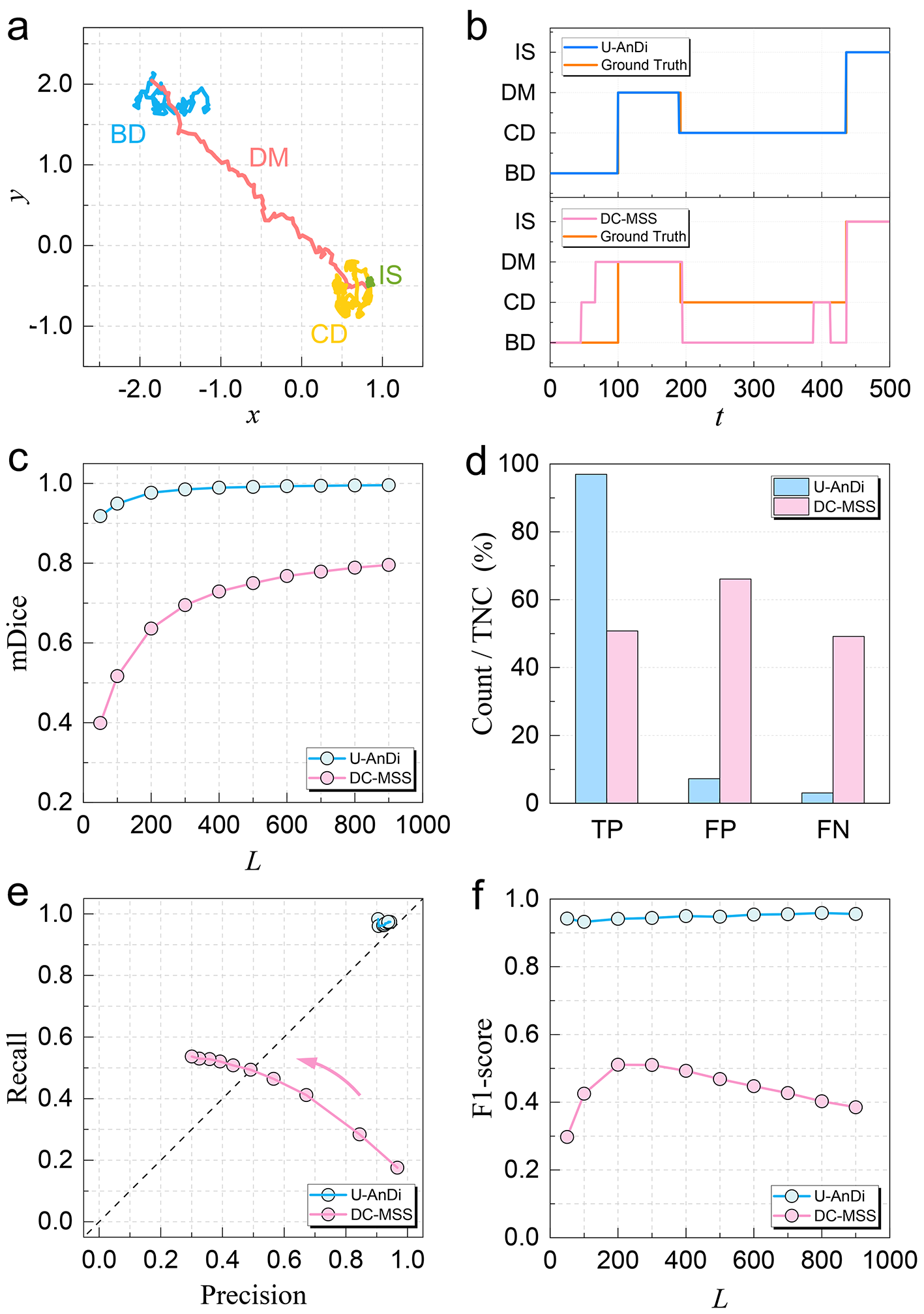}
\caption{{(a) Representative example of a simulated composite trajectory of length 500, containing the four dynamic states in biological experiment scenarios: Brownian diffusion (BD), confined diffusion (CD), directed motion (DM), and immobile state (IS). (b) Comparative segmentation results of U-AnDi and DC-MSS on the composite trajectory in (a). (c) mDice score comparisons between U-AnDi and DC-MSS for trajectories of varying lengths. (d)-(f): Analysis of changepoint detection for U-AnDi and DC-MSS. (d) Distribution of TP, FP, and FN as a percentage of TNC where $L=500$. (e) Recall versus precision for different trajectory lengths. (f) The variation of the F1-score with trajectory length.}}
\label{fig:fig10}
\end{figure}

\begin{table*}
\caption{\label{tab:table2}
{Summary of parameters, labels, and corresponding equations to calculate labels for segmentation tasks associated with the four long-time correlations.}}
\begin{ruledtabular}
\begin{tabular}{clccc}
 Correlation type & \multicolumn{1}{c}{Parameter} & Label & Label range & Label calculation\\
\hline
exp & change $A$, set $\tau = 10$ & $D$ &$[0.05,2)$ & $D=A\tau$\\
multi-exp & change $A$, set $\tau_1 = 10, \tau_2=20, \phi=0.5$ & $D$ &$[0.05,2)$ & $D=A\left[\phi \tau_1+(1-\phi) \tau_2\right]$\\
exp-cos & change $A$, set $\tau=10,\omega=0.1$ & $D$ &$[0.05,2)$ & $D=A \tau/{(1+\tau^2 \omega^2)}$\\
M-L &  change $\lambda$, set $\tau=10, A=1$ & $\alpha$ &$[0.05,2)$ & $\alpha = 2-\lambda$\\
\end{tabular}
\end{ruledtabular}
\end{table*}

{Similarly, U-AnDi's performance in detecting state transitions is markedly superior to that of DC-MSS. As with subtask 2, Figs. \ref{fig:fig10}(d)-\ref{fig:fig10}(f) detail the distribution of TP, FP, and FN as a percentage of TNC for $L=500$, plots of recall versus precision for different trajectory lengths, and the variation of the F1-score with trajectory length, respectively. Across various lengths, U-AnDi demonstrates high precision, recall, and F1-scores, while DC-MSS lags in performance. Additionally, as trajectory length increases, DC-MSS shows a gradual increase in recall but a decrease in precision, revealing a trade-off between the two metrics. [Figs. \ref{fig:fig10}(e)]. This trade-off results in uniformly low F1-scores for all lengths, which is evident from Figs. \ref{fig:fig10}(f).}

{These findings underscore U-AnDi's exceptional segmentation capabilities for anomalous diffusion trajectories in biological experimental scenarios. They also highlight the significant performance gap between unsupervised algorithms like DC-MSS and supervised approaches, reinforcing the superiority of the latter in trajectory segmentation tasks.}

\subsubsection{{Evaluating U-AnDi with more long-time correlations}}

{Considering that long-time correlations are a fundamental property inherent to anomalous diffusion behaviors, we aim to further validate the versatility and adaptability of U-AnDi in segmenting trajectories characterized by more correlations. To achieve this, we generate trajectories where the velocity autocorrelation functions (VACFs) conform to four distinct long-time correlations. Subsequently, U-AnDi is employed to segment these trajectories. The chosen correlation functions for this evaluation are listed in the following paragraph. To maintain clarity and without loss of generality, we present the mathematical expressions in a one-dimensional framework in this section.}

{We denote $C(t) = \langle v(0)v(t) \rangle$ as the VACF. Four long-time correlation functions are detailed below:
\begin{itemize}
\item Exponential decay (exp):
\begin{equation}
C(t) = Ae^{-t/\tau},
\end{equation}
where $A$ is the amplitude and $\tau$ is the time constant.
\item Multi-exponential decay (multi-exp):
\begin{equation}
C(t) = A\left[\phi e^{-t / \tau_1}+(1-\phi) e^{-t / \tau_2}\right].
\end{equation}
Here, $A$ is the amplitude, $\phi$ is the weighting factor. $\tau_1$ and $\tau_2$ are the time constants of two exponential decays, respectively.
\item Exponentially damped cosine wave (exp-cos):
\begin{equation}
C(t) = A e^{-t/\tau} \cos\omega t,
\end{equation}
where $A$ is the amplitude, $\tau$ is the time constant, and $\omega$ is the angular frequency.
\item Mittag-Leffler decay (M-L) \cite{2007anomalous}:
\begin{equation}
C(t) = \frac{A}{\tau^\lambda} E_\lambda\left[-(t / \tau)^\lambda\right],
\end{equation}
where
\begin{align}
E_\lambda(z)=\sum_{k=0}^{\infty} \frac{z^k}{\Gamma(\lambda k+1)}.
\end{align}
Here, $\tau$ acts as a characteristic memory time and $A$ is the amplitude coefficient. $\lambda$ is the exponent satisfying $\lambda>0$. $\Gamma(\cdot)$ denotes the Gamma function.
\end{itemize}
We generate these trajectories using a modified version of the ``fbm" Python package \cite{fbm, mittag-leffler}, which employs the Davies-Harte algorithm \cite{davies1987tests}. Associated codes are available in our GitHub repository \cite{U-AnDi}. The effectiveness of this generation method is validated in Appendix \ref{DH}.}

{Labels for the trajectory segmentation are determined based on the diffusion behaviors corresponding to their long-time correlations. Using the equation below \cite{Oxford}:
\begin{equation}
\label{eqmsd}
{\rm MSD} = 2\int_0^t{(t-s)C(s){\rm d}s},
\end{equation}
it is evident that trajectories exhibiting exp, multi-exp, and exp-cos correlations manifest long-time Fickian behaviors (see Appendix \ref{DH}), i.e., ${\rm MSD} \sim t$. Consequently, for these three types of correlations, we utilize the long-time diffusion coefficient $D$ as the target for trajectory segmentation. On the other hand, as outlined in Ref. \cite{Velocityauto}, the diffusion exponent $\alpha$ of trajectories characterized by the M-L correlation equals $2-\lambda$. Therefore, for trajectories with the M-L correlation, we choose $\alpha$ as the segmentation label. In Table \ref{tab:table2}, we provide a summary of the parameters, labels, and corresponding equations to calculate segmentation labels (see Appendix \ref{DH} for details) with respect to the four long-time correlations.}

\begin{figure}
\centering
\includegraphics[width=8.4cm]{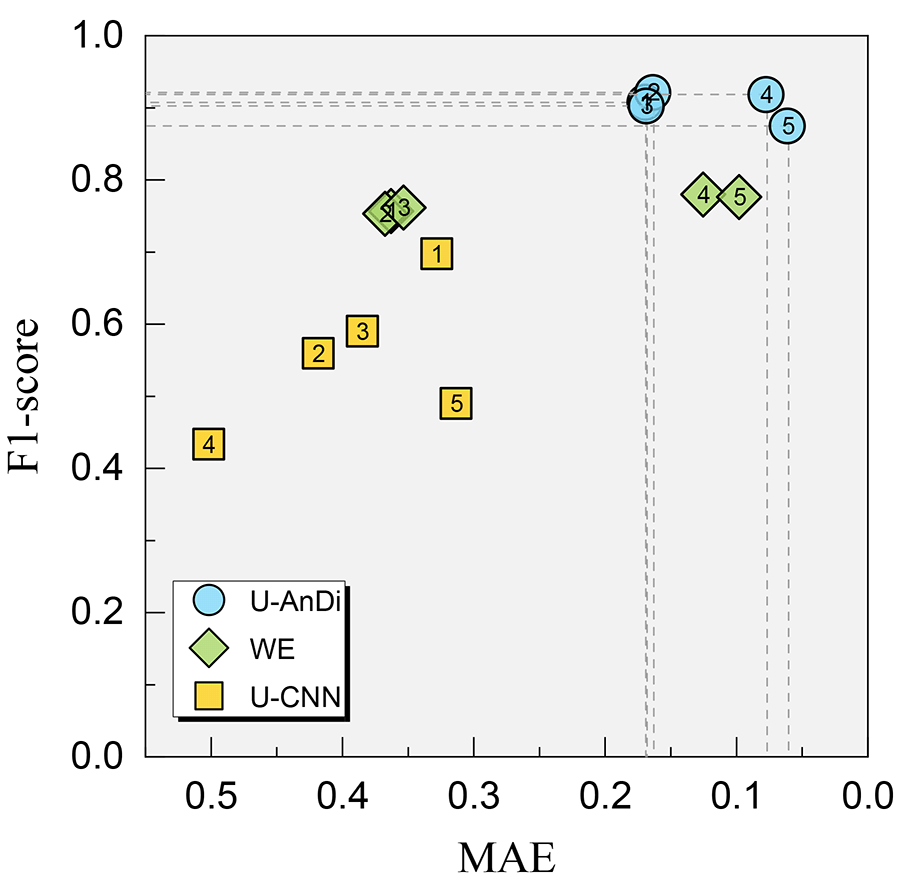}
\caption{{Comparative evaluation of U-AnDi, WE, and U-CNN using MAE and F1-score across five distinct long-time correlations: 1. exponential decay; 2. multi-exponential decay; 3. exponentially damped cosine wave; 4. Mittag-Leffler decay; 5. power-law decay. The dashed lines represent the best score for each correlation type, and the numbers on the dots correspond to the aforementioned correlation types.}}
\label{fig:fig11}
\end{figure}

{Trajectory segments are combined to form a composite trajectory using the same method as in subtask 1. Additionally, the model and training approach remain consistent with those applied in that subtask. We evaluate U-AnDi's segmentation performance using MAE and F1-score, and compare it with WE and U-CNN. The results, illustrated in Fig. \ref{fig:fig11}, encompass not only the performance of U-AnDi for the four distinct long-time correlations but also its evaluation scores in subtask 1, which pertains to the power-law correlation. A clear observation from the results is the superior performance of U-AnDi in comparison to both WE and U-CNN across all five types of long-time correlations, whether evaluated based on MAE or F1-score. Such consistent superiority across diverse correlation types highlights U-AnDi's robust adaptability, emphasizing its potential to serve as a universal scheme for the segmentation of time-series data.}

{Interestingly, as quantified by the MAE, both U-AnDi and WE exhibit a notably better performance in predicting the diffusion exponent compared to the diffusion coefficient. However, this distinction is not evident in the case of U-CNN.  A potential reason for this difference can be attributed to the inherent characteristics of diffusion coefficient and diffusion exponent. While the diffusion coefficient primarily captures the ``speed" of diffusion, the diffusion exponent provides deeper insights into the underlying mechanisms of diffusion process, inherently reflecting the long-time correlation information. Consequently, the diffusion exponent can be more readily predicted by models like U-AnDi and WE, both of which are adept at processing long-time correlations with the help of GAU. In contrast, U-CNN, which lacks the GAU, does not exhibit a similar advantage.}

\section{{Interpretability of U-AnDi's core components}}\label{inter}

{The quest for interpretability in machine learning, particularly within the realms of physical sciences, is not merely a pursuit of operational transparency but a foundational requirement for scientific validity. In this work, the comparative evaluation of U-AnDi against other models partially diminishes the ``black box" associated with our machine learning model, shedding light on the inner working of U-AnDi's core components: DCC, GAU, and U-Net. As long-time correlation is one of the fundamental properties of anomalous diffusion, these components faithfully encapsulate this underlying nature and are in strict accordance with our design principles. Within this section, we consolidate the interpretability insights that has been gained from the comparison among the performance of U-AnDi and its comparative models in the aforementioned tasks.}

\begin{itemize}
\item {\it Dilated causal convolution: Capturing long-time correlations}

{To effectively capture the pivotal long-time correlations inherent in anomalous diffusion, DCC, with its exponentially expanding receptive fields, is designed to encompass the full spectrum of these extensive dependencies. This enables DCC to acquire a comprehensive integration of temporal information across scales, ensuring that the long-time correlations are accurately represented within the model's predictions. Such an advantage is reflected in the enhanced performance of U-AnDi as well as WE's competitive superiority over other counterparts, where both models incorporate the DCC.}

\item {\it Gated activation unit: Processing long-range dependencies}

{To adeptly navigate the intricate long-time correlations in anomalous diffusion, the integration of GAU is engineered to leverage its gating mechanism to selectively emphasize salient temporal features. This selective filtration and prioritization by GAU are not merely structural benefits but are directly responsive to the characteristic memory effects and non-local interactions of anomalous diffusion. By modulating the flow of information, the GAU ensures that the model's focus is attuned to the most critical aspects of the data, which are indicative of the underlying long-time correlations. The superior performance of models equipped with GAU across all subtasks robustly demonstrates the proficiency of GAU in effectively processing long-time correlation information.}

\item {\it U-Net architecture: Enhancing detection of local changes}

{When segmenting the anomalous diffusion trajectories, the abrupt transitions and heterogeneous local dynamics demand a model's ability not only in emphasizing long-time correlations but also in precisely delineating local variations. The U-Net architecture excels in this regard, capturing fine details through skip connections, while concurrently grasping the broader context with its encoder-decoder framework. This dual capability makes U-Net exceptionally well-equipped to identify state transition points in anomalous diffusion. In this work, the integration of WE within the U-Net framework further refines the model's segmentation performance, covering all metrics. This result reinforces the utility of U-Net in the segmentation and changepoint detection of anomalous diffusion.}
\end{itemize}

{The interpretability of U-AnDi's core components is a tangible reflection of the model's alignment with the complex nature of anomalous diffusion. By capturing long-time correlations and enhancing the detection of local changes, these core components collectively contribute to the model's robust performance across varied segmentation tasks, bridging the gap between advanced machine learning techniques and the fundamental principles of anomalous diffusion.}

\section{Implementation of U-AnDi on Real-World Anomalous Diffusion Data}\label{realworld}

\begin{figure*}
\centering
\includegraphics[width=17.0cm]{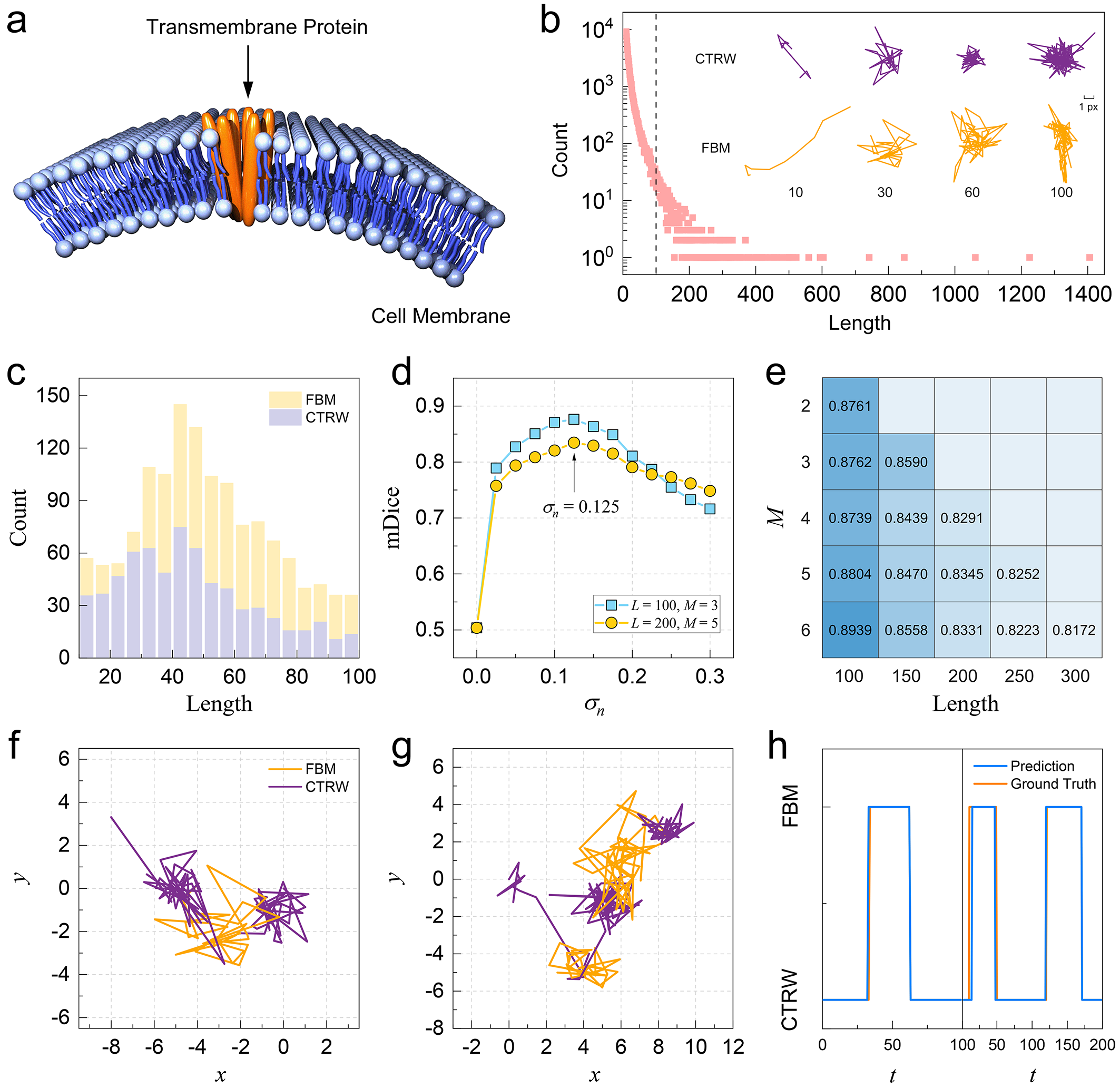}
\caption{(a) Schematic diagram of the selected real-world anomalous diffusion system: Diffusion of transmembrane proteins on cell membrane surfaces. (b) Logarithmically plotted distribution of trajectory lengths in the experimental dataset, with the dashed line denoting the length at 100. Inset: Representative instances of selected short trajectories of CTRW and FBM with different lengths, with labels assigned by WADNet. (c) The quantity distribution of short CTRW and FBM trajectories spanning various lengths from 10 to 100. (d) Segmentation performance of U-AnDi at different noise standard deviations $\sigma_n$. (e) mDice scores of U-AnDi on validation sets under diverse $L$ and $M$ parameters. (f)-(g) Trajectories from the validation sets with parameters $L=100, M=3$ in (f) and $L=200, M=5$ in (g). (h) Comparisons between the predictions (blue) and ground truths (orange) for the trajectory in (f) (left) and the trajectory in (g) (right), underscoring a high-level consistency.}
\label{fig:fig12}
\end{figure*}

In this section, we apply the U-AnDi model to real-world anomalous diffusion data and examine its performance on the semantic segmentation of experimental trajectories. The selected real-world system is the diffusion of transmembrane proteins on cell membrane surfaces, as illustrated by the schematic diagram in Fig. \ref{fig:fig12}(a). The pertinent data are obtained through single-particle tracking techniques \cite{Manzo1,Shen,Qian,Saxton,Torreno-Pina,TRamWAy} from experimental observations, and are provided by N. Granik {\it et al.} in Ref. \cite{Granik}. The content of this section is structured as follows. Sec. \ref{realworld1} provides an overview of the raw experimental trajectory dataset, and describes the segmentation tasks along with the approach for preparing the validation set. The training strategy is outlined in Sec. \ref{realworld2}. In Sec. \ref{realworld3}, we demonstrate and discuss the performance of the U-AnDi model on interpreting the anomalous diffusion dynamics of transmembrane proteins.

\subsection{Dataset description and segmentation tasks}\label{realworld1}

\begin{figure}[b]
\centering
\includegraphics[width=8.5cm]{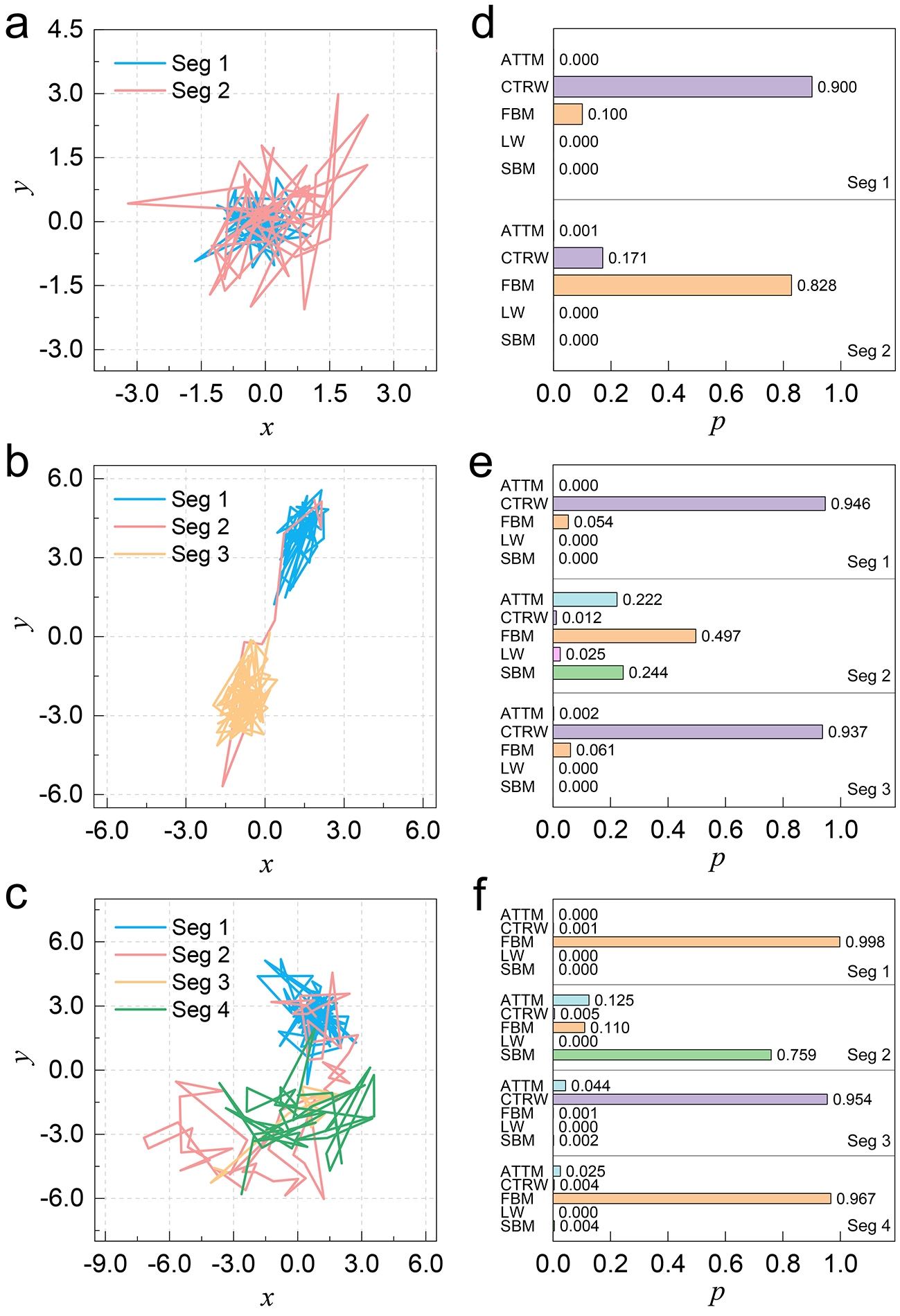}
\caption{(a)-(c) Representative natural trajectories from the experimental dataset with lengths surpassing 100. Segments identified by U-AnDi are distinguished by distinct colors. (d)-(f) Probabilities of diffusion models provided by WADNet for segments in (a)-(c), respectively.}
\label{fig:fig13}
\end{figure}

The dataset with respect to the diffusion of transmembrane proteins on membrane surfaces comprises a total of 70221 2D trajectories, in which the $x$ and $y$ pixel locations are provided. The distribution of trajectory lengths is displayed in Fig. \ref{fig:fig12}(b), with the counts plotted on a logarithmic scale to enhance clarity in visualization. An evident long tail distribution can be identified. In detail, trajectories of lengths no greater than 100 account for 98.6\% (69259) of the entire dataset , while the maximum length in this dataset reaches 1406. Diffusion models of these trajectories have been previously assessed by machine learning approaches in Ref. \cite{Granik}, with N. Granik {\it et al.} proposing that the predominant dynamics consist of both CTRW and FBM. Therefore, this dataset offers us an excellent opportunity to test the segmentation performance of U-AnDi on experimental data.

Two specific segmentation tasks are addressed here: one involving labeled data and the other utilizing unlabeled data. In the task with labeled data, the validation set is prepared as below. Trajectories to be evaluated are generated by {combining} short trajectories with lengths not exceeding 100 from the experimental dataset. Diffusion-model labels of these short trajectories are assigned by WADNet, a deep learning model capable of providing the probability of the diffusion model to which the input trajectory is most likely to belong. Further details regarding WADNet can be found in our previous work \cite{Li}. Leveraging the model probabilities provided by WADNet for these short trajectories, we carefully select CTRW and FBM instances that exhibit high confidence levels (with probabilities of CTRW or FBM exceeding 0.9) to generate extended trajectories for further evaluation. Representative samples of these selected short trajectories with different lengths are illustrated in the inset of Fig. \ref{fig:fig12}(b). The quantities of short CTRW and FBM trajectories spanning various lengths from 10 to 100 are presented in Fig. \ref{fig:fig12}(c). By setting the overall trajectory length $L$ and the number of segments $M$, we can utilize these short trajectories to create the validation set, thereby facilitating the evaluation of U-AnDi's performance. In particular, we establish a rule specifying that the average length of segments ($L/M$) should not exceed 50. This constraint ensures the prevention of trajectory generation errors or the dominance of overly long segments.

On the other hand, in the task concerning unlabeled data, we employ the U-AnDi model to segment those long trajectories with lengths exceeding 100. To examine the performance of our model, the segments distinguished by U-AnDi are classified by WADNet to identify their corresponding diffusion models. Due to the absence of ground truth labels, we cannot use a conventional evaluation metric like the mDice to assess the segmentation results. Instead, we rely on the confidence levels (probabilities) provided by WADNet to evaluate the effectiveness of our model.

\subsection{Training scheme}\label{realworld2}

The architecture of the U-AnDi model and training methods in this section remain almost identical to those in subtask 2. One deviation is a reduction in the filter number of the last 1D convolutional layer from 5 to 2. This modification is due to the task at hand, which requires semantic segmentations of just two diffusion models, CTRW and FBM. Another update is the initialization of model weights. To accelerate the training process, we adopt a transfer learning strategy where the optimal model weights from subtask 2 are employed to initialize the model. This strategy allows the model to reach optimal performance within approximately 10 epochs, significantly reducing the training time.

In addition, taking the inevitable errors and noise during experimental observations into account, Gaussian white noise with a zero mean and a standard deviation of $\sigma_n$ is added when generating simulated trajectories for training. In order to pinpoint the optimal $\sigma_n$ and better mimic the circumstances of real experimental data, we select two validation sets comprising of experimental trajectories to evaluate the performance of U-AnDi at different $\sigma_n$. The parameters for these two validation sets are $L=100, M=3$ and $L=200, M=5$, respectively, with each set consisting of 1000 trajectories. As demonstrated in Fig. \ref{fig:fig12}(d), for both validation sets, U-AnDi achieves the best performance (the highest mDice score) when $\sigma_n=0.125$. Consequently, simulated trajectories with the standard deviation of noise $\sigma_n=0.125$ are chosen to train the U-AnDi model in this section.

\subsection{Performance of U-AnDi on two tasks regarding experimental trajectories}\label{realworld3}

For the task involving labeled data, we investigate the performance of U-AnDi on validation sets with different parameters, as quantified by the mDice scores. The results are summarized in Fig. \ref{fig:fig12}(e), where each validation set contains 1000 trajectories. It is observed that all mDice scores are above 0.8 for trajectory lengths ranging from 100 to 300, indicating an exceptional capability of U-AnDi in segmenting experimental trajectories. We also observe that segmentation performance decreases as trajectory length increases. This trend may be due to the fact that trajectories with larger $L$ are composed of longer segments, which introduce more noise and thus can mislead the model's classification accuracy. Moreover, to illustrate the segmentation results intuitively, we present two representative samples from the validation sets, along with their corresponding predicted labels, in Figs. \ref{fig:fig12}(f) and \ref{fig:fig12}(g). Comparisons between the predictions and ground truths of these two samples are presented in the left and right panels of Fig. \ref{fig:fig12}(h), respectively, with a high degree of consistency being observed as expected.

Next, in the task with unlabeled data, we utilize the trained U-AnDi model to segment natural trajectories in the experimental dataset with lengths exceeding 100. In the absence of ground truth labels, we leverage WADNet to obtain probabilities of diffusion models for the segments identified by U-AnDi. To provide a concrete illustration of the model's effectiveness, three representative trajectories and their corresponding results are displayed in Fig. \ref{fig:fig13}. As distinguished by the distinct colors, trajectories in Figs. \ref{fig:fig13}(a)-\ref{fig:fig13}(c) are segmented by U-AnDi into 2, 3, and 4 segments, respectively. The corresponding probabilities of diffusion models for these segments are presented in Figs. \ref{fig:fig13}(d)-\ref{fig:fig13}(f). Observations suggest that, for the majority of segments, the diffusion models with the highest confidence levels (probabilities) remain CTRW and FBM. Nevertheless, for the second segment of the trajectory in Fig. \ref{fig:fig13}(c), SBM is identified by WADNet as the most probable diffusion model. The result derived from our analysis on the task involving unlabeled data affirms the conclusion of N. Granik {\it et al.} that the diffusion of transmembrane proteins on cell membranes primarily exhibits CTRW and FBM characteristics \cite{Granik}. However, it also hints that this conclusion may not be a comprehensive description of the entire diffusion process, which could potentially entail a higher degree of complexity.

\section{Conclusion}\label{conclusion}

{In summary, we have introduced U-AnDi, a novel deep learning model for the semantic segmentation of anomalous diffusion trajectories, which integrates the strengths of the dilated causal convolution (DCC), gated activation unit (GAU), and U-Net architecture. A variety of segmentation tasks have been addressed, demonstrating a superior performance of U-AnDi against other methods. Comparative evaluations underscore the interpretability of U-AnDi's core components and, to a certain extent, mitigate the ``black box" nature of deep learning models, aligning them with the complex physics of anomalous diffusion.} Further, U-AnDi also exhibits an excellent capability in dealing with the real-world anomalous diffusion data, which is examined using the experimental dataset regarding the diffusion of transmembrane proteins on cell membrane surfaces. The supporting codes in this work are accessible at our GitHub repository \cite{U-AnDi}.

Despite the promising performance of our model, there still exist opportunities for further refinements. The first is optimizing the detection of changepoints among different diffusion states, where reducing FP and FN can be a critical task for future training and finetuning. Secondly, the implementation of U-AnDi is trained to recognize only five diffusion models in this study. Extending the model's ability to identify a larger variety of diffusion models would enhance the utility of U-AnDi in future research. Such an expansion would enable researchers to apply U-AnDi more effectively in the segmentation and property prediction of experimentally observed heterogeneous diffusion dynamics.

Additionally, considering that U-AnDi does not require specific prior knowledge about anomalous diffusion, this model has the potential to be trained and perform segmentation tasks on other time-series datasets instead of anomalous diffusion. In other words, our method not only provides a powerful and versatile tool for better understanding the dynamics of anomalous diffusion, but it also holds promise to be generalized as a universal scheme for the segmentation of time-series data.

\begin{acknowledgments}
We thank Hui Zhao, Wentao Ju, and Gongyi Wang for helpful discussions. This work is supported by the National Natural Science Foundation of China (Grant No. 12104147) and the Fundamental Research Funds for the Central Universities.
\end{acknowledgments}

\appendix

\section{Implementation details of the U-AnDi model}\label{modeldetail}

The U-AnDi model is constructed by integrating the WADNet encoder into the U-Net architecture. The detailed structure of the WADNet encoder is depicted in Fig. 2(b). The dilation depth $d$ within the encoder is 5, and the filter number $f$ is consistent across all convolutional layers within a single encoder. The input tensor ${\bf x}$ initially passes through a standard causal convolution with a kernel size ($ks$) of 3. After that, the output ${\bf x}_1$ is processed by the subsequent layers, which consist of the gated activation unit (GAU) and 1D convolutional layer with $ks=1$, as highlighted by the rectangular dashed box in Fig. 2(b). Let ${\bf x}_k$ represent the input of the $k$th layer where $k$ ranges from 1 to $d$. The output of the $k$th layer ${\bf h}_k$ can be written as:
\begin{equation}
{\bf h}_k = W^1 * \left[\tanh \left(W_{f, k} * {\mathbf{x}}_k\right) \otimes \sigma\left(W_{g, k} * {\mathbf{x}}_k\right)\right].
\end{equation}
Here, * denotes the convolution operation, $\tanh(\cdot)$ and $\sigma(\cdot)$ refer to the hyperbolic tangent function and sigmoid function, respectively. $W^1$ represents the 1D convolution with $ks=1$. $W_{f, k}$ and $W_{g, k}$ are independent dilated causal 1D convolutions with $ks = 3$ and dilation $=2^{k-1}$. ${\bf h}_k$ is directly used as the input of next $(k+1)$th layer, meaning ${\bf x}_{k+1}={\bf h}_k$. The output ${\bf z}$ of an encoder can be finally obtained by ${\bf z} = {\bf x}_1 + \sum_{k=1}^{d}{\bf h}_k$. For convenience, we denote the operation of WADNet encoder as $\mathcal{W}(\cdot)$, i.e., ${\bf z} = \mathcal{W}({\bf x})$, in the following discussion. In particular, the encoder in the down-sampling block of U-Net is denoted as $\mathcal{W}^{\rm d}$, while the one in the up-sampling block is represented as $\mathcal{W}^{\rm u}$.

By incorporating this encoder into the U-Net structure, we construct the U-AnDi model, with its architecture displayed in Fig. 2(c). The down-sampling block consists of a WADNet encoder, a batch normalization layer, and a max-pooling layer. We employ four consecutive down-sampling blocks to perform feature extraction on the input tensor. In the $m$th block, the input ${\bf v}_m$ is processed as governed by:
\begin{align}
{\bf u}_m &= {\rm BN}\left[\mathcal{W}^{\rm d}_m({\bf v}_m)\right],\\
{\bf v}_{m+1}&= {\rm MaxPool}({\bf u}_m),
\end{align}
where $m=1,2,3,4$. BN and MaxPool represent the batch normalization \cite{Ioffe} and max-pooling operators, respectively. The filter number $f_m$ in encoder $\mathcal{W}^{\rm d}_m$ is $64\times 2^{m-1}$ for subtask 1 and $128\times 2^{m-1}$ for subtask 2. The pooling size of max-pooling layer is set to 2.

Next, the output from the 4th down-sampling block, ${\bf v}_5$, is processed through five consecutive up-sampling blocks to generate full-resolution feature maps, which are then used for point-wise predictions. The up-sampling block is comprised of a WADNet encoder, a batch normalization layer, an up-sampling layer using the nearest neighbor algorithm, and a 1D convolutional layer with $ks=1$. In particular, skip connections are applied to get the input ${\bf p}_n$ of the $n$th up-sampling block when $n \geq 2$, as displayed by the dashed lines in Fig. 2(c). These connections serve to reduce the loss of fine-grained details during the up-sampling process and can be mathematically described as:
\begin{align}
\tilde{{\bf p}}_n & = W_{n-1}^{1} * {\rm UP}\left\{{\rm BN}\left[\mathcal{W}^{\rm u}_{n-1}\left({\bf p}_{n-1}\right)\right]\right\}, \\ {\bf p}_n & =  {\rm Concat}\left(\tilde{{\bf p}}_n, {\bf u}_{6-n}\right), \label{concat}
\end{align}
where $n=2,3,4,5$ and ${\bf p}_1 = {\bf v}_5$. UP denotes the up-sampling operator, with a scale factor set as 2. {Concat is the concatenation operator, which is applied in the channel dimension}. The filter number $f_n$ in encoder $\mathcal{W}^{\rm u}_n$ is $64\times 2^{5-n}$ for subtask 1 and $128\times 2^{5-n}$ for subtask 2. The filter number of the 1D convolutional layer $W_n^1$ is half that of in encoder $\mathcal{W}^{\rm u}_n$. Note that when performing the skip connection in Eq. \eqref{concat}, the shapes of $\tilde{{\bf p}}_n$ and ${\bf u}_{6-n}$ might not match. In such cases, we apply padding to $\tilde{{\bf p}}_n$ to make it compatible with ${\bf u}_{6-n}$ for concatenation. In the last up-sampling block, the up-sampling layer is removed, and the filter number of the convolutional layer is adjusted (1 for subtask 1 and 5 for subtask 2). This yields a final output ${\bf p}_f$ of the U-AnDi model, written as:
\begin{equation}
{\bf p}_f = W_{5}^{1} * {\rm BN}\left[\mathcal{W}^{\rm u}_5\left({\bf p}_5\right)\right].
\end{equation}
To illustrate the details of the U-AnDi architecture more clearly, we take a trajectory with $L=500$ as the input and present the shapes of intermediate tensors for subtask 2 in Table \ref{tab:table3}. The shape follows the format [length, channel].

\begin{table}
\caption{\label{tab:table3}
The shapes of intermediate tensors for subtask 2 when length of input trajectory is 500.}
\begin{ruledtabular}
\begin{tabular}{cccc}
 Tensor & Shape & Tensor & Shape \\
\hline
${\bf v}_1$ & [500, 2] & ${\tilde{\bf p}}_2$ & [62, 1024] \\
${\bf v}_2$ & [250, 128] & ${\bf p}_2$ & [62, 2048]  \\
${\bf v}_3$ & [125, 256] & ${\tilde{\bf p}}_3$ & [124, 512] \\
${\bf v}_4$ & [62, 512] & ${\bf p}_3$ & [125, 1024] \\
${\bf v}_5$(${\bf p}_1$) & [31, 1024] & ${\tilde{\bf p}}_4$ & [250, 256]\\
${\bf u}_1$ & [500, 128] & ${\bf p}_4$ & [250, 512] \\
${\bf u}_2$ & [250, 256] & ${\tilde{\bf p}}_5$ & [500, 128] \\
${\bf u}_3$ & [125, 512] & ${\bf p}_5$  & [500, 256] \\
${\bf u}_4$ & [62, 1024] & ${\bf p}_f$  & [500, 5] \\
\end{tabular}
\end{ruledtabular}
\end{table}

The post-processing techniques used to obtain the final predictions for both subtasks are summarized below. In subtask 1, which is a point-wise regression task, the output ${\bf p}_f$ from a trained U-AnDi model serves as the predicted diffusion exponents. To ensure a reasonable prediction with exponents ranging from 0.05 to 2, values outside the interval are clipped to the interval edges in ${\bf p}_f$. On the other hand, ${\bf p}_f$ is transformed using the following equations in subtask 2 to address this point-wise classification task:
\begin{align}
{\bf X}^{\rm Prob}_i &= {\rm Softmax}\left({\bf p}_{f,i}\right),\\
{\bf X}^{\rm Pred}_i &= \arg \max\left({\bf X}^{\rm Prob}_i\right). \label{argmax}
\end{align}
Here, $i$ denotes the $i$th trajectory. Softmax represents the softmax activation function. ${\bf X}^{\rm Prob}_i$ and ${\bf X}^{\rm Pred}_i$ are the predicted probability and label, respectively.

\section{Data preprocessing and training scheme of the U-AnDi model}\label{training}

\begin{figure}
\centering
\includegraphics[width=8.6cm]{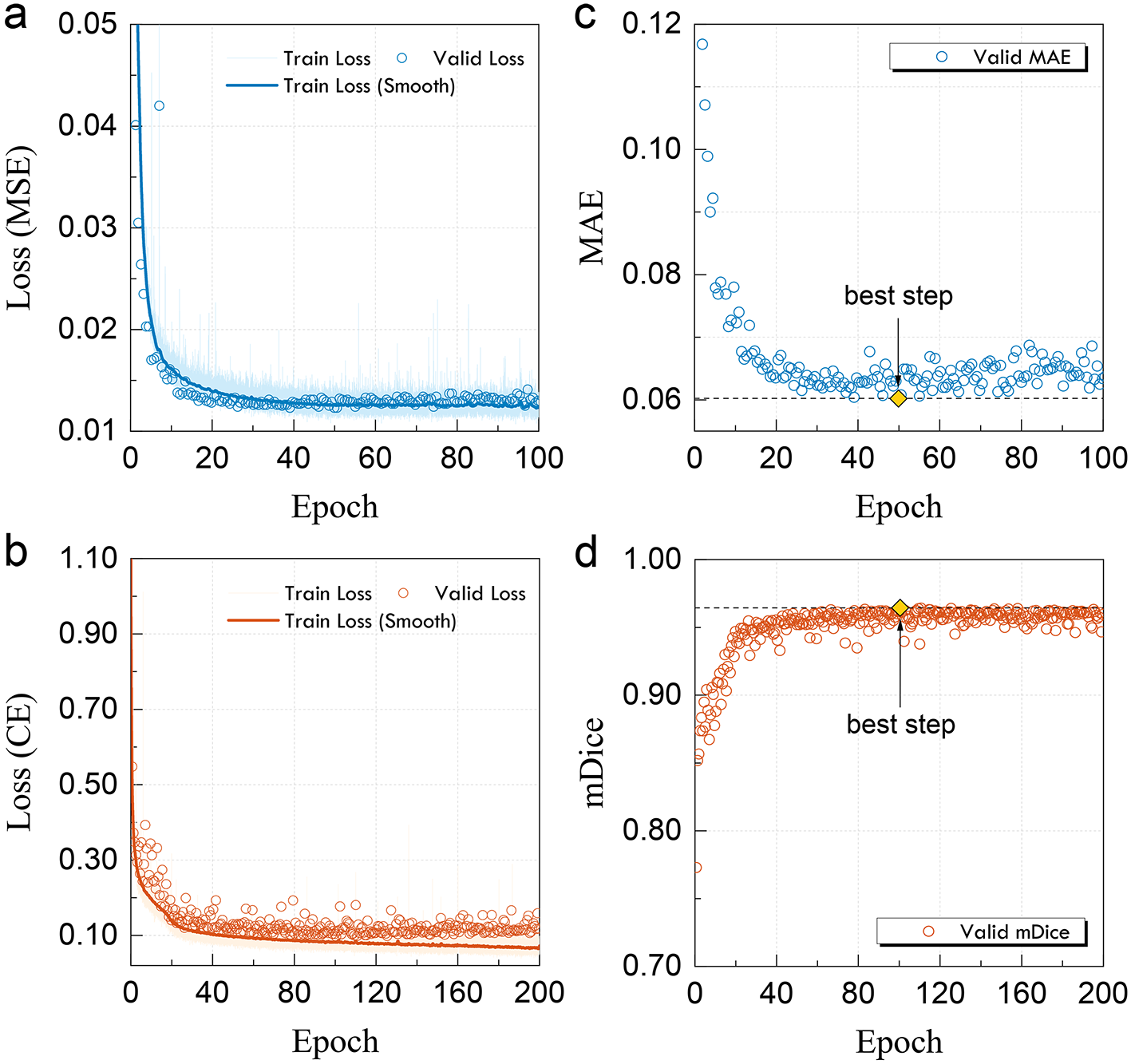}
\caption{(a)-(b) Evolution of the loss function during training for subtask 1 (a) and subtask 2 (b). The thick line represents a smoothed version of the original train loss to provide better visual guidance. (c)-(d) Epoch dependence of the valid metric for subtask 1 (c) and 2 (d). The dashed line denotes the value of the best metric, and the yellow dot highlights the location of the best step.}
\label{fig:figA1}
\end{figure}

To train our deep learning model, 1 000 000 trajectories are generated as the dataset for a fixed length parameter $L$. This dataset is randomly split into two parts: 80\% for the training set and 20\% for the validation set. The raw trajectory data is independently normalized along the $x$ and $y$ dimensions, achieving a mean of 0 and a standard deviation of 1 for each dimension. {Then, the normalized data from both dimensions, $\left[x(1), x(2), x(3), \cdots, x(L)\right]$ and $\left[y(1), y(2), y(3), \cdots, y(L)\right]$, are concatenated to form a sequence:
\begin{equation}
\left\{\begin{array}{l}
{\left[x(1), x(2), x(3), \cdots, x(L)\right]} \\
{\left[y(1), y(2), y(3), \cdots, y(L)\right]}
\end{array}\right\},
\end{equation}}
which is utilized as the input of the U-AnDi model.

The selected loss functions are mean squared error (MSE) for subtask 1 and cross entropy (CE) for subtask 2. The model is trained using the back-propagation algorithm with a batch size of 512 on a single NVIDIA A100 GPU. The learning rate is 0.0002, and the optimizer is Adam. In particular, an L2 penalty is applied for subtask 2 with a weight decay of $2\times 10^{-5}$. The training process involves 100 epochs for subtask 1 and 200 epochs for subtask 2. The model performance on the validation set is evaluated every 1000 steps.

We use the trajectory data with $L = 500$ as the example to illustrate the evolution of loss function during training, as shown in Fig. \ref{fig:figA1}(a) for subtask 1 and Fig. \ref{fig:figA1}(b) for subtask 2. A rapid convergence of the train loss can be observed for both tasks, with the valid loss exhibiting a similar trend. This indicates that our model is fully capable of accomplishing both tasks without encountering over-fitting issues. Furthermore, the optimal weight of the U-AnDi model is determined based on the best metric achieved on the validation set. The epoch dependences of valid metrics for subtask 1 and 2 are presented in Figs. \ref{fig:figA1}(c) and \ref{fig:figA1}(d), respectively. As indicated by the yellow dot, the selected weight corresponds to the step where the best metric is reached.

\section{{Details of comparative models}}\label{compare}

{In this appendix, we provide a comprehensive description of the models utilized for comparative analysis in our study. These models are specifically chosen and designed to enable a robust comparative evaluation against our proposed U-AnDi model regarding the semantic segmentation of anomalous diffusion trajectories.}

{\begin{itemize}
\item Long short-term memory (LSTM) \cite{Hochreiter}: The LSTM model we use in this work is structured as a three-layered stack, where the dimensionality of hidden state is set as 64.
\item Gated recurrent unit (GRU) \cite{Kyunghyun}: In this work, the GRU model is also architecturally configured as a three-layered stack, wherein each layer possesses a hidden state with a dimensionality of 64.
\item Transformer (TFM) \cite{Ashish}: Analogous to the LSTM and GRU, the TFM model is structured with three layers of Transformer encoders, where each encoder employs an embedding dimensionality of 128.
\item WADNet encoder (WE): The WE model is derived from the encoder part of our previously proposed WADNet structure \cite{Li}. The model's hyperparameters align with the WADNet encoder in the first down-sampling block of U-AnDi.
\item {U-CNN: The U-CNN model represents a modified architecture developed from U-AnDi, wherein WADNet encoders are substituted with simpler dilated causal 1D CNN layers. Specifically, the input and output channels, kernel size, and dilation rate of a CNN layer are maintained consistent with the corresponding WADNet encoder.}
\end{itemize}}

{In addition, we provide a concise overview of the divide-and-conquer moment scaling spectrum (DC-MSS) here as described in Ref. \cite{multistep}. DC-MSS is a technique designed for analyzing single-particle tracking data. It works by dividing trajectories into shorter segments and computing their moment scaling spectrum. Through detecting shifts in the scaling exponents of these segments, DC-MSS pinpoints changes in diffusion behavior. Notably, it excels at distinguishing transitions among different modes of motion, including Brownian diffusion, confined diffusion, directed motion, and the immobile state. In this work, we employ the DC-MSS technique with the default hyper-parameters as outlined in Ref. \cite{multistep}.}

\section{Post-processing technique in subtask 1}\label{postprocess}

Given a continuous point-wise prediction of diffusion exponents, denoted as:
\begin{equation}
\alpha_1, \alpha_2, \alpha_3, \cdots, \alpha_L,
\end{equation}
for a trajectory of length $L$, we first calculate the differences between adjacent elements and get the sequence:
\begin{equation}
\beta_1, \beta_2, \beta_3, \cdots, \beta_{L-1},
\end{equation}
where $\beta_i = \alpha_{i+1}-\alpha_i$. Next, we set a threshold value $\epsilon$ and identify all indices $i$ satisfying $\beta_i > \epsilon$. These indices form a new sequence:
\begin{equation}
i_1, i_2, i_3, \cdots, i_k.
\end{equation}
If two consecutive indices $i_m$ and $i_{m+1}$ in this sequence meet $i_{m+1} - i_m \geq 10$, we denote $i_m+1$ as the starting point of the current segment, and $i_{m+1}$ as the endpoint of this segment. Then, the changepoint between two successive predicted segments $\left[i_a+1, i_{a+1}\right]$ and $\left[i_b+1, i_{b+1}\right]$ can be identified in the interval $\left[i_{a+1}+1, i_b\right]$. We select the point within this interval that exhibits the maximum absolute value of the derivative as the changepoint. Applying this rule, we can obtain the sequence of changepoints, given by:
\begin{equation}
c_1, c_2, c_3, \cdots, c_n.
\end{equation}
Here, $c_2,c_3,\cdots,c_{n-1}$ are the predicted changepoints, and $c_1$ and $c_n$ are set as 1 and $L$, respectively. Based on the detection result of changepoints, the predicted diffusion exponent $\tilde{\alpha}_i$ of the $i$th segment can be given as:
\begin{equation}
\tilde{\alpha}_i = \frac{1}{c_{i+1}-c_i}\sum_{j=c_i}^{c_{i+1}}{\alpha_{j}},
\end{equation}
where $i=1,2,\cdots,n-1$.

\begin{figure}
\centering
\includegraphics[width=8.5cm]{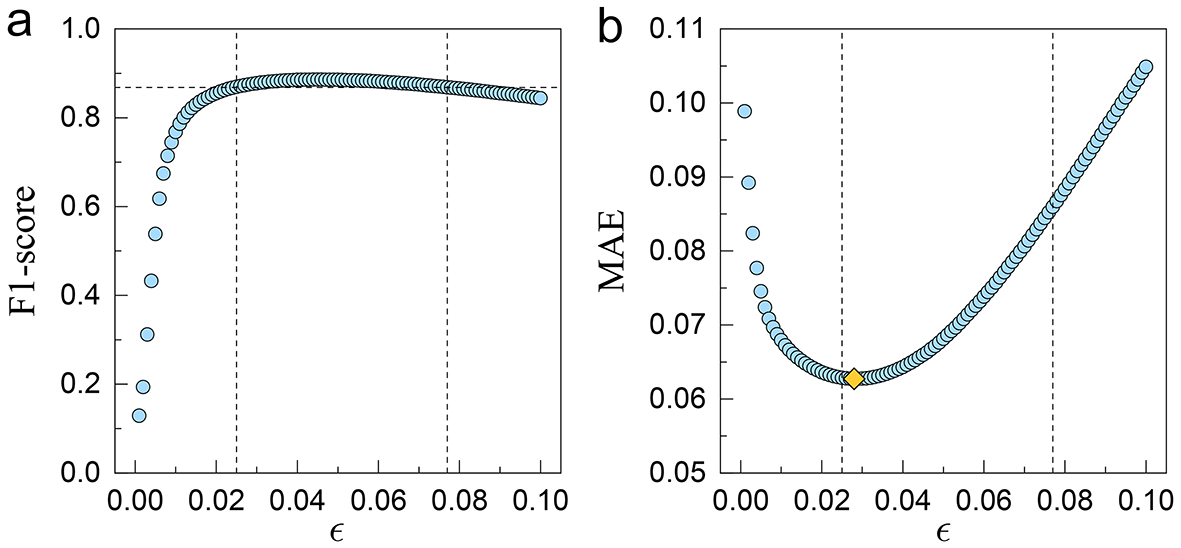}
\caption{{(a) F1-score for changepoint detection as a function of $\epsilon$. The horizontal dashed line signifies 98\% of the maximum F1-score, and the vertical dashed lines indicate the candidate interval for $\epsilon$, within which the F1-score retains at least 98\% of its peak value. (b) MAE as a function of $\epsilon$. The optimal $\epsilon$ is selected to minimize the MAE within the candidate interval (vertical dashed lines). The yellow dot highlights the chosen optimal $\epsilon$ value of 0.028 for this work.}}
\label{fig:figA2}
\end{figure}

{In particular, the method for determining the threshold value $\epsilon$ is summarized here. A larger $\epsilon$ tends to yield higher precision in detecting changepoints, albeit with a lower recall, while a smaller $\epsilon$ may compromise the precision but enhance recall. Therefore, as $\epsilon$ increases, the F1-score for changepoint detection initially rises and subsequently diminishes, as depicted in Fig. \ref{fig:figA2}(a). Concurrently, alterations in $\epsilon$ also induce variation in the model's MAE, as illustrated in Fig. \ref{fig:figA2}(b). Under these circumstances, the selection of $\epsilon$ emerges as a result of balancing a trade-off between MAE and F1-score on the validation set. In more detail, we select the optimal $\epsilon$ from a candidate interval where the F1-score is at least 98\% of its maximum value, as indicated by the vertical dashed lines in Fig. \ref{fig:figA2}(a). Within this interval, we seek the $\epsilon$ that minimizes the MAE value. In this work, 0.028 is identified as the optimal $\epsilon$, as highlighted by the yellow dot in Fig. \ref{fig:figA2}(b).}

\section{{Addressing concurrent exponent and model variations using multi-task learning}}\label{bothchange}

{The versatility of U-AnDi has been previously highlighted for its effectiveness at segmenting trajectories based on variations in either the diffusion exponent or the dynamic model. However, real-world scenarios often present more intricate challenges where both exponent and model might concurrently vary within the same trajectory. To address this issue, we have further enhanced U-AnDi's capabilities through the integration of a multi-task learning strategy in this section.}

\begin{figure}
\centering
\includegraphics[width=8.6cm]{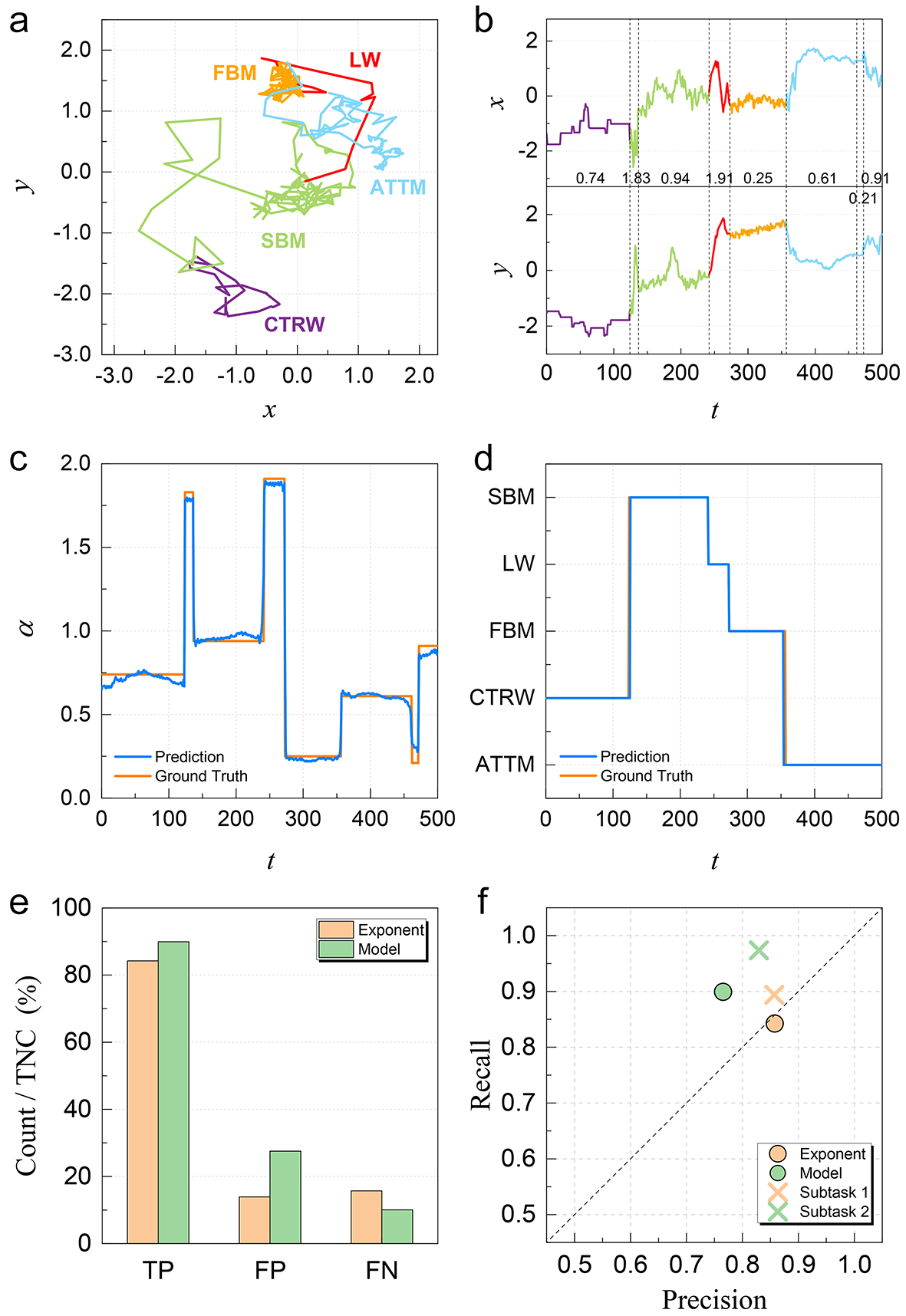}
\caption{{(a) Illustration of a composite trajectory of length 500 with different diffusion models. (b) Time evolution of trajectories in (a) for both $x$ and $y$ dimensions, showing variations of the diffusion exponent within a single model segment. (c)-(d) Segmentation results for the trajectory in (a) based on the diffusion exponent (c) and the diffusion model (d). (e) Distribution of TP, FP, and FN as a percentage of TNC for both segmentation tasks. (f) Comparative analysis regarding precision and recall of changepoint detection results with those from subtasks 1 and 2.}}
\label{fig:figA3}
\end{figure}

{The method for simulating trajectories with concurrent variations in both exponent and model is an extension of the approach used in subtask 2. We combine segments from different diffusion models to generate a composite trajectory, as illustrated in Fig. \ref{fig:figA3}(a). A notable distinction in this enhanced simulation is the allowance for variations of the diffusion exponent within a segment characterized by a single dynamic model, as depicted in Fig. \ref{fig:figA3}(b). To enrich the diversity of simulated trajectories, we expand the number of segments, $M$, from the initial range of 2 to 5 to an extended range of 3 to 8.}

{Subsequently, we employ a multi-task learning approach to enable U-AnDi to effectively segment these trajectories. In detail, we utilize the intermediate output of U-AnDi, just prior to the final 1D CNN layer, which is denoted as ${\bf p}_o$ and expressed as ${\bf p}_o = \mathrm{BN}\left[\mathcal{W}^{\rm u}_5\left(\mathbf{p}_5\right)\right]$. This intermediate output is directed into two separate 1D CNN head layers, $W_{f_1}^1$ and $W_{f_2}^1$. The layer $W_{f_1}^1$, with an output ${\bf p}_{f_1}=W_{f_1}^1 * {\bf p}_o$, is dedicated to a point-wise regression task, aiming to segment based on the diffusion exponent. In contrast, the other layer $W_{f_2}^1$ focuses on a point-wise classification task, targeting the segmentation of the diffusion model and leading to an output ${\bf p}_{f_2}=W_{f_2}^1 * {\bf p}_o$. After that, we compute the loss functions $\mathcal{L}_1$ and $\mathcal{L}_2$ for ${\bf p}_{f_1}$ and ${\bf p}_{f_2}$ using MSE and CE respectively. The final loss function $\mathcal{L}$ for this task is given as $\mathcal{L} = \mathcal{L}_1 + \mathcal{L}_2$. The data size and training method are consistent with those in subtask 2.}

{In line with the presentation style of subtask 1 and 2, we demonstrate U-AnDi's segmentation performance on trajectories of length 500 here. To visualize this intuitively, segmentation results for the diffusion exponent and dynamic model of the trajectory in Fig. \ref{fig:figA3}(a). are illustrated in Figs. \ref{fig:figA3}(c) and \ref{fig:figA3}(d), respectively. As expected, both predictions display a high degree of alignment with the ground truth. The results on the validation set yield an MAE of 0.1376 and an mDice of 0.9474, indicating that U-AnDi can effectively segment trajectories where the exponent and model concurrently vary.}

{On the other hand, when it comes to the detection of changepoints, Fig. \ref{fig:figA3}(e) reveals that TP consistently ranks highest for both segmentation tasks, whether based on the exponent or the model. F1-scores achieved for these two detections are 0.8501 and 0.8269, respectively, underscoring a robust capability of U-AnDi in identifying changepoints. Furthermore, as depicted in Fig. \ref{fig:figA3}(f), we compare the detection results with those from subtasks 1 and 2. For both exponent and model changepoint detections, there's a decline in recall, with the latter also experiencing a noticeable decrease in precision. The common reduction in recall can be attributed to the increased complexity of the task. The diminished precision in detecting model changepoints might stem from variations of diffusion exponent within a single model segment. These variations potentially result in occasional misinterpretations and consequently lead the model to detect more transition points than actual.}

\section{{Simulation of multi-state trajectories in biological experiment scenarios}}\label{4statesgen}

{In this section, we introduce the simulation techniques in detail for generating trajectories that represent different states commonly observed in biological experiments. These techniques are primarily adapted from the procedures described in Ref. \cite{multistep}. While the simulation parameters for each individual trajectory remain constant, they differ across the various trajectories. The chosen parameters align closely with those in Ref. \cite{multistep} to ensure an equitable evaluation of the DC-MSS method's segmentation performance.
\begin{itemize}
\item Brownian diffusion (BD): Simulating the free Brownian diffusion requires a series of $x$ and $y$ displacements. Each displacement is drawn from a Gaussian distribution $\mathcal{N}(0, \sqrt{2D\delta t})$, where $D$ is the diffusion coefficient and $\delta t$ is the time interval. We set $\delta t=1$ and select $D$ uniformly from the range 1.0 to 3.0.
\item Confined diffusion (CD): The generation of confined diffusion trajectories is akin to that of free Brownian diffusion trajectories. However, a reflection occurs once the subsequent coordinate of the random walker exceeds the boundary of the confined region. This confined region is defined as a circle with a radius $R$. Here, the radius $R$ ranges from 3.0 to 5.0, while the diffusion coefficient $D$ varies between 1.0 and 3.0.
\item Directed motion (DM): Each displacement in a trajectory of the directed motion consists of two components: a Brownian diffusive component with a diffusion coefficient $D \in [1.0, 3.0]$ and a directed component. The directed component is defined by $v \cos\theta \cdot \delta t$ for $x$-displacement and $v \sin \theta \cdot \delta t$ for $y$-displacement. The drift velocity, $v$, varies between 2.0 and 3.0, while the drift angle, $\theta$, spans from 0 to $2\pi$.
\item Immobile state (IS): The simulation of immobile state is achieved by generating $x$ and $y$ coordinates from a Gaussian distribution $\mathcal{N}(0, \sigma_p)$. Here, $\sigma_p$ represents the positional localization error, with values ranging from 0.8 to 1.6.
\end{itemize}}

{Leveraging these simulation techniques, we employ the same approach as in subtask 2 to combine trajectory segments of four diffusion states, ultimately producing the multi-state trajectories in biological experiment scenarios. The codes for trajectory generation can be found in our GitHub repository \cite{U-AnDi}.}

\begin{table*}
\caption{\label{tab:table4}
{Theoretical expressions of MSD and long-time MSD corresponding to the exp, multi-exp, and exp-cos correlations.}}
\begin{ruledtabular}
\begin{tabular}{clc}
 Correlation type & \multicolumn{1}{c}{Theoretical expression of MSD} & \multicolumn{1}{c} {Long-time MSD} \\
\hline
exp & $2A\tau\left[t+\tau\left(e^{-t / \tau}-1\right)\right]$ & $2A\tau t$\\
multi-exp &$2 A\left[\phi \tau_1 t+\phi \tau_1^2\left(e^{-t / \tau_1}-1\right)+(1-\phi) \tau_2 t+(1-\phi) \tau_2^2\left(e^{-t / \tau_2}-1\right)\right]$ & $2 A\left[\phi \tau_1+(1-\phi) \tau_2\right] t$\\
exp-cos & $\frac{2 A \tau}{\left(1+\tau^2 \omega^2\right)^2}\left[\left(1+\tau^2 \omega^2\right) t-\tau\left(1-\tau^2 \omega^2\right)+\tau e^{-t / \tau}\left[\left(1-\tau^2 \omega^2\right) \cos \omega t-2 \tau \omega \sin \omega t\right]\right]$ & $\frac{2 A \tau}{1+\tau^2 \omega^2} t$
\end{tabular}
\end{ruledtabular}
\end{table*}

\section{{Validating the method for generating trajectories with defined correlation functions}} \label{DH}

\begin{figure}
\centering
\includegraphics[width=8.6cm]{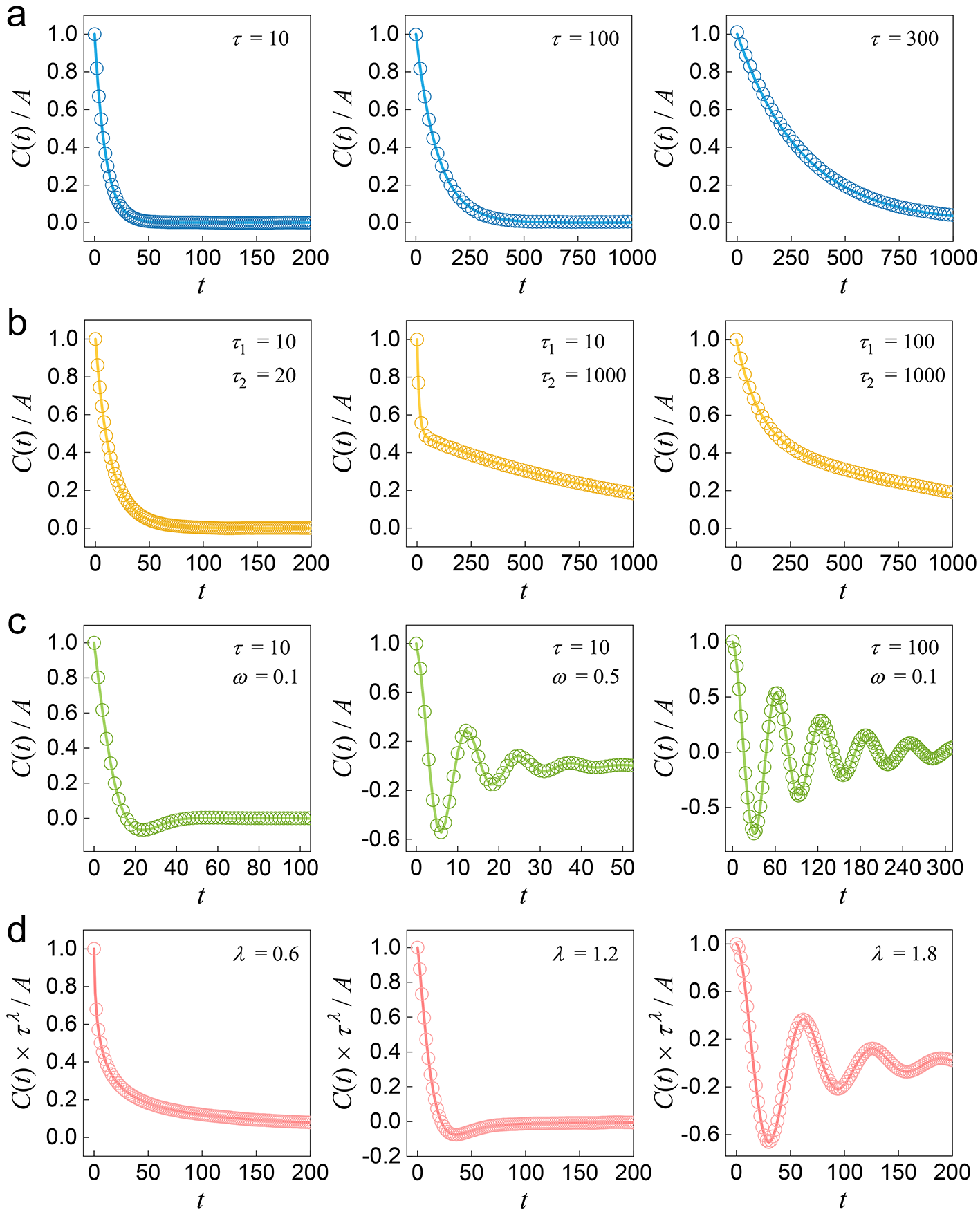}
\caption{{Comparisons of the velocity autocorrelation functions of simulated trajectories (dots) with their theoretical values (lines) under various parameters for the four types of long-time correlations: (a) exponential decay (exp); (b) multi-exponential decay (multi-exp) where $\phi=0.5$; (c) exponentially damped cosine wave (exp-cos); (d) Mittag-Leffler decay (M-L) where $\tau=10$.}}
\label{fig:figA4}
\end{figure}

\begin{figure}[b]
\centering
\includegraphics[width=8.6cm]{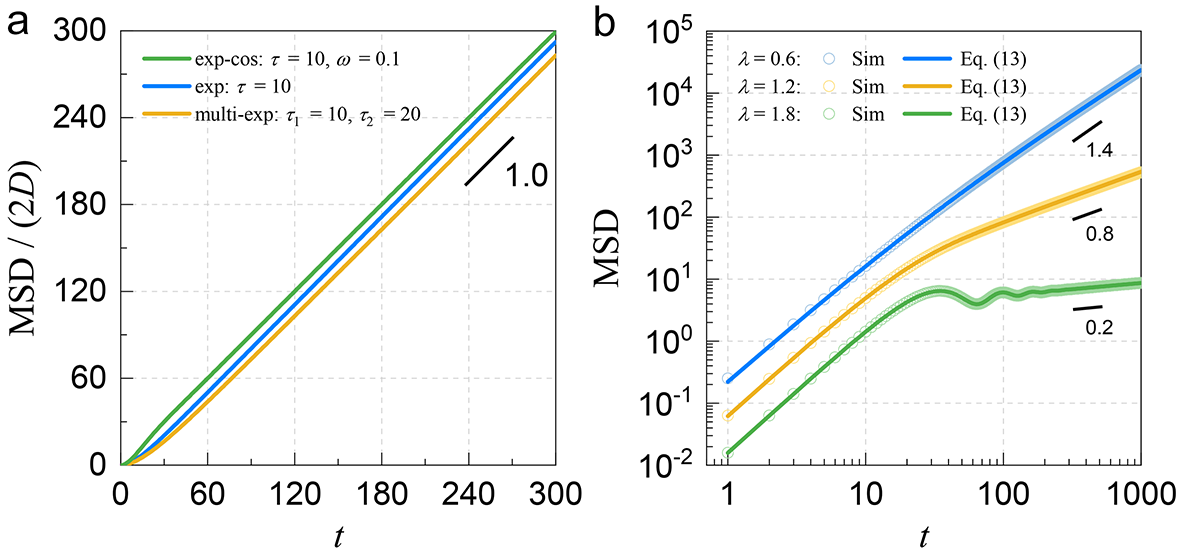}
\caption{{(a) Relationship between ${\rm MSD}/(2D)$ and time $t$ for simulated trajectories associated with exp, multi-exp, and exp-cos correlations. (b) Log-log plot of MSDs of simulated trajectories governed by a M-L correlation for specific $\lambda$ values of 0.6, 1.2, and 1.8 (represented by dots). The corresponding theoretical predictions, derived from Eq. \eqref{eqmsd}, are also depicted as lines.}}
\label{fig:figA5}
\end{figure}

{Within this appendix, we will validate the effectiveness of our trajectory generation method based on the Davies-Harte algorithm. For that purpose, we initially compare the velocity autocorrelation functions [VACFs, $C(t)$] of trajectories generated using this method under various parameters with their theoretical values. The results of this comparison are summarized in Fig. \ref{fig:figA4}. As observed, for the four types of long-time correlations, i.e., exp, multi-exp, exp-cos, and M-L, VACFs of simulated trajectories (represented by dots) align closely with their respective theoretical values (depicted by lines). This consistency strongly attests to the validity of our trajectory generation method.}

{On the other hand, from the perspective of MSD, we further validate the effectiveness of this method. Using the equation ${\rm MSD} = 2\int_0^t{(t-s)C(s){\rm d}s}$, we derive the theoretical expressions for exp, multi-exp, and exp-cos correlations, as detailed in Table \ref{tab:table4}. From the long-time behavior of MSD, we can discern that the trajectories associated with exp, multi-exp, and exp-cos correlations exhibit long-time Fickian diffusion theoretically. This diffusion type is typically represented by ${\rm MSD} \approx 2Dt$, where $D$ denotes the long-time diffusion coefficient. To substantiate this, we present the relationship between ${\rm MSD}/(2D)$ and time $t$ for simulated trajectories of these three types of correlations in Fig. \ref{fig:figA5}(a). The slope of 1.0 across these plots confirms that MSDs for these trajectories are consistent with their theoretical descriptions. For the M-L correlation, the theoretical MSD value is determined via numerical integration. In Fig. \ref{fig:figA5}(b), we present the MSDs of simulated trajectories for specific $\lambda$ values of 0.6, 1.2, and 1.8 in a log-log form. These are juxtaposed with their theoretical predictions, underscoring a pronounced congruence between them. Furthermore, as indicated by the slope, the long-time diffusion exponent $\alpha$ derived from our measurements relates to $\lambda$ following the relationship $\alpha = 2-\lambda$. This observation agrees well with the descriptions provided in Ref. \cite{Velocityauto}.}

{In summary, the alignment of the VACF and MSD with their theoretical values underscores the effectiveness of our trajectory generation method. This method offers a solid framework for studies focusing on segmenting anomalous diffusion trajectories that exhibit defined long-time correlations.}

\bibliography{Bibliography}

\providecommand{\noopsort}[1]{}\providecommand{\singleletter}[1]{#1}%
\begin{thebibliography}{81}%
\makeatletter
\providecommand \@ifxundefined [1]{%
 \@ifx{#1\undefined}
}%
\providecommand \@ifnum [1]{%
 \ifnum #1\expandafter \@firstoftwo
 \else \expandafter \@secondoftwo
 \fi
}%
\providecommand \@ifx [1]{%
 \ifx #1\expandafter \@firstoftwo
 \else \expandafter \@secondoftwo
 \fi
}%
\providecommand \natexlab [1]{#1}%
\providecommand \enquote  [1]{``#1''}%
\providecommand \bibnamefont  [1]{#1}%
\providecommand \bibfnamefont [1]{#1}%
\providecommand \citenamefont [1]{#1}%
\providecommand \href@noop [0]{\@secondoftwo}%
\providecommand \href [0]{\begingroup \@sanitize@url \@href}%
\providecommand \@href[1]{\@@startlink{#1}\@@href}%
\providecommand \@@href[1]{\endgroup#1\@@endlink}%
\providecommand \@sanitize@url [0]{\catcode `\\12\catcode `\$12\catcode
  `\&12\catcode `\#12\catcode `\^12\catcode `\_12\catcode `\%12\relax}%
\providecommand \@@startlink[1]{}%
\providecommand \@@endlink[0]{}%
\providecommand \url  [0]{\begingroup\@sanitize@url \@url }%
\providecommand \@url [1]{\endgroup\@href {#1}{\urlprefix }}%
\providecommand \urlprefix  [0]{URL }%
\providecommand \Eprint [0]{\href }%
\providecommand \doibase [0]{http://dx.doi.org/}%
\providecommand \selectlanguage [0]{\@gobble}%
\providecommand \bibinfo  [0]{\@secondoftwo}%
\providecommand \bibfield  [0]{\@secondoftwo}%
\providecommand \translation [1]{[#1]}%
\providecommand \BibitemOpen [0]{}%
\providecommand \bibitemStop [0]{}%
\providecommand \bibitemNoStop [0]{.\EOS\space}%
\providecommand \EOS [0]{\spacefactor3000\relax}%
\providecommand \BibitemShut  [1]{\csname bibitem#1\endcsname}%
\let\auto@bib@innerbib\@empty
\bibitem [{\citenamefont {Metzler}\ \emph {et~al.}(2014)\citenamefont
  {Metzler}, \citenamefont {Jeon}, \citenamefont {Cherstvya},\ and\
  \citenamefont {Barkaid}}]{Metzler}%
  \BibitemOpen
  \bibfield  {author} {\bibinfo {author} {\bibfnamefont {R.}~\bibnamefont
  {Metzler}}, \bibinfo {author} {\bibfnamefont {J.-H.}\ \bibnamefont {Jeon}},
  \bibinfo {author} {\bibfnamefont {A.~G.}\ \bibnamefont {Cherstvya}}, \ and\
  \bibinfo {author} {\bibfnamefont {E.}~\bibnamefont {Barkaid}},\ }\href@noop
  {} {\bibfield  {journal} {\bibinfo  {journal} {Phys. Chem. Chem. Phys.}\
  }\textbf {\bibinfo {volume} {16}},\ \bibinfo {pages} {24128} (\bibinfo {year}
  {2014})}\BibitemShut {NoStop}%
\bibitem [{\citenamefont {Klafter}\ and\ \citenamefont
  {Sokolov}(2005)}]{Klafter}%
  \BibitemOpen
  \bibfield  {author} {\bibinfo {author} {\bibfnamefont {J.}~\bibnamefont
  {Klafter}}\ and\ \bibinfo {author} {\bibfnamefont {I.~M.}\ \bibnamefont
  {Sokolov}},\ }\href@noop {} {\bibfield  {journal} {\bibinfo  {journal} {Phys.
  World}\ }\textbf {\bibinfo {volume} {18}},\ \bibinfo {pages} {29} (\bibinfo
  {year} {2005})}\BibitemShut {NoStop}%
\bibitem [{\citenamefont {Mason}\ and\ \citenamefont {Weitz}(1995)}]{Mason}%
  \BibitemOpen
  \bibfield  {author} {\bibinfo {author} {\bibfnamefont {T.~G.}\ \bibnamefont
  {Mason}}\ and\ \bibinfo {author} {\bibfnamefont {D.~A.}\ \bibnamefont
  {Weitz}},\ }\href@noop {} {\bibfield  {journal} {\bibinfo  {journal} {Phys.
  Rev. Lett.}\ }\textbf {\bibinfo {volume} {74}},\ \bibinfo {pages} {1250}
  (\bibinfo {year} {1995})}\BibitemShut {NoStop}%
\bibitem [{\citenamefont {Reis}(2016)}]{Reis}%
  \BibitemOpen
  \bibfield  {author} {\bibinfo {author} {\bibfnamefont {F.~D. A.~A.}\
  \bibnamefont {Reis}},\ }\href@noop {} {\bibfield  {journal} {\bibinfo
  {journal} {Phys. Rev. E}\ }\textbf {\bibinfo {volume} {94}},\ \bibinfo
  {pages} {052124} (\bibinfo {year} {2016})}\BibitemShut {NoStop}%
\bibitem [{\citenamefont {Volpe}\ and\ \citenamefont {Volpe}(2017)}]{Volpe}%
  \BibitemOpen
  \bibfield  {author} {\bibinfo {author} {\bibfnamefont {G.}~\bibnamefont
  {Volpe}}\ and\ \bibinfo {author} {\bibfnamefont {G.}~\bibnamefont {Volpe}},\
  }\href@noop {} {\bibfield  {journal} {\bibinfo  {journal} {Proc. Natl. Acad.
  Sci. USA}\ }\textbf {\bibinfo {volume} {114}},\ \bibinfo {pages} {11350}
  (\bibinfo {year} {2017})}\BibitemShut {NoStop}%
\bibitem [{\citenamefont {Sagi}\ \emph {et~al.}(2012)\citenamefont {Sagi},
  \citenamefont {Brook}, \citenamefont {Almog},\ and\ \citenamefont
  {Davidson}}]{Sagi}%
  \BibitemOpen
  \bibfield  {author} {\bibinfo {author} {\bibfnamefont {Y.}~\bibnamefont
  {Sagi}}, \bibinfo {author} {\bibfnamefont {M.}~\bibnamefont {Brook}},
  \bibinfo {author} {\bibfnamefont {I.}~\bibnamefont {Almog}}, \ and\ \bibinfo
  {author} {\bibfnamefont {N.}~\bibnamefont {Davidson}},\ }\href@noop {}
  {\bibfield  {journal} {\bibinfo  {journal} {Phys. Rev. Lett.}\ }\textbf
  {\bibinfo {volume} {108}},\ \bibinfo {pages} {093002} (\bibinfo {year}
  {2012})}\BibitemShut {NoStop}%
\bibitem [{\citenamefont {Hartich}\ and\ \citenamefont
  {Godec}(2021)}]{Hartich}%
  \BibitemOpen
  \bibfield  {author} {\bibinfo {author} {\bibfnamefont {D.}~\bibnamefont
  {Hartich}}\ and\ \bibinfo {author} {\bibfnamefont {A.}~\bibnamefont
  {Godec}},\ }\href@noop {} {\bibfield  {journal} {\bibinfo  {journal} {Phys.
  Rev. Lett.}\ }\textbf {\bibinfo {volume} {127}},\ \bibinfo {pages} {080601}
  (\bibinfo {year} {2021})}\BibitemShut {NoStop}%
\bibitem [{\citenamefont {Barkai}\ \emph {et~al.}(2012)\citenamefont {Barkai},
  \citenamefont {Garini},\ and\ \citenamefont {Metzler}}]{Barkai}%
  \BibitemOpen
  \bibfield  {author} {\bibinfo {author} {\bibfnamefont {E.}~\bibnamefont
  {Barkai}}, \bibinfo {author} {\bibfnamefont {Y.}~\bibnamefont {Garini}}, \
  and\ \bibinfo {author} {\bibfnamefont {R.}~\bibnamefont {Metzler}},\
  }\href@noop {} {\bibfield  {journal} {\bibinfo  {journal} {Phys. Today}\
  }\textbf {\bibinfo {volume} {65}},\ \bibinfo {pages} {29} (\bibinfo {year}
  {2012})}\BibitemShut {NoStop}%
\bibitem [{\citenamefont {Ding}\ \emph {et~al.}(2014)\citenamefont {Ding},
  \citenamefont {Hassanali},\ and\ \citenamefont {Parrinello}}]{Ding}%
  \BibitemOpen
  \bibfield  {author} {\bibinfo {author} {\bibfnamefont {Y.}~\bibnamefont
  {Ding}}, \bibinfo {author} {\bibfnamefont {A.~A.}\ \bibnamefont {Hassanali}},
  \ and\ \bibinfo {author} {\bibfnamefont {M.}~\bibnamefont {Parrinello}},\
  }\href@noop {} {\bibfield  {journal} {\bibinfo  {journal} {Proc. Natl. Acad.
  Sci. USA}\ }\textbf {\bibinfo {volume} {111}},\ \bibinfo {pages} {3310}
  (\bibinfo {year} {2014})}\BibitemShut {NoStop}%
\bibitem [{\citenamefont {Song}\ \emph {et~al.}(2019)\citenamefont {Song},
  \citenamefont {Bhattacharya}, \citenamefont {Kim}, \citenamefont {Chang},
  \citenamefont {Tang}, \citenamefont {Guo}, \citenamefont {Ghosh},
  \citenamefont {Yang}, \citenamefont {Jiang}, \citenamefont {Kim},
  \citenamefont {Russell}, \citenamefont {Arya}, \citenamefont {Narayanan},\
  and\ \citenamefont {Sinha}}]{Song}%
  \BibitemOpen
  \bibfield  {author} {\bibinfo {author} {\bibfnamefont {J.-J.}\ \bibnamefont
  {Song}}, \bibinfo {author} {\bibfnamefont {R.}~\bibnamefont {Bhattacharya}},
  \bibinfo {author} {\bibfnamefont {H.}~\bibnamefont {Kim}}, \bibinfo {author}
  {\bibfnamefont {J.}~\bibnamefont {Chang}}, \bibinfo {author} {\bibfnamefont
  {T.-Y.}\ \bibnamefont {Tang}}, \bibinfo {author} {\bibfnamefont
  {H.}~\bibnamefont {Guo}}, \bibinfo {author} {\bibfnamefont {S.~K.}\
  \bibnamefont {Ghosh}}, \bibinfo {author} {\bibfnamefont {Y.}~\bibnamefont
  {Yang}}, \bibinfo {author} {\bibfnamefont {Z.}~\bibnamefont {Jiang}},
  \bibinfo {author} {\bibfnamefont {H.}~\bibnamefont {Kim}}, \bibinfo {author}
  {\bibfnamefont {T.~P.}\ \bibnamefont {Russell}}, \bibinfo {author}
  {\bibfnamefont {G.}~\bibnamefont {Arya}}, \bibinfo {author} {\bibfnamefont
  {S.}~\bibnamefont {Narayanan}}, \ and\ \bibinfo {author} {\bibfnamefont
  {S.~K.}\ \bibnamefont {Sinha}},\ }\href@noop {} {\bibfield  {journal}
  {\bibinfo  {journal} {Phys. Rev. Lett.}\ }\textbf {\bibinfo {volume} {122}},\
  \bibinfo {pages} {107802} (\bibinfo {year} {2019})}\BibitemShut {NoStop}%
\bibitem [{\citenamefont {Xu}\ \emph {et~al.}(2021)\citenamefont {Xu},
  \citenamefont {Dai}, \citenamefont {Bu}, \citenamefont {Yang}, \citenamefont
  {Zhang}, \citenamefont {Man}, \citenamefont {Zhang}, \citenamefont {Doi},\
  and\ \citenamefont {Yan}}]{Xu}%
  \BibitemOpen
  \bibfield  {author} {\bibinfo {author} {\bibfnamefont {Z.}~\bibnamefont
  {Xu}}, \bibinfo {author} {\bibfnamefont {X.}~\bibnamefont {Dai}}, \bibinfo
  {author} {\bibfnamefont {X.}~\bibnamefont {Bu}}, \bibinfo {author}
  {\bibfnamefont {Y.}~\bibnamefont {Yang}}, \bibinfo {author} {\bibfnamefont
  {X.}~\bibnamefont {Zhang}}, \bibinfo {author} {\bibfnamefont
  {X.}~\bibnamefont {Man}}, \bibinfo {author} {\bibfnamefont {X.}~\bibnamefont
  {Zhang}}, \bibinfo {author} {\bibfnamefont {M.}~\bibnamefont {Doi}}, \ and\
  \bibinfo {author} {\bibfnamefont {L.-T.}\ \bibnamefont {Yan}},\ }\href@noop
  {} {\bibfield  {journal} {\bibinfo  {journal} {ACS Nano}\ }\textbf {\bibinfo
  {volume} {15}},\ \bibinfo {pages} {4608} (\bibinfo {year}
  {2021})}\BibitemShut {NoStop}%
\bibitem [{\citenamefont {H\"{o}fling}\ and\ \citenamefont
  {Franosch}(2013)}]{Hofling}%
  \BibitemOpen
  \bibfield  {author} {\bibinfo {author} {\bibfnamefont {F.}~\bibnamefont
  {H\"{o}fling}}\ and\ \bibinfo {author} {\bibfnamefont {T.}~\bibnamefont
  {Franosch}},\ }\href@noop {} {\bibfield  {journal} {\bibinfo  {journal} {Rep.
  Prog. Phys.}\ }\textbf {\bibinfo {volume} {76}},\ \bibinfo {pages} {046602}
  (\bibinfo {year} {2013})}\BibitemShut {NoStop}%
\bibitem [{\citenamefont {Wu}\ and\ \citenamefont {Libchaber}(2000)}]{Wu}%
  \BibitemOpen
  \bibfield  {author} {\bibinfo {author} {\bibfnamefont {X.-L.}\ \bibnamefont
  {Wu}}\ and\ \bibinfo {author} {\bibfnamefont {A.}~\bibnamefont {Libchaber}},\
  }\href@noop {} {\bibfield  {journal} {\bibinfo  {journal} {Phys. Rev. Lett.}\
  }\textbf {\bibinfo {volume} {84}},\ \bibinfo {pages} {3017} (\bibinfo {year}
  {2000})}\BibitemShut {NoStop}%
\bibitem [{\citenamefont {Gonz\'{a}lez}\ \emph {et~al.}(2008)\citenamefont
  {Gonz\'{a}lez}, \citenamefont {Hidalgo},\ and\ \citenamefont
  {Barab\'{a}si}}]{Gonzalez}%
  \BibitemOpen
  \bibfield  {author} {\bibinfo {author} {\bibfnamefont {M.~C.}\ \bibnamefont
  {Gonz\'{a}lez}}, \bibinfo {author} {\bibfnamefont {C.~A.}\ \bibnamefont
  {Hidalgo}}, \ and\ \bibinfo {author} {\bibfnamefont {A.-L.}\ \bibnamefont
  {Barab\'{a}si}},\ }\href@noop {} {\bibfield  {journal} {\bibinfo  {journal}
  {Nature}\ }\textbf {\bibinfo {volume} {453}},\ \bibinfo {pages} {779}
  (\bibinfo {year} {2008})}\BibitemShut {NoStop}%
\bibitem [{\citenamefont {Wang}\ \emph {et~al.}(2013)\citenamefont {Wang},
  \citenamefont {Kuo},\ and\ \citenamefont {Granick}}]{Wang}%
  \BibitemOpen
  \bibfield  {author} {\bibinfo {author} {\bibfnamefont {B.}~\bibnamefont
  {Wang}}, \bibinfo {author} {\bibfnamefont {J.}~\bibnamefont {Kuo}}, \ and\
  \bibinfo {author} {\bibfnamefont {S.}~\bibnamefont {Granick}},\ }\href@noop
  {} {\bibfield  {journal} {\bibinfo  {journal} {Phys. Rev. Lett.}\ }\textbf
  {\bibinfo {volume} {111}},\ \bibinfo {pages} {208102} (\bibinfo {year}
  {2013})}\BibitemShut {NoStop}%
\bibitem [{\citenamefont {Chen}\ \emph {et~al.}(2016)\citenamefont {Chen},
  \citenamefont {Huang}, \citenamefont {Liang}, \citenamefont {Cui},
  \citenamefont {Zhang}, \citenamefont {Miao},\ and\ \citenamefont
  {Yan}}]{Chen}%
  \BibitemOpen
  \bibfield  {author} {\bibinfo {author} {\bibfnamefont {P.}~\bibnamefont
  {Chen}}, \bibinfo {author} {\bibfnamefont {Z.}~\bibnamefont {Huang}},
  \bibinfo {author} {\bibfnamefont {J.}~\bibnamefont {Liang}}, \bibinfo
  {author} {\bibfnamefont {T.}~\bibnamefont {Cui}}, \bibinfo {author}
  {\bibfnamefont {X.}~\bibnamefont {Zhang}}, \bibinfo {author} {\bibfnamefont
  {B.}~\bibnamefont {Miao}}, \ and\ \bibinfo {author} {\bibfnamefont {L.-T.}\
  \bibnamefont {Yan}},\ }\href@noop {} {\bibfield  {journal} {\bibinfo
  {journal} {ACS Nano}\ }\textbf {\bibinfo {volume} {10}},\ \bibinfo {pages}
  {11541} (\bibinfo {year} {2016})}\BibitemShut {NoStop}%
\bibitem [{\citenamefont {Plerou}\ \emph {et~al.}(2000)\citenamefont {Plerou},
  \citenamefont {Gopikrishnan}, \citenamefont {Amaral}, \citenamefont
  {Gabaix},\ and\ \citenamefont {Stanley}}]{Plerou}%
  \BibitemOpen
  \bibfield  {author} {\bibinfo {author} {\bibfnamefont {V.}~\bibnamefont
  {Plerou}}, \bibinfo {author} {\bibfnamefont {P.}~\bibnamefont
  {Gopikrishnan}}, \bibinfo {author} {\bibfnamefont {L.~A.~N.}\ \bibnamefont
  {Amaral}}, \bibinfo {author} {\bibfnamefont {X.}~\bibnamefont {Gabaix}}, \
  and\ \bibinfo {author} {\bibfnamefont {H.~E.}\ \bibnamefont {Stanley}},\
  }\href@noop {} {\bibfield  {journal} {\bibinfo  {journal} {Phys. Rev. E}\
  }\textbf {\bibinfo {volume} {62}},\ \bibinfo {pages} {R3023} (\bibinfo {year}
  {2000})}\BibitemShut {NoStop}%
\bibitem [{\citenamefont {Masoliver}\ \emph {et~al.}(2003)\citenamefont
  {Masoliver}, \citenamefont {Montero},\ and\ \citenamefont
  {Weiss}}]{Masoliver}%
  \BibitemOpen
  \bibfield  {author} {\bibinfo {author} {\bibfnamefont {J.}~\bibnamefont
  {Masoliver}}, \bibinfo {author} {\bibfnamefont {M.}~\bibnamefont {Montero}},
  \ and\ \bibinfo {author} {\bibfnamefont {G.~H.}\ \bibnamefont {Weiss}},\
  }\href@noop {} {\bibfield  {journal} {\bibinfo  {journal} {Phys. Rev. E}\
  }\textbf {\bibinfo {volume} {67}},\ \bibinfo {pages} {021112} (\bibinfo
  {year} {2003})}\BibitemShut {NoStop}%
\bibitem [{\citenamefont {Jiang}\ \emph {et~al.}(2019)\citenamefont {Jiang},
  \citenamefont {Xie}, \citenamefont {Zhou},\ and\ \citenamefont
  {Sornette}}]{Jiang}%
  \BibitemOpen
  \bibfield  {author} {\bibinfo {author} {\bibfnamefont {Z.-Q.}\ \bibnamefont
  {Jiang}}, \bibinfo {author} {\bibfnamefont {W.-J.}\ \bibnamefont {Xie}},
  \bibinfo {author} {\bibfnamefont {W.-X.}\ \bibnamefont {Zhou}}, \ and\
  \bibinfo {author} {\bibfnamefont {D.}~\bibnamefont {Sornette}},\ }\href@noop
  {} {\bibfield  {journal} {\bibinfo  {journal} {Rep. Prog. Phys.}\ }\textbf
  {\bibinfo {volume} {82}},\ \bibinfo {pages} {125901} (\bibinfo {year}
  {2019})}\BibitemShut {NoStop}%
\bibitem [{\citenamefont {Sposini}\ \emph {et~al.}(2022)\citenamefont
  {Sposini}, \citenamefont {Krapf}, \citenamefont {Marinari}, \citenamefont
  {Sunyer}, \citenamefont {Ritort}, \citenamefont {Taheri}, \citenamefont
  {Selhuber-Unkel}, \citenamefont {Benelli}, \citenamefont {Weiss},
  \citenamefont {Metzler} \emph {et~al.}}]{Sposini}%
  \BibitemOpen
  \bibfield  {author} {\bibinfo {author} {\bibfnamefont {V.}~\bibnamefont
  {Sposini}}, \bibinfo {author} {\bibfnamefont {D.}~\bibnamefont {Krapf}},
  \bibinfo {author} {\bibfnamefont {E.}~\bibnamefont {Marinari}}, \bibinfo
  {author} {\bibfnamefont {R.}~\bibnamefont {Sunyer}}, \bibinfo {author}
  {\bibfnamefont {F.}~\bibnamefont {Ritort}}, \bibinfo {author} {\bibfnamefont
  {F.}~\bibnamefont {Taheri}}, \bibinfo {author} {\bibfnamefont
  {C.}~\bibnamefont {Selhuber-Unkel}}, \bibinfo {author} {\bibfnamefont
  {R.}~\bibnamefont {Benelli}}, \bibinfo {author} {\bibfnamefont
  {M.}~\bibnamefont {Weiss}}, \bibinfo {author} {\bibfnamefont
  {R.}~\bibnamefont {Metzler}},  \emph {et~al.},\ }\href@noop {} {\bibfield
  {journal} {\bibinfo  {journal} {Commun. Phys.}\ }\textbf {\bibinfo {volume}
  {5}},\ \bibinfo {pages} {305} (\bibinfo {year} {2022})}\BibitemShut {NoStop}%
\bibitem [{\citenamefont {Vilk}\ \emph {et~al.}(2022)\citenamefont {Vilk},
  \citenamefont {Aghion}, \citenamefont {Avgar}, \citenamefont {Beta},
  \citenamefont {Nagel}, \citenamefont {Sabri}, \citenamefont {Sarfati},
  \citenamefont {Schwartz}, \citenamefont {Weiss}, \citenamefont {Krapf},
  \citenamefont {Nathan}, \citenamefont {Metzler},\ and\ \citenamefont
  {Assaf}}]{Vilk}%
  \BibitemOpen
  \bibfield  {author} {\bibinfo {author} {\bibfnamefont {O.}~\bibnamefont
  {Vilk}}, \bibinfo {author} {\bibfnamefont {E.}~\bibnamefont {Aghion}},
  \bibinfo {author} {\bibfnamefont {T.}~\bibnamefont {Avgar}}, \bibinfo
  {author} {\bibfnamefont {C.}~\bibnamefont {Beta}}, \bibinfo {author}
  {\bibfnamefont {O.}~\bibnamefont {Nagel}}, \bibinfo {author} {\bibfnamefont
  {A.}~\bibnamefont {Sabri}}, \bibinfo {author} {\bibfnamefont
  {R.}~\bibnamefont {Sarfati}}, \bibinfo {author} {\bibfnamefont {D.~K.}\
  \bibnamefont {Schwartz}}, \bibinfo {author} {\bibfnamefont {M.}~\bibnamefont
  {Weiss}}, \bibinfo {author} {\bibfnamefont {D.}~\bibnamefont {Krapf}},
  \bibinfo {author} {\bibfnamefont {R.}~\bibnamefont {Nathan}}, \bibinfo
  {author} {\bibfnamefont {R.}~\bibnamefont {Metzler}}, \ and\ \bibinfo
  {author} {\bibfnamefont {M.}~\bibnamefont {Assaf}},\ }\href@noop {}
  {\bibfield  {journal} {\bibinfo  {journal} {Phys. Rev. Res.}\ }\textbf
  {\bibinfo {volume} {4}},\ \bibinfo {pages} {033055} (\bibinfo {year}
  {2022})}\BibitemShut {NoStop}%
\bibitem [{\citenamefont {Wang}\ \emph {et~al.}(2022)\citenamefont {Wang},
  \citenamefont {Cherstvy}, \citenamefont {Metzler},\ and\ \citenamefont
  {Sokolov}}]{Wang1}%
  \BibitemOpen
  \bibfield  {author} {\bibinfo {author} {\bibfnamefont {W.}~\bibnamefont
  {Wang}}, \bibinfo {author} {\bibfnamefont {A.~G.}\ \bibnamefont {Cherstvy}},
  \bibinfo {author} {\bibfnamefont {R.}~\bibnamefont {Metzler}}, \ and\
  \bibinfo {author} {\bibfnamefont {I.~M.}\ \bibnamefont {Sokolov}},\
  }\href@noop {} {\bibfield  {journal} {\bibinfo  {journal} {Phys. Rev. Res.}\
  }\textbf {\bibinfo {volume} {4}},\ \bibinfo {pages} {013161} (\bibinfo {year}
  {2022})}\BibitemShut {NoStop}%
\bibitem [{\citenamefont {Seckler}\ and\ \citenamefont
  {Metzler}(2022)}]{Seckler}%
  \BibitemOpen
  \bibfield  {author} {\bibinfo {author} {\bibfnamefont {H.}~\bibnamefont
  {Seckler}}\ and\ \bibinfo {author} {\bibfnamefont {R.}~\bibnamefont
  {Metzler}},\ }\href@noop {} {\bibfield  {journal} {\bibinfo  {journal} {Nat.
  Commun.}\ }\textbf {\bibinfo {volume} {13}},\ \bibinfo {pages} {6717}
  (\bibinfo {year} {2022})}\BibitemShut {NoStop}%
\bibitem [{\citenamefont {Requena}\ \emph {et~al.}(2023)\citenamefont
  {Requena}, \citenamefont {Mas{\'o}}, \citenamefont {Bertran}, \citenamefont
  {Lewenstein}, \citenamefont {Manzo},\ and\ \citenamefont
  {Mu{\~n}oz-Gil}}]{Borja}%
  \BibitemOpen
  \bibfield  {author} {\bibinfo {author} {\bibfnamefont {B.}~\bibnamefont
  {Requena}}, \bibinfo {author} {\bibfnamefont {S.}~\bibnamefont {Mas{\'o}}},
  \bibinfo {author} {\bibfnamefont {J.}~\bibnamefont {Bertran}}, \bibinfo
  {author} {\bibfnamefont {M.}~\bibnamefont {Lewenstein}}, \bibinfo {author}
  {\bibfnamefont {C.}~\bibnamefont {Manzo}}, \ and\ \bibinfo {author}
  {\bibfnamefont {G.}~\bibnamefont {Mu{\~n}oz-Gil}},\ }\href@noop {} {\bibfield
   {journal} {\bibinfo  {journal} {arXiv:2302.00410}\ } (\bibinfo {year}
  {2023})}\BibitemShut {NoStop}%
\bibitem [{\citenamefont {Mu{\~n}oz-Gil}\ \emph
  {et~al.}(2021{\natexlab{a}})\citenamefont {Mu{\~n}oz-Gil}, \citenamefont
  {Volpe}, \citenamefont {Garcia-March}, \citenamefont {Aghion}, \citenamefont
  {Argun}, \citenamefont {Hong}, \citenamefont {Bland}, \citenamefont {Bo},
  \citenamefont {Conejero}, \citenamefont {Firbas} \emph {et~al.}}]{Munoz-Gil}%
  \BibitemOpen
  \bibfield  {author} {\bibinfo {author} {\bibfnamefont {G.}~\bibnamefont
  {Mu{\~n}oz-Gil}}, \bibinfo {author} {\bibfnamefont {G.}~\bibnamefont
  {Volpe}}, \bibinfo {author} {\bibfnamefont {M.~A.}\ \bibnamefont
  {Garcia-March}}, \bibinfo {author} {\bibfnamefont {E.}~\bibnamefont
  {Aghion}}, \bibinfo {author} {\bibfnamefont {A.}~\bibnamefont {Argun}},
  \bibinfo {author} {\bibfnamefont {C.~B.}\ \bibnamefont {Hong}}, \bibinfo
  {author} {\bibfnamefont {T.}~\bibnamefont {Bland}}, \bibinfo {author}
  {\bibfnamefont {S.}~\bibnamefont {Bo}}, \bibinfo {author} {\bibfnamefont
  {J.~A.}\ \bibnamefont {Conejero}}, \bibinfo {author} {\bibfnamefont
  {N.}~\bibnamefont {Firbas}},  \emph {et~al.},\ }\href@noop {} {\bibfield
  {journal} {\bibinfo  {journal} {Nat. Commun.}\ }\textbf {\bibinfo {volume}
  {12}},\ \bibinfo {pages} {6253} (\bibinfo {year}
  {2021}{\natexlab{a}})}\BibitemShut {NoStop}%
\bibitem [{\citenamefont {Manzo}\ \emph {et~al.}(2023)\citenamefont {Manzo},
  \citenamefont {Mu{\~n}oz-Gil}, \citenamefont {Volpe}, \citenamefont
  {Garcia-March}, \citenamefont {Lewenstein},\ and\ \citenamefont
  {Metzler}}]{Manzo}%
  \BibitemOpen
  \bibfield  {author} {\bibinfo {author} {\bibfnamefont {C.}~\bibnamefont
  {Manzo}}, \bibinfo {author} {\bibfnamefont {G.}~\bibnamefont
  {Mu{\~n}oz-Gil}}, \bibinfo {author} {\bibfnamefont {G.}~\bibnamefont
  {Volpe}}, \bibinfo {author} {\bibfnamefont {M.~A.}\ \bibnamefont
  {Garcia-March}}, \bibinfo {author} {\bibfnamefont {M.}~\bibnamefont
  {Lewenstein}}, \ and\ \bibinfo {author} {\bibfnamefont {R.}~\bibnamefont
  {Metzler}},\ }\href@noop {} {\bibfield  {journal} {\bibinfo  {journal} {J.
  Phys. A: Math. Theor.}\ }\textbf {\bibinfo {volume} {56}},\ \bibinfo {pages}
  {010401} (\bibinfo {year} {2023})}\BibitemShut {NoStop}%
\bibitem [{\citenamefont {Bo}\ \emph {et~al.}(2019)\citenamefont {Bo},
  \citenamefont {Schmidt}, \citenamefont {Eichhorn},\ and\ \citenamefont
  {Volpe}}]{Bo}%
  \BibitemOpen
  \bibfield  {author} {\bibinfo {author} {\bibfnamefont {S.}~\bibnamefont
  {Bo}}, \bibinfo {author} {\bibfnamefont {F.}~\bibnamefont {Schmidt}},
  \bibinfo {author} {\bibfnamefont {R.}~\bibnamefont {Eichhorn}}, \ and\
  \bibinfo {author} {\bibfnamefont {G.}~\bibnamefont {Volpe}},\ }\href@noop {}
  {\bibfield  {journal} {\bibinfo  {journal} {Phys. Rev. E}\ }\textbf {\bibinfo
  {volume} {100}},\ \bibinfo {pages} {010102} (\bibinfo {year}
  {2019})}\BibitemShut {NoStop}%
\bibitem [{\citenamefont {Kowalek}\ \emph {et~al.}(2019)\citenamefont
  {Kowalek}, \citenamefont {Loch-Olszewska},\ and\ \citenamefont
  {Szwabi\ifmmode~\'{n}\else \'{n}\fi{}ski}}]{Kowalek}%
  \BibitemOpen
  \bibfield  {author} {\bibinfo {author} {\bibfnamefont {P.}~\bibnamefont
  {Kowalek}}, \bibinfo {author} {\bibfnamefont {H.}~\bibnamefont
  {Loch-Olszewska}}, \ and\ \bibinfo {author} {\bibfnamefont {J.}~\bibnamefont
  {Szwabi\ifmmode~\'{n}\else \'{n}\fi{}ski}},\ }\href@noop {} {\bibfield
  {journal} {\bibinfo  {journal} {Phys. Rev. E}\ }\textbf {\bibinfo {volume}
  {100}},\ \bibinfo {pages} {032410} (\bibinfo {year} {2019})}\BibitemShut
  {NoStop}%
\bibitem [{\citenamefont {Firbas}\ \emph {et~al.}(2023)\citenamefont {Firbas},
  \citenamefont {\`{O}scar Garibo-i Orts}, \citenamefont {\'{A}ngel
  Garcia-March},\ and\ \citenamefont {Conejero}}]{Firbas}%
  \BibitemOpen
  \bibfield  {author} {\bibinfo {author} {\bibfnamefont {N.}~\bibnamefont
  {Firbas}}, \bibinfo {author} {\bibnamefont {\`{O}scar Garibo-i Orts}},
  \bibinfo {author} {\bibfnamefont {M.}~\bibnamefont {\'{A}ngel Garcia-March}},
  \ and\ \bibinfo {author} {\bibfnamefont {J.~A.}\ \bibnamefont {Conejero}},\
  }\href@noop {} {\bibfield  {journal} {\bibinfo  {journal} {J. Phys. A: Math.
  Theor.}\ }\textbf {\bibinfo {volume} {56}},\ \bibinfo {pages} {014001}
  (\bibinfo {year} {2023})}\BibitemShut {NoStop}%
\bibitem [{\citenamefont {Mu{\~n}oz-Gil}\ \emph
  {et~al.}(2021{\natexlab{b}})\citenamefont {Mu{\~n}oz-Gil}, \citenamefont
  {i~Corominas},\ and\ \citenamefont {Lewenstein}}]{Munoz-Gil1}%
  \BibitemOpen
  \bibfield  {author} {\bibinfo {author} {\bibfnamefont {G.}~\bibnamefont
  {Mu{\~n}oz-Gil}}, \bibinfo {author} {\bibfnamefont {G.~G.}\ \bibnamefont
  {i~Corominas}}, \ and\ \bibinfo {author} {\bibfnamefont {M.}~\bibnamefont
  {Lewenstein}},\ }\href@noop {} {\bibfield  {journal} {\bibinfo  {journal} {J.
  Phys. A: Math. Theor.}\ }\textbf {\bibinfo {volume} {54}},\ \bibinfo {pages}
  {504001} (\bibinfo {year} {2021}{\natexlab{b}})}\BibitemShut {NoStop}%
\bibitem [{\citenamefont {Gentili}\ and\ \citenamefont
  {Volpe}(2021)}]{Gentili}%
  \BibitemOpen
  \bibfield  {author} {\bibinfo {author} {\bibfnamefont {A.}~\bibnamefont
  {Gentili}}\ and\ \bibinfo {author} {\bibfnamefont {G.}~\bibnamefont
  {Volpe}},\ }\href@noop {} {\bibfield  {journal} {\bibinfo  {journal} {J.
  Phys. A: Math. Theor.}\ }\textbf {\bibinfo {volume} {54}},\ \bibinfo {pages}
  {314003} (\bibinfo {year} {2021})}\BibitemShut {NoStop}%
\bibitem [{\citenamefont {Argun}\ \emph {et~al.}(2021)\citenamefont {Argun},
  \citenamefont {Volpe},\ and\ \citenamefont {Bo}}]{Argun}%
  \BibitemOpen
  \bibfield  {author} {\bibinfo {author} {\bibfnamefont {A.}~\bibnamefont
  {Argun}}, \bibinfo {author} {\bibfnamefont {G.}~\bibnamefont {Volpe}}, \ and\
  \bibinfo {author} {\bibfnamefont {S.}~\bibnamefont {Bo}},\ }\href@noop {}
  {\bibfield  {journal} {\bibinfo  {journal} {J. Phys. A: Math. Theor.}\
  }\textbf {\bibinfo {volume} {54}},\ \bibinfo {pages} {294003} (\bibinfo
  {year} {2021})}\BibitemShut {NoStop}%
\bibitem [{\citenamefont {Verdier}\ \emph {et~al.}(2021)\citenamefont
  {Verdier}, \citenamefont {Duval}, \citenamefont {Laurent}, \citenamefont
  {Cass{\'e}}, \citenamefont {Vestergaard},\ and\ \citenamefont
  {Masson}}]{LearningVerdier}%
  \BibitemOpen
  \bibfield  {author} {\bibinfo {author} {\bibfnamefont {H.}~\bibnamefont
  {Verdier}}, \bibinfo {author} {\bibfnamefont {M.}~\bibnamefont {Duval}},
  \bibinfo {author} {\bibfnamefont {F.}~\bibnamefont {Laurent}}, \bibinfo
  {author} {\bibfnamefont {A.}~\bibnamefont {Cass{\'e}}}, \bibinfo {author}
  {\bibfnamefont {C.~L.}\ \bibnamefont {Vestergaard}}, \ and\ \bibinfo {author}
  {\bibfnamefont {J.-B.}\ \bibnamefont {Masson}},\ }\href@noop {} {\bibfield
  {journal} {\bibinfo  {journal} {J. Phys. A: Math. Theor.}\ }\textbf {\bibinfo
  {volume} {54}},\ \bibinfo {pages} {234001} (\bibinfo {year}
  {2021})}\BibitemShut {NoStop}%
\bibitem [{\citenamefont {Li}\ \emph {et~al.}(2021)\citenamefont {Li},
  \citenamefont {Yao},\ and\ \citenamefont {Huang}}]{Li}%
  \BibitemOpen
  \bibfield  {author} {\bibinfo {author} {\bibfnamefont {D.}~\bibnamefont
  {Li}}, \bibinfo {author} {\bibfnamefont {Q.}~\bibnamefont {Yao}}, \ and\
  \bibinfo {author} {\bibfnamefont {Z.}~\bibnamefont {Huang}},\ }\href@noop {}
  {\bibfield  {journal} {\bibinfo  {journal} {J. Phys. A: Math. Theor.}\
  }\textbf {\bibinfo {volume} {54}},\ \bibinfo {pages} {404003} (\bibinfo
  {year} {2021})}\BibitemShut {NoStop}%
\bibitem [{\citenamefont {Verdier}\ \emph {et~al.}(2022)\citenamefont
  {Verdier}, \citenamefont {Laurent}, \citenamefont {Cass{\'e}}, \citenamefont
  {Vestergaard},\ and\ \citenamefont {Masson}}]{verdier2022}%
  \BibitemOpen
  \bibfield  {author} {\bibinfo {author} {\bibfnamefont {H.}~\bibnamefont
  {Verdier}}, \bibinfo {author} {\bibfnamefont {F.}~\bibnamefont {Laurent}},
  \bibinfo {author} {\bibfnamefont {A.}~\bibnamefont {Cass{\'e}}}, \bibinfo
  {author} {\bibfnamefont {C.~L.}\ \bibnamefont {Vestergaard}}, \ and\ \bibinfo
  {author} {\bibfnamefont {J.-B.}\ \bibnamefont {Masson}},\ }\href@noop {}
  {\bibfield  {journal} {\bibinfo  {journal} {Phys. Rev. E}\ }\textbf {\bibinfo
  {volume} {106}},\ \bibinfo {pages} {055311} (\bibinfo {year}
  {2022})}\BibitemShut {NoStop}%
\bibitem [{\citenamefont {Verdier}\ \emph {et~al.}(2023)\citenamefont
  {Verdier}, \citenamefont {Laurent}, \citenamefont {Cass{\'e}}, \citenamefont
  {Vestergaard}, \citenamefont {Specht},\ and\ \citenamefont
  {Masson}}]{Simulation-based}%
  \BibitemOpen
  \bibfield  {author} {\bibinfo {author} {\bibfnamefont {H.}~\bibnamefont
  {Verdier}}, \bibinfo {author} {\bibfnamefont {F.}~\bibnamefont {Laurent}},
  \bibinfo {author} {\bibfnamefont {A.}~\bibnamefont {Cass{\'e}}}, \bibinfo
  {author} {\bibfnamefont {C.~L.}\ \bibnamefont {Vestergaard}}, \bibinfo
  {author} {\bibfnamefont {C.~G.}\ \bibnamefont {Specht}}, \ and\ \bibinfo
  {author} {\bibfnamefont {J.-B.}\ \bibnamefont {Masson}},\ }\href@noop {}
  {\bibfield  {journal} {\bibinfo  {journal} {PLoS Comput. Biol.}\ }\textbf
  {\bibinfo {volume} {19}},\ \bibinfo {pages} {e1010088} (\bibinfo {year}
  {2023})}\BibitemShut {NoStop}%
\bibitem [{\citenamefont {Kabbech}\ and\ \citenamefont {Smal}(2022)}]{Kabbech}%
  \BibitemOpen
  \bibfield  {author} {\bibinfo {author} {\bibfnamefont {H.}~\bibnamefont
  {Kabbech}}\ and\ \bibinfo {author} {\bibfnamefont {I.}~\bibnamefont {Smal}},\
  }in\ \href@noop {} {\emph {\bibinfo {booktitle} {2022 IEEE 19th International
  Symposium on Biomedical Imaging (ISBI)}}}\ (\bibinfo {address} {Kolkata,
  India},\ \bibinfo {year} {2022})\ pp.\ \bibinfo {pages} {1--4}\BibitemShut
  {NoStop}%
\bibitem [{\citenamefont {Vega}\ \emph {et~al.}(2018)\citenamefont {Vega},
  \citenamefont {Freeman}, \citenamefont {Grinstein},\ and\ \citenamefont
  {Jaqaman}}]{multistep}%
  \BibitemOpen
  \bibfield  {author} {\bibinfo {author} {\bibfnamefont {A.~R.}\ \bibnamefont
  {Vega}}, \bibinfo {author} {\bibfnamefont {S.~A.}\ \bibnamefont {Freeman}},
  \bibinfo {author} {\bibfnamefont {S.}~\bibnamefont {Grinstein}}, \ and\
  \bibinfo {author} {\bibfnamefont {K.}~\bibnamefont {Jaqaman}},\ }\href@noop
  {} {\bibfield  {journal} {\bibinfo  {journal} {Biophys. J.}\ }\textbf
  {\bibinfo {volume} {114}},\ \bibinfo {pages} {1018} (\bibinfo {year}
  {2018})}\BibitemShut {NoStop}%
\bibitem [{\citenamefont {Arts}\ \emph {et~al.}(2019)\citenamefont {Arts},
  \citenamefont {Smal}, \citenamefont {Paul}, \citenamefont {Wyman},\ and\
  \citenamefont {Meijering}}]{ParticleM}%
  \BibitemOpen
  \bibfield  {author} {\bibinfo {author} {\bibfnamefont {M.}~\bibnamefont
  {Arts}}, \bibinfo {author} {\bibfnamefont {I.}~\bibnamefont {Smal}}, \bibinfo
  {author} {\bibfnamefont {M.~W.}\ \bibnamefont {Paul}}, \bibinfo {author}
  {\bibfnamefont {C.}~\bibnamefont {Wyman}}, \ and\ \bibinfo {author}
  {\bibfnamefont {E.}~\bibnamefont {Meijering}},\ }\href@noop {} {\bibfield
  {journal} {\bibinfo  {journal} {Sci. Rep.}\ }\textbf {\bibinfo {volume}
  {9}},\ \bibinfo {pages} {17160} (\bibinfo {year} {2019})}\BibitemShut
  {NoStop}%
\bibitem [{\citenamefont {Seckler}\ \emph {et~al.}(2023)\citenamefont
  {Seckler}, \citenamefont {Szwabin\'{n}ski},\ and\ \citenamefont
  {Metzler}}]{seckler2023machine}%
  \BibitemOpen
  \bibfield  {author} {\bibinfo {author} {\bibfnamefont {H.}~\bibnamefont
  {Seckler}}, \bibinfo {author} {\bibfnamefont {J.}~\bibnamefont
  {Szwabin\'{n}ski}}, \ and\ \bibinfo {author} {\bibfnamefont {R.}~\bibnamefont
  {Metzler}},\ }\href@noop {} {\bibfield  {journal} {\bibinfo  {journal} {Phys.
  Chem. Chem. Phys.}\ }\textbf {\bibinfo {volume} {14}},\ \bibinfo {pages}
  {7910} (\bibinfo {year} {2023})}\BibitemShut {NoStop}%
\bibitem [{\citenamefont {Meijering}\ \emph {et~al.}(2023)\citenamefont
  {Meijering}, \citenamefont {Smal}, \citenamefont {Dzyubachyk},\ and\
  \citenamefont {Olivo-Marin}}]{MEIJERING}%
  \BibitemOpen
  \bibfield  {author} {\bibinfo {author} {\bibfnamefont {E.}~\bibnamefont
  {Meijering}}, \bibinfo {author} {\bibfnamefont {I.}~\bibnamefont {Smal}},
  \bibinfo {author} {\bibfnamefont {O.}~\bibnamefont {Dzyubachyk}}, \ and\
  \bibinfo {author} {\bibfnamefont {J.-C.}\ \bibnamefont {Olivo-Marin}},\ }in\
  \href@noop {} {\emph {\bibinfo {booktitle} {Microscope Image Processing}}},\
  \bibinfo {editor} {edited by\ \bibinfo {editor} {\bibfnamefont {F.~A.}\
  \bibnamefont {Merchant}}\ and\ \bibinfo {editor} {\bibfnamefont {K.~R.}\
  \bibnamefont {Castleman}}}\ (\bibinfo  {publisher} {Academic Press},\
  \bibinfo {year} {2023})\ \bibinfo {edition} {2nd}\ ed.,\ Chap.~\bibinfo
  {chapter} {14}, pp.\ \bibinfo {pages} {393--430}\BibitemShut {NoStop}%
\bibitem [{\citenamefont {Chen}\ \emph {et~al.}(2015)\citenamefont {Chen},
  \citenamefont {Wang},\ and\ \citenamefont {Granick}}]{Chen1}%
  \BibitemOpen
  \bibfield  {author} {\bibinfo {author} {\bibfnamefont {K.}~\bibnamefont
  {Chen}}, \bibinfo {author} {\bibfnamefont {B.}~\bibnamefont {Wang}}, \ and\
  \bibinfo {author} {\bibfnamefont {S.}~\bibnamefont {Granick}},\ }\href@noop
  {} {\bibfield  {journal} {\bibinfo  {journal} {Nat. Mater.}\ }\textbf
  {\bibinfo {volume} {14}},\ \bibinfo {pages} {589} (\bibinfo {year}
  {2015})}\BibitemShut {NoStop}%
\bibitem [{\citenamefont {Chen}\ \emph {et~al.}(2013)\citenamefont {Chen},
  \citenamefont {Wang}, \citenamefont {Guan},\ and\ \citenamefont
  {Granick}}]{Chen2}%
  \BibitemOpen
  \bibfield  {author} {\bibinfo {author} {\bibfnamefont {K.}~\bibnamefont
  {Chen}}, \bibinfo {author} {\bibfnamefont {B.}~\bibnamefont {Wang}}, \bibinfo
  {author} {\bibfnamefont {J.}~\bibnamefont {Guan}}, \ and\ \bibinfo {author}
  {\bibfnamefont {S.}~\bibnamefont {Granick}},\ }\href@noop {} {\bibfield
  {journal} {\bibinfo  {journal} {ACS Nano}\ }\textbf {\bibinfo {volume} {7}},\
  \bibinfo {pages} {8634} (\bibinfo {year} {2013})}\BibitemShut {NoStop}%
\bibitem [{\citenamefont {Persson}\ \emph {et~al.}(2013)\citenamefont
  {Persson}, \citenamefont {Lind{\'e}n}, \citenamefont {Unoson},\ and\
  \citenamefont {Elf}}]{Persson}%
  \BibitemOpen
  \bibfield  {author} {\bibinfo {author} {\bibfnamefont {F.}~\bibnamefont
  {Persson}}, \bibinfo {author} {\bibfnamefont {M.}~\bibnamefont {Lind{\'e}n}},
  \bibinfo {author} {\bibfnamefont {C.}~\bibnamefont {Unoson}}, \ and\ \bibinfo
  {author} {\bibfnamefont {J.}~\bibnamefont {Elf}},\ }\href@noop {} {\bibfield
  {journal} {\bibinfo  {journal} {Nat. Methods}\ }\textbf {\bibinfo {volume}
  {10}},\ \bibinfo {pages} {265} (\bibinfo {year} {2013})}\BibitemShut
  {NoStop}%
\bibitem [{\citenamefont {Monnier}\ \emph {et~al.}(2015)\citenamefont
  {Monnier}, \citenamefont {Barry}, \citenamefont {Park}, \citenamefont {Su},
  \citenamefont {Katz}, \citenamefont {English}, \citenamefont {Dey},
  \citenamefont {Pan}, \citenamefont {Cheeseman}, \citenamefont {Singer} \emph
  {et~al.}}]{Monnier}%
  \BibitemOpen
  \bibfield  {author} {\bibinfo {author} {\bibfnamefont {N.}~\bibnamefont
  {Monnier}}, \bibinfo {author} {\bibfnamefont {Z.}~\bibnamefont {Barry}},
  \bibinfo {author} {\bibfnamefont {H.~Y.}\ \bibnamefont {Park}}, \bibinfo
  {author} {\bibfnamefont {K.-C.}\ \bibnamefont {Su}}, \bibinfo {author}
  {\bibfnamefont {Z.}~\bibnamefont {Katz}}, \bibinfo {author} {\bibfnamefont
  {B.~P.}\ \bibnamefont {English}}, \bibinfo {author} {\bibfnamefont
  {A.}~\bibnamefont {Dey}}, \bibinfo {author} {\bibfnamefont {K.}~\bibnamefont
  {Pan}}, \bibinfo {author} {\bibfnamefont {I.~M.}\ \bibnamefont {Cheeseman}},
  \bibinfo {author} {\bibfnamefont {R.~H.}\ \bibnamefont {Singer}},  \emph
  {et~al.},\ }\href@noop {} {\bibfield  {journal} {\bibinfo  {journal} {Nat.
  Methods}\ }\textbf {\bibinfo {volume} {12}},\ \bibinfo {pages} {838}
  (\bibinfo {year} {2015})}\BibitemShut {NoStop}%
\bibitem [{\citenamefont {Mo}\ \emph {et~al.}(2022)\citenamefont {Mo},
  \citenamefont {Wu}, \citenamefont {Yang}, \citenamefont {Liu},\ and\
  \citenamefont {Liao}}]{Mo}%
  \BibitemOpen
  \bibfield  {author} {\bibinfo {author} {\bibfnamefont {Y.}~\bibnamefont
  {Mo}}, \bibinfo {author} {\bibfnamefont {Y.}~\bibnamefont {Wu}}, \bibinfo
  {author} {\bibfnamefont {X.}~\bibnamefont {Yang}}, \bibinfo {author}
  {\bibfnamefont {F.}~\bibnamefont {Liu}}, \ and\ \bibinfo {author}
  {\bibfnamefont {Y.}~\bibnamefont {Liao}},\ }\href@noop {} {\bibfield
  {journal} {\bibinfo  {journal} {Neurocomputing}\ }\textbf {\bibinfo {volume}
  {493}},\ \bibinfo {pages} {626} (\bibinfo {year} {2022})}\BibitemShut
  {NoStop}%
\bibitem [{\citenamefont {Yu}\ and\ \citenamefont
  {Koltun}(2016)}]{yu2015multi}%
  \BibitemOpen
  \bibfield  {author} {\bibinfo {author} {\bibfnamefont {F.}~\bibnamefont
  {Yu}}\ and\ \bibinfo {author} {\bibfnamefont {V.}~\bibnamefont {Koltun}},\
  }in\ \href@noop {} {\emph {\bibinfo {booktitle} {4th International Conference
  on Learning Representations, {ICLR}}}}\ (\bibinfo {year} {2016})\BibitemShut
  {NoStop}%
\bibitem [{\citenamefont {He}\ \emph {et~al.}(2017)\citenamefont {He},
  \citenamefont {Gkioxari}, \citenamefont {Dollor},\ and\ \citenamefont
  {Girshick}}]{He_2017_ICCV1}%
  \BibitemOpen
  \bibfield  {author} {\bibinfo {author} {\bibfnamefont {K.}~\bibnamefont
  {He}}, \bibinfo {author} {\bibfnamefont {G.}~\bibnamefont {Gkioxari}},
  \bibinfo {author} {\bibfnamefont {P.}~\bibnamefont {Dollor}}, \ and\ \bibinfo
  {author} {\bibfnamefont {R.}~\bibnamefont {Girshick}},\ }in\ \href@noop {}
  {\emph {\bibinfo {booktitle} {Proceedings of the IEEE International
  Conference on Computer Vision (ICCV)}}}\ (\bibinfo  {publisher} {IEEE
  Computer Society},\ \bibinfo {address} {Venice},\ \bibinfo {year} {2017})\
  pp.\ \bibinfo {pages} {2961--2969}\BibitemShut {NoStop}%
\bibitem [{\citenamefont {Lin}\ \emph {et~al.}(2017)\citenamefont {Lin},
  \citenamefont {Doll{\'a}r}, \citenamefont {Girshick}, \citenamefont {He},
  \citenamefont {Hariharan},\ and\ \citenamefont {Belongie}}]{Lin_2017_CVPR1}%
  \BibitemOpen
  \bibfield  {author} {\bibinfo {author} {\bibfnamefont {T.-Y.}\ \bibnamefont
  {Lin}}, \bibinfo {author} {\bibfnamefont {P.}~\bibnamefont {Doll{\'a}r}},
  \bibinfo {author} {\bibfnamefont {R.}~\bibnamefont {Girshick}}, \bibinfo
  {author} {\bibfnamefont {K.}~\bibnamefont {He}}, \bibinfo {author}
  {\bibfnamefont {B.}~\bibnamefont {Hariharan}}, \ and\ \bibinfo {author}
  {\bibfnamefont {S.}~\bibnamefont {Belongie}},\ }in\ \href@noop {} {\emph
  {\bibinfo {booktitle} {Proceedings of the IEEE Conference on Computer Vision
  and Pattern Recognition (CVPR)}}}\ (\bibinfo  {publisher} {IEEE Computer
  Society},\ \bibinfo {address} {Honolulu, Hawaii},\ \bibinfo {year} {2017})\
  pp.\ \bibinfo {pages} {2117--2125}\BibitemShut {NoStop}%
\bibitem [{\citenamefont {Badrinarayanan}\ \emph {et~al.}(2017)\citenamefont
  {Badrinarayanan}, \citenamefont {Kendall},\ and\ \citenamefont
  {Cipolla}}]{Badrinarayanan}%
  \BibitemOpen
  \bibfield  {author} {\bibinfo {author} {\bibfnamefont {V.}~\bibnamefont
  {Badrinarayanan}}, \bibinfo {author} {\bibfnamefont {A.}~\bibnamefont
  {Kendall}}, \ and\ \bibinfo {author} {\bibfnamefont {R.}~\bibnamefont
  {Cipolla}},\ }\href@noop {} {\bibfield  {journal} {\bibinfo  {journal} {IEEE
  Trans. Pattern Anal. Mach. Intell.}\ }\textbf {\bibinfo {volume} {39}},\
  \bibinfo {pages} {2481} (\bibinfo {year} {2017})}\BibitemShut {NoStop}%
\bibitem [{\citenamefont {Li}\ \emph {et~al.}(2009)\citenamefont {Li},
  \citenamefont {Socher},\ and\ \citenamefont {Li}}]{Li-Jia}%
  \BibitemOpen
  \bibfield  {author} {\bibinfo {author} {\bibfnamefont {L.-J.}\ \bibnamefont
  {Li}}, \bibinfo {author} {\bibfnamefont {R.}~\bibnamefont {Socher}}, \ and\
  \bibinfo {author} {\bibfnamefont {F.-F.}\ \bibnamefont {Li}},\ }in\
  \href@noop {} {\emph {\bibinfo {booktitle} {2009 IEEE Conference on Computer
  Vision and Pattern Recognition}}}\ (\bibinfo  {publisher} {IEEE Computer
  Society},\ \bibinfo {address} {Miami, Florida},\ \bibinfo {year} {2009})\
  pp.\ \bibinfo {pages} {2036--2043}\BibitemShut {NoStop}%
\bibitem [{\citenamefont {Ronneberger}\ \emph {et~al.}(2015)\citenamefont
  {Ronneberger}, \citenamefont {Fischer},\ and\ \citenamefont {Brox}}]{Olaf}%
  \BibitemOpen
  \bibfield  {author} {\bibinfo {author} {\bibfnamefont {O.}~\bibnamefont
  {Ronneberger}}, \bibinfo {author} {\bibfnamefont {P.}~\bibnamefont
  {Fischer}}, \ and\ \bibinfo {author} {\bibfnamefont {T.}~\bibnamefont
  {Brox}},\ }in\ \href@noop {} {\emph {\bibinfo {booktitle} {Medical Image
  Computing and Computer-Assisted Intervention -- MICCAI 2015}}}\ (\bibinfo
  {publisher} {Springer International Publishing},\ \bibinfo {address} {Cham},\
  \bibinfo {year} {2015})\ pp.\ \bibinfo {pages} {234--241}\BibitemShut
  {NoStop}%
\bibitem [{\citenamefont {Antonelli}\ \emph {et~al.}(2022)\citenamefont
  {Antonelli}, \citenamefont {Reinke}, \citenamefont {Bakas}, \citenamefont
  {Farahani}, \citenamefont {Kopp-Schneider}, \citenamefont {Landman},
  \citenamefont {Litjens}, \citenamefont {Menze}, \citenamefont {Ronneberger},
  \citenamefont {Summers} \emph {et~al.}}]{Antonelli}%
  \BibitemOpen
  \bibfield  {author} {\bibinfo {author} {\bibfnamefont {M.}~\bibnamefont
  {Antonelli}}, \bibinfo {author} {\bibfnamefont {A.}~\bibnamefont {Reinke}},
  \bibinfo {author} {\bibfnamefont {S.}~\bibnamefont {Bakas}}, \bibinfo
  {author} {\bibfnamefont {K.}~\bibnamefont {Farahani}}, \bibinfo {author}
  {\bibfnamefont {A.}~\bibnamefont {Kopp-Schneider}}, \bibinfo {author}
  {\bibfnamefont {B.~A.}\ \bibnamefont {Landman}}, \bibinfo {author}
  {\bibfnamefont {G.}~\bibnamefont {Litjens}}, \bibinfo {author} {\bibfnamefont
  {B.}~\bibnamefont {Menze}}, \bibinfo {author} {\bibfnamefont
  {O.}~\bibnamefont {Ronneberger}}, \bibinfo {author} {\bibfnamefont {R.~M.}\
  \bibnamefont {Summers}},  \emph {et~al.},\ }\href@noop {} {\bibfield
  {journal} {\bibinfo  {journal} {Nat. Commun.}\ }\textbf {\bibinfo {volume}
  {13}},\ \bibinfo {pages} {4128} (\bibinfo {year} {2022})}\BibitemShut
  {NoStop}%
\bibitem [{\citenamefont {Du}\ \emph {et~al.}(2020)\citenamefont {Du},
  \citenamefont {Cao}, \citenamefont {Liang}, \citenamefont {Chen},\ and\
  \citenamefont {Zhan}}]{Getao}%
  \BibitemOpen
  \bibfield  {author} {\bibinfo {author} {\bibfnamefont {G.}~\bibnamefont
  {Du}}, \bibinfo {author} {\bibfnamefont {X.}~\bibnamefont {Cao}}, \bibinfo
  {author} {\bibfnamefont {J.}~\bibnamefont {Liang}}, \bibinfo {author}
  {\bibfnamefont {X.}~\bibnamefont {Chen}}, \ and\ \bibinfo {author}
  {\bibfnamefont {Y.}~\bibnamefont {Zhan}},\ }\href@noop {} {\bibfield
  {journal} {\bibinfo  {journal} {J. Imaging Sci. Technol.}\ }\textbf {\bibinfo
  {volume} {64}},\ \bibinfo {pages} {020508} (\bibinfo {year}
  {2020})}\BibitemShut {NoStop}%
\bibitem [{\citenamefont {Granik}\ \emph {et~al.}(2019)\citenamefont {Granik},
  \citenamefont {Weiss}, \citenamefont {Nehme}, \citenamefont {Levin},
  \citenamefont {Chein}, \citenamefont {Perlson}, \citenamefont {Roichman},\
  and\ \citenamefont {Shechtman}}]{Granik}%
  \BibitemOpen
  \bibfield  {author} {\bibinfo {author} {\bibfnamefont {N.}~\bibnamefont
  {Granik}}, \bibinfo {author} {\bibfnamefont {L.~E.}\ \bibnamefont {Weiss}},
  \bibinfo {author} {\bibfnamefont {E.}~\bibnamefont {Nehme}}, \bibinfo
  {author} {\bibfnamefont {M.}~\bibnamefont {Levin}}, \bibinfo {author}
  {\bibfnamefont {M.}~\bibnamefont {Chein}}, \bibinfo {author} {\bibfnamefont
  {E.}~\bibnamefont {Perlson}}, \bibinfo {author} {\bibfnamefont
  {Y.}~\bibnamefont {Roichman}}, \ and\ \bibinfo {author} {\bibfnamefont
  {Y.}~\bibnamefont {Shechtman}},\ }\href@noop {} {\bibfield  {journal}
  {\bibinfo  {journal} {Biophys. J.}\ }\textbf {\bibinfo {volume} {117}},\
  \bibinfo {pages} {185} (\bibinfo {year} {2019})}\BibitemShut {NoStop}%
\bibitem [{\citenamefont {Mu{\~n}oz-Gil}\ \emph
  {et~al.}(2021{\natexlab{c}})\citenamefont {Mu{\~n}oz-Gil}, \citenamefont
  {Requena}, \citenamefont {Volpe}, \citenamefont {Garcia-March},\ and\
  \citenamefont {Manzo}}]{Munoz-Gil2}%
  \BibitemOpen
  \bibfield  {author} {\bibinfo {author} {\bibfnamefont {G.}~\bibnamefont
  {Mu{\~n}oz-Gil}}, \bibinfo {author} {\bibfnamefont {B.}~\bibnamefont
  {Requena}}, \bibinfo {author} {\bibfnamefont {G.}~\bibnamefont {Volpe}},
  \bibinfo {author} {\bibfnamefont {M.~A.}\ \bibnamefont {Garcia-March}}, \
  and\ \bibinfo {author} {\bibfnamefont {C.}~\bibnamefont {Manzo}},\ }\href
  {\doibase 10.5281/zenodo.4775311} {\enquote {\bibinfo {title}
  {Andichallenge/andi\_datasets: Challenge 2020 release (v.1.0). zenodo.}}\ }
  (\bibinfo {year} {2021}{\natexlab{c}})\BibitemShut {NoStop}%
\bibitem [{\citenamefont {Mandelbrot}\ and\ \citenamefont
  {Ness}(1968)}]{Mandelbrot}%
  \BibitemOpen
  \bibfield  {author} {\bibinfo {author} {\bibfnamefont {B.~B.}\ \bibnamefont
  {Mandelbrot}}\ and\ \bibinfo {author} {\bibfnamefont {J.~W.~V.}\ \bibnamefont
  {Ness}},\ }\href@noop {} {\bibfield  {journal} {\bibinfo  {journal} {SIAM
  Rev.}\ }\textbf {\bibinfo {volume} {10}},\ \bibinfo {pages} {422} (\bibinfo
  {year} {1968})}\BibitemShut {NoStop}%
\bibitem [{\citenamefont {Chein}\ \emph {et~al.}(2019)\citenamefont {Chein},
  \citenamefont {Perlson},\ and\ \citenamefont {Roichman}}]{Chein}%
  \BibitemOpen
  \bibfield  {author} {\bibinfo {author} {\bibfnamefont {M.}~\bibnamefont
  {Chein}}, \bibinfo {author} {\bibfnamefont {E.}~\bibnamefont {Perlson}}, \
  and\ \bibinfo {author} {\bibfnamefont {Y.}~\bibnamefont {Roichman}},\
  }\href@noop {} {\bibfield  {journal} {\bibinfo  {journal} {Biophys. J.}\
  }\textbf {\bibinfo {volume} {117}},\ \bibinfo {pages} {810} (\bibinfo {year}
  {2019})}\BibitemShut {NoStop}%
\bibitem [{\citenamefont {Scher}\ and\ \citenamefont {Montroll}(1975)}]{Scher}%
  \BibitemOpen
  \bibfield  {author} {\bibinfo {author} {\bibfnamefont {H.}~\bibnamefont
  {Scher}}\ and\ \bibinfo {author} {\bibfnamefont {E.~W.}\ \bibnamefont
  {Montroll}},\ }\href@noop {} {\bibfield  {journal} {\bibinfo  {journal}
  {Phys. Rev. B}\ }\textbf {\bibinfo {volume} {12}},\ \bibinfo {pages} {2455}
  (\bibinfo {year} {1975})}\BibitemShut {NoStop}%
\bibitem [{\citenamefont {Massignan}\ \emph {et~al.}(2014)\citenamefont
  {Massignan}, \citenamefont {Manzo}, \citenamefont {Torreno-Pina},
  \citenamefont {Garc\'{\i}a-Parajo}, \citenamefont {Lewenstein},\ and\
  \citenamefont {Lapeyre}}]{Massignan}%
  \BibitemOpen
  \bibfield  {author} {\bibinfo {author} {\bibfnamefont {P.}~\bibnamefont
  {Massignan}}, \bibinfo {author} {\bibfnamefont {C.}~\bibnamefont {Manzo}},
  \bibinfo {author} {\bibfnamefont {J.~A.}\ \bibnamefont {Torreno-Pina}},
  \bibinfo {author} {\bibfnamefont {M.~F.}\ \bibnamefont {Garc\'{\i}a-Parajo}},
  \bibinfo {author} {\bibfnamefont {M.}~\bibnamefont {Lewenstein}}, \ and\
  \bibinfo {author} {\bibfnamefont {G.~J.}\ \bibnamefont {Lapeyre}},\
  }\href@noop {} {\bibfield  {journal} {\bibinfo  {journal} {Phys. Rev. Lett.}\
  }\textbf {\bibinfo {volume} {112}},\ \bibinfo {pages} {150603} (\bibinfo
  {year} {2014})}\BibitemShut {NoStop}%
\bibitem [{\citenamefont {Klafter}\ and\ \citenamefont
  {Zumofen}(1994)}]{Klafter1}%
  \BibitemOpen
  \bibfield  {author} {\bibinfo {author} {\bibfnamefont {J.}~\bibnamefont
  {Klafter}}\ and\ \bibinfo {author} {\bibfnamefont {G.}~\bibnamefont
  {Zumofen}},\ }\href@noop {} {\bibfield  {journal} {\bibinfo  {journal} {Phys.
  Rev. E}\ }\textbf {\bibinfo {volume} {49}},\ \bibinfo {pages} {4873}
  (\bibinfo {year} {1994})}\BibitemShut {NoStop}%
\bibitem [{\citenamefont {Lim}\ and\ \citenamefont {Muniandy}(2002)}]{Lim}%
  \BibitemOpen
  \bibfield  {author} {\bibinfo {author} {\bibfnamefont {S.~C.}\ \bibnamefont
  {Lim}}\ and\ \bibinfo {author} {\bibfnamefont {S.~V.}\ \bibnamefont
  {Muniandy}},\ }\href@noop {} {\bibfield  {journal} {\bibinfo  {journal}
  {Phys. Rev. E}\ }\textbf {\bibinfo {volume} {66}},\ \bibinfo {pages} {021114}
  (\bibinfo {year} {2002})}\BibitemShut {NoStop}%
\bibitem [{\citenamefont {Dice}(1945)}]{Dice}%
  \BibitemOpen
  \bibfield  {author} {\bibinfo {author} {\bibfnamefont {L.~R.}\ \bibnamefont
  {Dice}},\ }\href@noop {} {\bibfield  {journal} {\bibinfo  {journal}
  {Ecology}\ }\textbf {\bibinfo {volume} {26}},\ \bibinfo {pages} {297}
  (\bibinfo {year} {1945})}\BibitemShut {NoStop}%
\bibitem [{\citenamefont {Hochreiter}\ and\ \citenamefont
  {Schmidhuber}(1997)}]{Hochreiter}%
  \BibitemOpen
  \bibfield  {author} {\bibinfo {author} {\bibfnamefont {S.}~\bibnamefont
  {Hochreiter}}\ and\ \bibinfo {author} {\bibfnamefont {J.}~\bibnamefont
  {Schmidhuber}},\ }\href@noop {} {\bibfield  {journal} {\bibinfo  {journal}
  {Neural Comput.}\ }\textbf {\bibinfo {volume} {9}},\ \bibinfo {pages} {1735}
  (\bibinfo {year} {1997})}\BibitemShut {NoStop}%
\bibitem [{\citenamefont {Cho}\ \emph {et~al.}(2014)\citenamefont {Cho},
  \citenamefont {van Merri{\"e}nboer}, \citenamefont {Gulcehre}, \citenamefont
  {Bahdanau}, \citenamefont {Bougares}, \citenamefont {Schwenk},\ and\
  \citenamefont {Bengio}}]{Kyunghyun}%
  \BibitemOpen
  \bibfield  {author} {\bibinfo {author} {\bibfnamefont {K.}~\bibnamefont
  {Cho}}, \bibinfo {author} {\bibfnamefont {B.}~\bibnamefont {van
  Merri{\"e}nboer}}, \bibinfo {author} {\bibfnamefont {C.}~\bibnamefont
  {Gulcehre}}, \bibinfo {author} {\bibfnamefont {D.}~\bibnamefont {Bahdanau}},
  \bibinfo {author} {\bibfnamefont {F.}~\bibnamefont {Bougares}}, \bibinfo
  {author} {\bibfnamefont {H.}~\bibnamefont {Schwenk}}, \ and\ \bibinfo
  {author} {\bibfnamefont {Y.}~\bibnamefont {Bengio}},\ }in\ \href@noop {}
  {\emph {\bibinfo {booktitle} {Proceedings of the 2014 Conference on Empirical
  Methods in Natural Language Processing ({EMNLP})}}}\ (\bibinfo  {publisher}
  {Association for Computational Linguistics},\ \bibinfo {address} {Doha,
  Qatar},\ \bibinfo {year} {2014})\ pp.\ \bibinfo {pages}
  {1724--1734}\BibitemShut {NoStop}%
\bibitem [{\citenamefont {Vaswani}\ \emph {et~al.}(2017)\citenamefont
  {Vaswani}, \citenamefont {Shazeer}, \citenamefont {Parmar}, \citenamefont
  {Uszkoreit}, \citenamefont {Jones}, \citenamefont {Gomez}, \citenamefont
  {Kaiser},\ and\ \citenamefont {Polosukhin}}]{Ashish}%
  \BibitemOpen
  \bibfield  {author} {\bibinfo {author} {\bibfnamefont {A.}~\bibnamefont
  {Vaswani}}, \bibinfo {author} {\bibfnamefont {N.}~\bibnamefont {Shazeer}},
  \bibinfo {author} {\bibfnamefont {N.}~\bibnamefont {Parmar}}, \bibinfo
  {author} {\bibfnamefont {J.}~\bibnamefont {Uszkoreit}}, \bibinfo {author}
  {\bibfnamefont {L.}~\bibnamefont {Jones}}, \bibinfo {author} {\bibfnamefont
  {A.~N.}\ \bibnamefont {Gomez}}, \bibinfo {author} {\bibfnamefont
  {L.}~\bibnamefont {Kaiser}}, \ and\ \bibinfo {author} {\bibfnamefont
  {I.}~\bibnamefont {Polosukhin}},\ }in\ \href@noop {} {\emph {\bibinfo
  {booktitle} {Proceedings of the 31st International Conference on Neural
  Information Processing Systems}}},\ Vol.~\bibinfo {volume} {30}\ (\bibinfo
  {publisher} {Curran Associates Inc.},\ \bibinfo {address} {Red Hook, NY},\
  \bibinfo {year} {2017})\ p.\ \bibinfo {pages} {6000–6010}\BibitemShut
  {NoStop}%
\bibitem [{\citenamefont {van~den Oord}\ \emph {et~al.}(2016)\citenamefont
  {van~den Oord}, \citenamefont {Dieleman}, \citenamefont {Zen}, \citenamefont
  {Simonyan}, \citenamefont {Vinyals}, \citenamefont {Graves}, \citenamefont
  {Kalchbrenner}, \citenamefont {Senior},\ and\ \citenamefont
  {Kavukcuoglu}}]{Aaron}%
  \BibitemOpen
  \bibfield  {author} {\bibinfo {author} {\bibfnamefont {A.}~\bibnamefont
  {van~den Oord}}, \bibinfo {author} {\bibfnamefont {S.}~\bibnamefont
  {Dieleman}}, \bibinfo {author} {\bibfnamefont {H.}~\bibnamefont {Zen}},
  \bibinfo {author} {\bibfnamefont {K.}~\bibnamefont {Simonyan}}, \bibinfo
  {author} {\bibfnamefont {O.}~\bibnamefont {Vinyals}}, \bibinfo {author}
  {\bibfnamefont {A.}~\bibnamefont {Graves}}, \bibinfo {author} {\bibfnamefont
  {N.}~\bibnamefont {Kalchbrenner}}, \bibinfo {author} {\bibfnamefont
  {A.}~\bibnamefont {Senior}}, \ and\ \bibinfo {author} {\bibfnamefont
  {K.}~\bibnamefont {Kavukcuoglu}},\ }\href@noop {} {\bibfield  {journal}
  {\bibinfo  {journal} {arXiv:1609.03499}\ } (\bibinfo {year}
  {2016})}\BibitemShut {NoStop}%
\bibitem [{\citenamefont {Manzo}\ and\ \citenamefont
  {Garcia-Parajo}(2015)}]{Manzo1}%
  \BibitemOpen
  \bibfield  {author} {\bibinfo {author} {\bibfnamefont {C.}~\bibnamefont
  {Manzo}}\ and\ \bibinfo {author} {\bibfnamefont {M.~F.}\ \bibnamefont
  {Garcia-Parajo}},\ }\href@noop {} {\bibfield  {journal} {\bibinfo  {journal}
  {Rep. Prog. Phys.}\ }\textbf {\bibinfo {volume} {78}},\ \bibinfo {pages}
  {124601} (\bibinfo {year} {2015})}\BibitemShut {NoStop}%
\bibitem [{\citenamefont {Shen}\ \emph {et~al.}(2017)\citenamefont {Shen},
  \citenamefont {Tauzin}, \citenamefont {Baiyasi}, \citenamefont {Wang},
  \citenamefont {Moringo}, \citenamefont {Shuang},\ and\ \citenamefont
  {Landes}}]{Shen}%
  \BibitemOpen
  \bibfield  {author} {\bibinfo {author} {\bibfnamefont {H.}~\bibnamefont
  {Shen}}, \bibinfo {author} {\bibfnamefont {L.~J.}\ \bibnamefont {Tauzin}},
  \bibinfo {author} {\bibfnamefont {R.}~\bibnamefont {Baiyasi}}, \bibinfo
  {author} {\bibfnamefont {W.}~\bibnamefont {Wang}}, \bibinfo {author}
  {\bibfnamefont {N.}~\bibnamefont {Moringo}}, \bibinfo {author} {\bibfnamefont
  {B.}~\bibnamefont {Shuang}}, \ and\ \bibinfo {author} {\bibfnamefont {C.~F.}\
  \bibnamefont {Landes}},\ }\href@noop {} {\bibfield  {journal} {\bibinfo
  {journal} {Chem. Rev.}\ }\textbf {\bibinfo {volume} {117}},\ \bibinfo {pages}
  {7331} (\bibinfo {year} {2017})}\BibitemShut {NoStop}%
\bibitem [{\citenamefont {Qian}\ \emph {et~al.}(1991)\citenamefont {Qian},
  \citenamefont {Sheetz},\ and\ \citenamefont {Elson}}]{Qian}%
  \BibitemOpen
  \bibfield  {author} {\bibinfo {author} {\bibfnamefont {H.}~\bibnamefont
  {Qian}}, \bibinfo {author} {\bibfnamefont {M.}~\bibnamefont {Sheetz}}, \ and\
  \bibinfo {author} {\bibfnamefont {E.}~\bibnamefont {Elson}},\ }\href@noop {}
  {\bibfield  {journal} {\bibinfo  {journal} {Biophys. J.}\ }\textbf {\bibinfo
  {volume} {60}},\ \bibinfo {pages} {910} (\bibinfo {year} {1991})}\BibitemShut
  {NoStop}%
\bibitem [{\citenamefont {Saxton}(2008)}]{Saxton}%
  \BibitemOpen
  \bibfield  {author} {\bibinfo {author} {\bibfnamefont {M.~J.}\ \bibnamefont
  {Saxton}},\ }\href@noop {} {\bibfield  {journal} {\bibinfo  {journal} {Nat.
  Methods}\ }\textbf {\bibinfo {volume} {5}},\ \bibinfo {pages} {671} (\bibinfo
  {year} {2008})}\BibitemShut {NoStop}%
\bibitem [{\citenamefont {Torreno-Pina}\ \emph {et~al.}(2016)\citenamefont
  {Torreno-Pina}, \citenamefont {Manzo},\ and\ \citenamefont
  {Garcia-Parajo}}]{Torreno-Pina}%
  \BibitemOpen
  \bibfield  {author} {\bibinfo {author} {\bibfnamefont {J.~A.}\ \bibnamefont
  {Torreno-Pina}}, \bibinfo {author} {\bibfnamefont {C.}~\bibnamefont {Manzo}},
  \ and\ \bibinfo {author} {\bibfnamefont {M.~F.}\ \bibnamefont
  {Garcia-Parajo}},\ }\href@noop {} {\bibfield  {journal} {\bibinfo  {journal}
  {J. Phys. D: Appl. Phys.}\ }\textbf {\bibinfo {volume} {49}},\ \bibinfo
  {pages} {104002} (\bibinfo {year} {2016})}\BibitemShut {NoStop}%
\bibitem [{\citenamefont {Laurent}\ \emph {et~al.}(2022)\citenamefont
  {Laurent}, \citenamefont {Verdier}, \citenamefont {Duval}, \citenamefont
  {Serov}, \citenamefont {Vestergaard},\ and\ \citenamefont
  {Masson}}]{TRamWAy}%
  \BibitemOpen
  \bibfield  {author} {\bibinfo {author} {\bibfnamefont {F.}~\bibnamefont
  {Laurent}}, \bibinfo {author} {\bibfnamefont {H.}~\bibnamefont {Verdier}},
  \bibinfo {author} {\bibfnamefont {M.}~\bibnamefont {Duval}}, \bibinfo
  {author} {\bibfnamefont {A.}~\bibnamefont {Serov}}, \bibinfo {author}
  {\bibfnamefont {C.~L.}\ \bibnamefont {Vestergaard}}, \ and\ \bibinfo {author}
  {\bibfnamefont {J.-B.}\ \bibnamefont {Masson}},\ }\href@noop {} {\bibfield
  {journal} {\bibinfo  {journal} {Bioinformatics}\ }\textbf {\bibinfo {volume}
  {38}},\ \bibinfo {pages} {3149} (\bibinfo {year} {2022})}\BibitemShut
  {NoStop}%
\bibitem [{\citenamefont {Vi{\~n}ales}\ and\ \citenamefont
  {Desp{\'o}sito}(2007)}]{2007anomalous}%
  \BibitemOpen
  \bibfield  {author} {\bibinfo {author} {\bibfnamefont {A.~D.}\ \bibnamefont
  {Vi{\~n}ales}}\ and\ \bibinfo {author} {\bibfnamefont {M.~A.}\ \bibnamefont
  {Desp{\'o}sito}},\ }\href@noop {} {\bibfield  {journal} {\bibinfo  {journal}
  {Phys. Rev. E}\ }\textbf {\bibinfo {volume} {75}},\ \bibinfo {pages} {042102}
  (\bibinfo {year} {2007})}\BibitemShut {NoStop}%
\bibitem [{fbm()}]{fbm}%
  \BibitemOpen
  \href@noop {} {\bibinfo  {journal} {\url{https://github.com/crflynn/fbm},
  2019}\ }\BibitemShut {NoStop}%
\bibitem [{mit()}]{mittag-leffler}%
  \BibitemOpen
\bibfield  {journal} {  }\href@noop {} {\bibinfo  {journal}
  {\url{https://github.com/khinsen/mittag-leffler}, 2017}\ }\BibitemShut
  {NoStop}%
\bibitem [{\citenamefont {Davies}\ and\ \citenamefont
  {Harte}(1987)}]{davies1987tests}%
  \BibitemOpen
\bibfield  {journal} {  }\bibfield  {author} {\bibinfo {author} {\bibfnamefont
  {R.~B.}\ \bibnamefont {Davies}}\ and\ \bibinfo {author} {\bibfnamefont
  {D.~S.}\ \bibnamefont {Harte}},\ }\href@noop {} {\bibfield  {journal}
  {\bibinfo  {journal} {Biometrika}\ }\textbf {\bibinfo {volume} {74}},\
  \bibinfo {pages} {95} (\bibinfo {year} {1987})}\BibitemShut {NoStop}%
\bibitem [{U-A()}]{U-AnDi}%
  \BibitemOpen
  \href@noop {} {\bibinfo  {journal} {\url{https://github.com/huangzih/U-AnDi},
  2023}\ }\BibitemShut {NoStop}%
\bibitem [{\citenamefont {Allen}\ and\ \citenamefont
  {Tildesley}(1987)}]{Oxford}%
  \BibitemOpen
\bibfield  {journal} {  }\bibfield  {author} {\bibinfo {author} {\bibfnamefont
  {M.~P.}\ \bibnamefont {Allen}}\ and\ \bibinfo {author} {\bibfnamefont
  {D.~J.}\ \bibnamefont {Tildesley}},\ }\href@noop {} {\emph {\bibinfo {title}
  {Computer simulation of liquids}}}\ (\bibinfo  {publisher} {Clarendon,
  Oxford},\ \bibinfo {year} {1987})\BibitemShut {NoStop}%
\bibitem [{\citenamefont {Vi{\~n}ales}\ and\ \citenamefont
  {Paissan}(2014)}]{Velocityauto}%
  \BibitemOpen
  \bibfield  {author} {\bibinfo {author} {\bibfnamefont {A.~D.}\ \bibnamefont
  {Vi{\~n}ales}}\ and\ \bibinfo {author} {\bibfnamefont {G.~H.}\ \bibnamefont
  {Paissan}},\ }\href@noop {} {\bibfield  {journal} {\bibinfo  {journal} {Phys.
  Rev. E}\ }\textbf {\bibinfo {volume} {90}},\ \bibinfo {pages} {062103}
  (\bibinfo {year} {2014})}\BibitemShut {NoStop}%
\bibitem [{\citenamefont {Ioffe}\ and\ \citenamefont {Szegedy}(2015)}]{Ioffe}%
  \BibitemOpen
  \bibfield  {author} {\bibinfo {author} {\bibfnamefont {S.}~\bibnamefont
  {Ioffe}}\ and\ \bibinfo {author} {\bibfnamefont {C.}~\bibnamefont
  {Szegedy}},\ }in\ \href@noop {} {\emph {\bibinfo {booktitle} {Proceedings of
  the 32nd International Conference on International Conference on Machine
  Learning - Volume 37}}}\ (\bibinfo  {publisher} {JMLR.org},\ \bibinfo
  {address} {Lille, France},\ \bibinfo {year} {2015})\ pp.\ \bibinfo {pages}
  {448--456}\BibitemShut {NoStop}%
\end{thebibliography}%

\end{document}